\documentclass{aa}  

\usepackage{graphicx}
\usepackage{txfonts}
\usepackage{subcaption}
\usepackage{tabularx}
\usepackage{ulem}
\usepackage{placeins}
\usepackage{xcolor}
\usepackage{hyperref}
\usepackage[version=4]{mhchem}
\newcommand\standardstate{{\circ\kern-0.420em-}}

\begin{document}

   \title{Revisiting fundamental properties of TiO\textsubscript{2} nanoclusters as condensation seeds in astrophysical environments}

   \subtitle{}

   \author{J.P. Sindel
          \inst{1,2,3,4}
          \and
          D. Gobrecht\inst{4,5}
          \and
          Ch. Helling \inst{1,6}
          \and
          L. Decin \inst{4}
          }

   \institute{Space Research Institute, Austrian Academy of Sciences, Schmiedlstrasse 6, A-8042 Graz, Austria\\
   \email{JanPhilip.Sindel@oeaw.ac.at}
   \and 
   Centre for Exoplanet Science, University of St Andrews, North Haugh, St Andrews, KY169SS, UK
        \and
            SUPA, School of Physics \& Astronomy, University of St Andrews, North Haugh, St Andrews, KY169SS, UK
            \and
             Institute for Astronomy, KU Leuven, Celestijnenlaan 200D, 3001 Leuven, Belgium
            \and 
            Department of Chemistry and Molecular Biology, University of Gothenburg, Kemig\aa rden 4, 412 96 Gothenburg, Sweden
         \and
         TU Graz, Fakult\"at f\"ur Mathematik, Physik und Geod\"asie, Petersgasse 16, 8010 Graz, Austria
         }

   \date{Received -----; accepted -----}

 
  \abstract
   {The formation of inorganic cloud particles takes place in several atmospheric environments including those of warm, hot, rocky and gaseous exoplanets, brown dwarfs, and AGB stars. The cloud particle formation needs to be triggered by the in-situ formation of condensation seeds since it can not be reasonably assumed that such condensation seeds preexist in these chemically complex gas-phase environments.} 
   {We aim to develop a methodology to calculate the thermochemical properties of clusters as key inputs to model the formation of condensation nuclei in gases of changing chemical composition. TiO$_2$ is used as benchmark species for cluster sizes $N$ = 1 - 15. }
   {We create a total of 90000 candidate (TiO$_2$)$_N$ geometries, for cluster sizes $N$ = 3 - 15.  We employ a hierarchical optimisation approach, consisting of a force field description, density functional based tight binding (DFTB) and all-electron density functional theory (DFT)
   to obtain accurate zero-point energies and thermochemical properties for the clusters.}
   {Among 129 functional/basis set combinations we find B3LYP/cc-pVTZ including Grimmes empirical dispersion to perform most accurately with respect to experimentally derived thermochemical properties of the TiO$_2$ molecule. We present a hitherto unreported global minimum candidate for size N = 13. The DFT derived thermochemical cluster data are used to evaluate the nucleation rates for a given temperature-pressure profile of a model hot Jupiter atmosphere. We find that with the updated and refined cluster data, nucleation becomes unfeasible at slightly lower temperatures, raising the lower boundary for seed formation in the atmosphere.}
  {The approach presented in this paper allows to find stable isomers for small (TiO$_2$)$_N$ clusters. The choice of functional and basis set for the all-electron DFT calculations have a measurable impact on the resulting surface tension and nucleation rate and the updated thermochemical data is recommended for future considerations.}
  
   \keywords{Molecular data, astrochemistry, molecular processes, planets and satellites: atmospheres, planets and satellites: gaseous planets}

   \maketitle
%

\section{Introduction}
The quest for the first (or primary) astrophysical 
condensate that triggers the formation of cosmic dust is 
as old as the discovery of 
newly-formed condensates
in 
astrophysical environments 
(e.g. \cite{Blander1967CondensationDust, Andriesse1978TheCarinae, Gail1986TheStars, Patzer1995PrimaryDesiderata, Sloan2009DustAbundances, Goumans2013StardustSiO+TiO2}). More recently, this quest was refreshed by the need to understand the formation of clouds in the chemically diverse atmospheres of extrasolar planets and brown dwarfs (e.g. \cite{Helling2008AAtmospheres, Helling2016The189733b, Lee2018DustRegimes, Samra2020MineralShape, Charnay2021AAriel, Min2020TheRetrieval}).

Cloud particles form when a supersaturated gas condenses on the surface of ultra-small particles.
These nano-sized particles are called cloud condensation nuclei (CCN) \citep{CloudCondensationNuclei}.
The formation of CCN is a crucial step within cloud formation, however its formation rate needs to be determined from first principles, as attempts to derive it from models using the nucleation rate as a free parameter are unable to constrain it accurately \citep{Ormel2019ARCiSModel}.
Several efforts were undertaken to model the formation of CCN from a quantum-mechanical 
bottom-up approach including nucleating species like titanates 
\citep{Jeong2000ElectronicTiSUBx/SUBOSUBy/SUB, Plane2013OnOutflows, Patzer2014AProperties}, SiO \citep{Bromley2016UnderEvaluation}, Fe- and Kr-bearing molecules \citep{Chang2005InorganicStudy, Chang2013SmallSpectra} and alumina \citep{Patzer2005AProperties}, and
Al\textsubscript{2}O\textsubscript{3} \citep{Lam2015AtomisticNanoparticles, Gobrecht2021Bottom-upClusters}. Recently, \cite{Kohn2021DustApproach} introduced a 3D Monte Carlo approach for nucleation of TiO$_2$, that agrees with results from kinetic nucleation approaches reasonably well. Such comparison studies require the knowledge of thermochemical cluster data of the most favourable isomers.

On cool rocky planets, CCN can be formed from external sources, i.e. sulfites from volcanic activity \citep{Andres1998AEmissions}, sea salt from ocean spray, condensing meteoritic dust and dust particles from sand storms. As these sources do not exist for gaseous planets and are not guaranteed to exist for hot rocky planets, 
CCN need to be produced by chemical reactions out of the gas-phase within the atmosphere to allow cloud formation. The existence of clouds and hazes has been predicted by models and has been observed in hot Jupiters, for example HD189733b \citep{Pont2013TheObservations, Barstow2014CloudsSpectrum}, HAT-P-7b \citep{Helling2019UnderstandingMapping}, or WASP-43b \citep{Helling2020MineralWASP-43b}, warm Saturns (e.g. \cite{Nikolov2021Ground-basedWASP-110b}, super earths \citep{Kreidberg2013Clouds1214b} and brown dwarfs \citep{Apai2013HSTVARIATIONS}. The process that is considered to produce CCN in the atmospheres of gaseous planets is the formation of small 
clusters through nucleation. One of the species that has been considered for the formation of CCN in gas giant atmospheres is TiO$_2$ in addition to less refractory species, for example SiO or KCl \citep{Helling2018ExoplanetClouds}. 
Different descriptions have been utilised to describe the formation of  condensation seeds in exoplanet, brown dwarf and AGB star research: Classical nucleation theory \citep{Gail1986TheStars}, modified classical nucleation theory \citep{Gail2013a}, kinetic nucleation networks \citep{Patzer1998DustPhase} and chemical-kinetic nucleation descriptions \citep{Gobrecht2016DustTauri, Boulangier2019DevelopingMixture}. 
All rely at some point in the modelling process on thermochemical data of the species and their small nano-sized clusters. Since experimental data exists often for condensed phases and simple gas phase molecules only, quantum-chemical calculations provide the possibility to address the cluster size space in between. Each cluster size 
appears with different geometrical structures (i.e. its isomers) and it is not a priori known which of the isomers is the most favourable at each reaction step to eventually form a CCN. Experimental input would be required here. In lieu of that, we assume that the thermodynamically most
favourable
isomer takes this role of a key reactant. We therefore search for the isomer at the global
minimum 
in potential energy
for each cluster size, as it ideally will be the configuration that any cluster of that size will relax towards. 
Global minimum candidate isomers were found in
previous studies on 
TiO$_2$ 
clusters \citep{Lamiel-Garcia2017, Berardo2014ModelingDescription, Jeong2000ElectronicTiSUBx/SUBOSUBy/SUB}
as well as the analysis of their thermochemical properties \citep{Lee2015}. In this work, a hierarchical approach is applied, utilising three different levels of complexity: force fields, Density-Functional based Tight Binding (DFTB), and Density Functional Theory (DFT) calculations. This approach is developed to globally search for potential geometries of small 
(TiO$_2$)$_N$ clusters ($N = 3-15$) and their thermochemical properties are analysed using quantum-chemical DFT calculations.
In Section \ref{methods}, we describe the methods used to create possible (TiO$_2$)$_N$ cluster structures and the approximations used to describe their interatomic interactions when searching for a geometry that is 
energetically favourable
i.e. is located at a potential minimum. 
This approach is tested against previous results for small TiO$_2$ clusters $N$=3-6. For the DFT calculations, 129 combinations of functionals and basis-sets are 
benchmarked
against experimental data and we find the B3LYP functional with the cc-pVTZ basis-set including Grimmes empirical dispersion to most accurately predict the potential energies and thermochemical properties of the TiO$_2$ molecule. In Section \ref{sec:results} the results for the different steps in our multi-level approach are evaluated. This section also evaluates the quality of the approach, by comparing the found cluster isomers with known isomers from literature. Section \ref{sec:tio2astrophysics} analyses the impact of the updated cluster potential energies and the related thermochemical data on nucleation rates for a model hot Jupiter atmosphere. Finally, we discuss our results and possible further work in Section \ref{sec:discussion}.
\section{Methods}
\label{methods}

\begin{figure}[ht]
  \label{fig:funnel}
    \includegraphics[width=\linewidth]{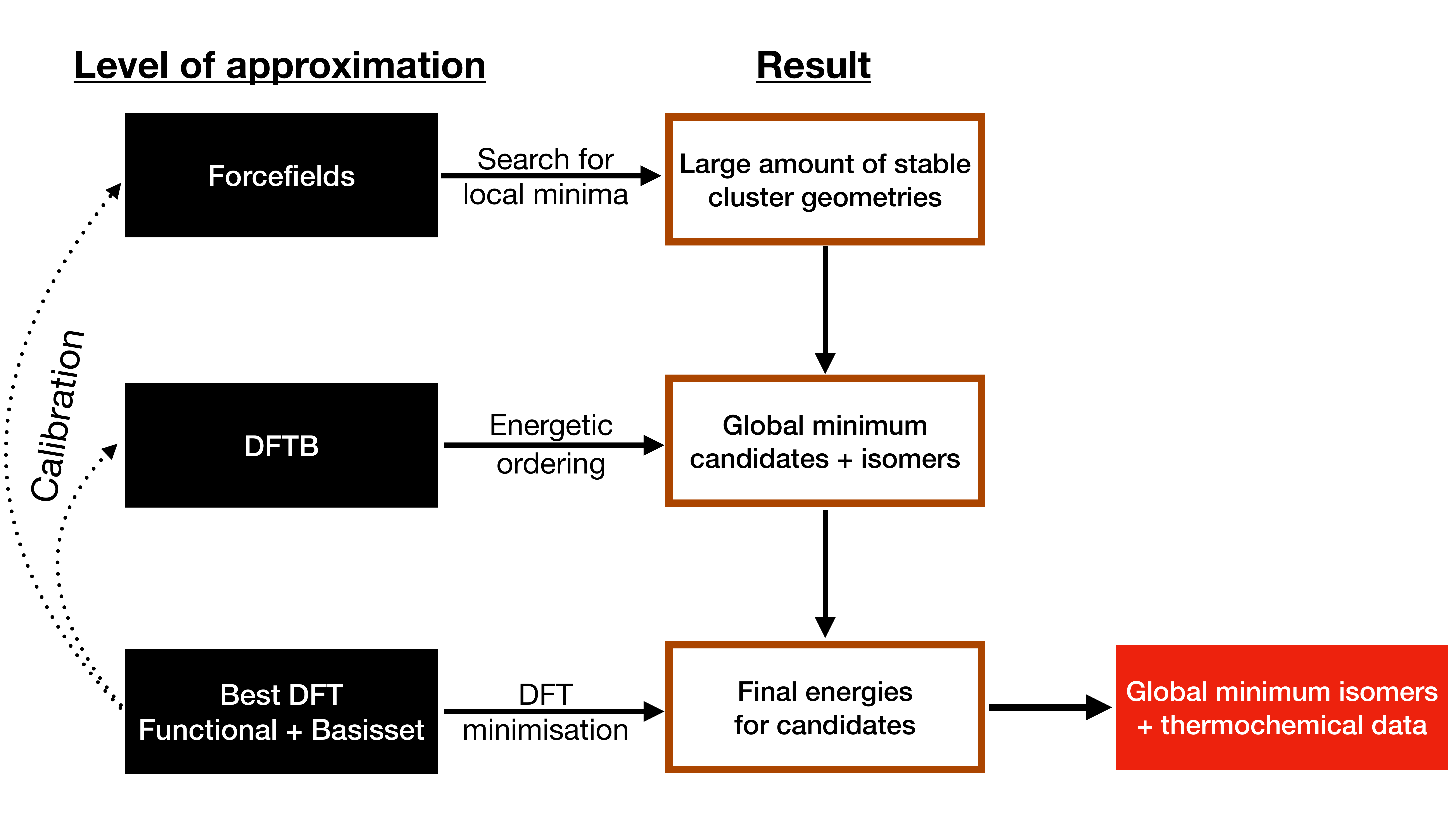}
    \caption{\label{fig:flowdiagram} A hierarchical approach to find global minimum candidate structures for 
    clusters:
1.) The geometries and binding energies
    of small clusters and isomers
    are used to calibrate the accuracy of the force field and DFTB methods. 2.) A force field description of interatomic interactions is used to locally optimise the geometries of a large amount of generated clusters for each size $N$ towards a potential energy minimum.
    3.) The geometries of these locally 
    and semi-classically
    optimised cluster candidates are then further refined in a third step, using DFTB methods, optimising the geometries of the cluster candidates for the lowest possible potential energy. This provides an energetic ordering of the candidate geometries. 4.) The energetically most
    favourable
    candidate geometries are then used as inputs for DFT calculations resulting in the final, and most accurate geometries, binding energies, vibrational and rotational frequencies for each cluster within this approach.}
    
    \end{figure}

   The goal of this paper is to evaluate small clusters of titanium dioxide (TiO\textsubscript{2})$_N$, $N = 1-15$, with regards to their geometry, binding energy and thermochemical properties and the impact of those parameters on cloud nucleation in exoplanet atmospheres. A search is conducted for the most 
   favourable
   isomer of each size $N$ that represents the global minimum of the the potential energy surface (PES), which characterises the energy of the system. Additionally, other
   favourable isomers with potential energies close to the minimum are explored.
   In order to achieve this, the geometries of the clusters are varied, and brought towards a geometry that is located at a potential minimum.  
   A hierarchical method is employed, using three different levels of complexity to describe the 
   PES: force fields, Density-Functional based Tight Binding (DFTB), and Density Functional Theory (DFT) calculations (see Figure \ref{fig:flowdiagram}). Since the DFTB approach takes into account 
   electron-electron interactions of the valence electrons, compared to the purely 
   ionic
   force field description, it better approximates the binding energies for the isomers, resulting in a more accurate energetic ordering of the candidate clusters for each size. For every cluster size N,  candidate cluster isomers are created using different methods and optimised towards a local minimum in potential energy 
   using a basin-hopping algorithm. The energy evaluation in this optimisation procedure is based on a description of the inter-atomic interactions by the Coulomb-Buckingham force field approximation. The resulting cluster geometries are then used as inputs for further geometry optimisation using the DFTB approach, because it describes  
   the PES and interatomic interactions more accurately than the force fields. The DFTB approach is 
   used as a second step only as its higher accuracy comes at higher computational cost, making it more economical to only use
   pre-optimised clusters from the force field description as 
   inputs.
    Finally the candidates at the lowest minima in the potential, which in turn have the 
   highest
   binding energies, from this second step are used as inputs for all-electron DFT optimisations. With this approach, we aim to achieve highly accurate geometries and energies for the thermodynamically most stable
   isomers of each cluster size $N$.

\subsection{Construction of seed structures for cluster geometries}
\label{section:seedcreation}

\begin{figure}
    \centering
     \begin{subfigure}[b]{0.49\hsize}
         \centering
         \includegraphics[width=\hsize]{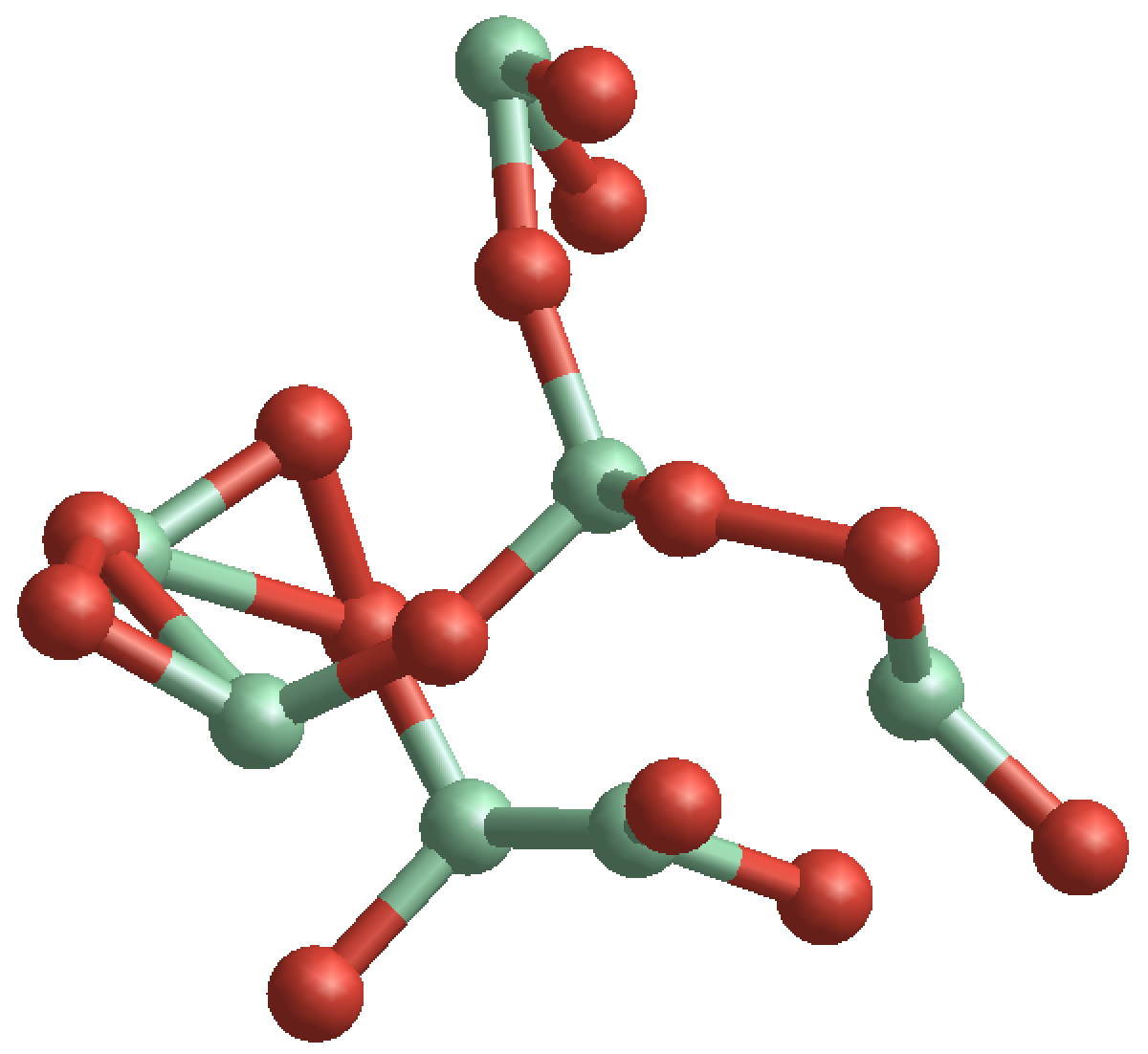}
         \caption{}
         \label{fig:seed_random}
     \end{subfigure}
     \hfill
     \begin{subfigure}[b]{0.49\hsize}
         \centering
         \includegraphics[width=\hsize]{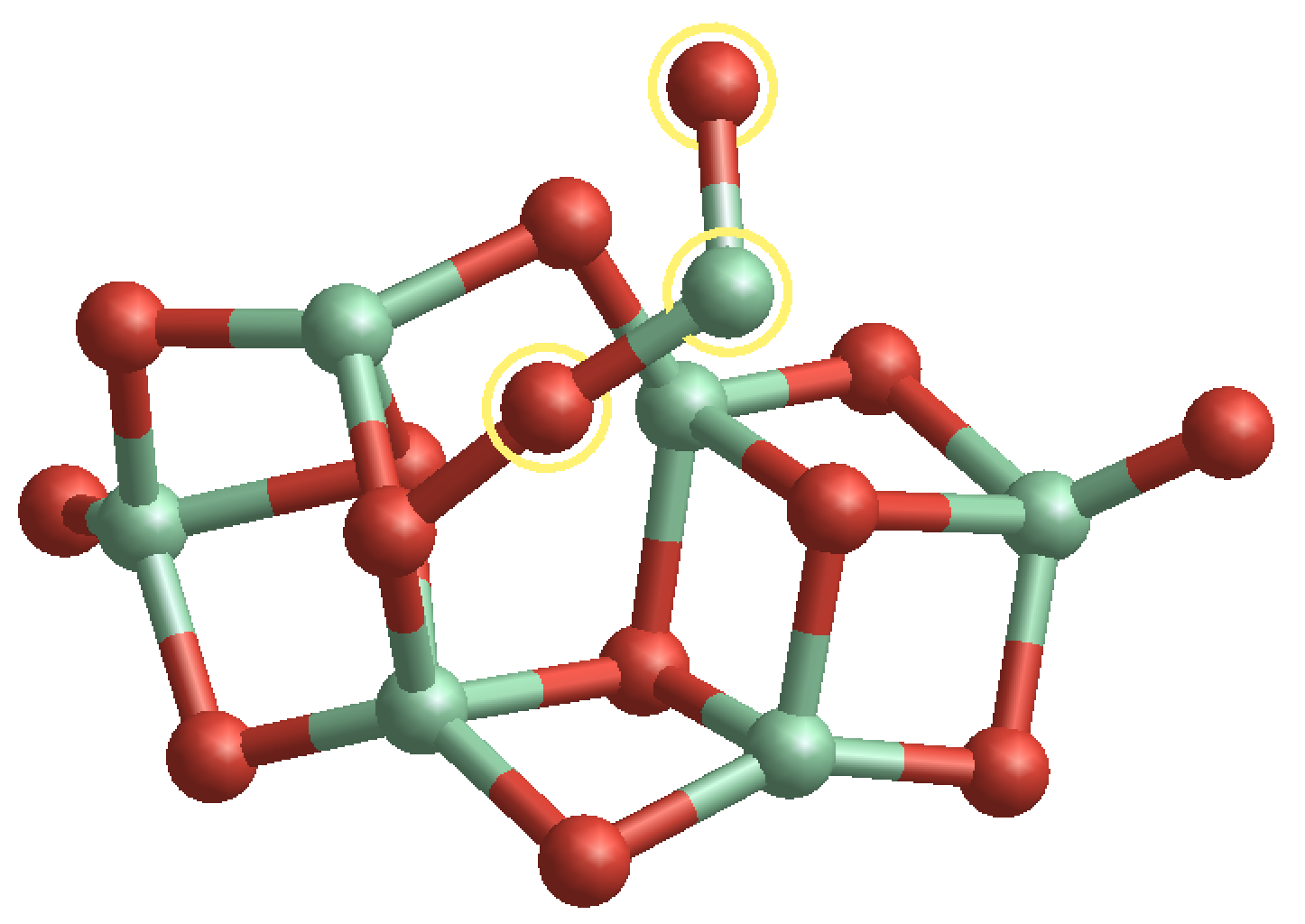}
         \caption{}
         \label{fig:seed_knownplusone}
     \end{subfigure}
     \hfill
     \begin{subfigure}[b]{0.49\hsize}
         \centering
         \includegraphics[width=\hsize]{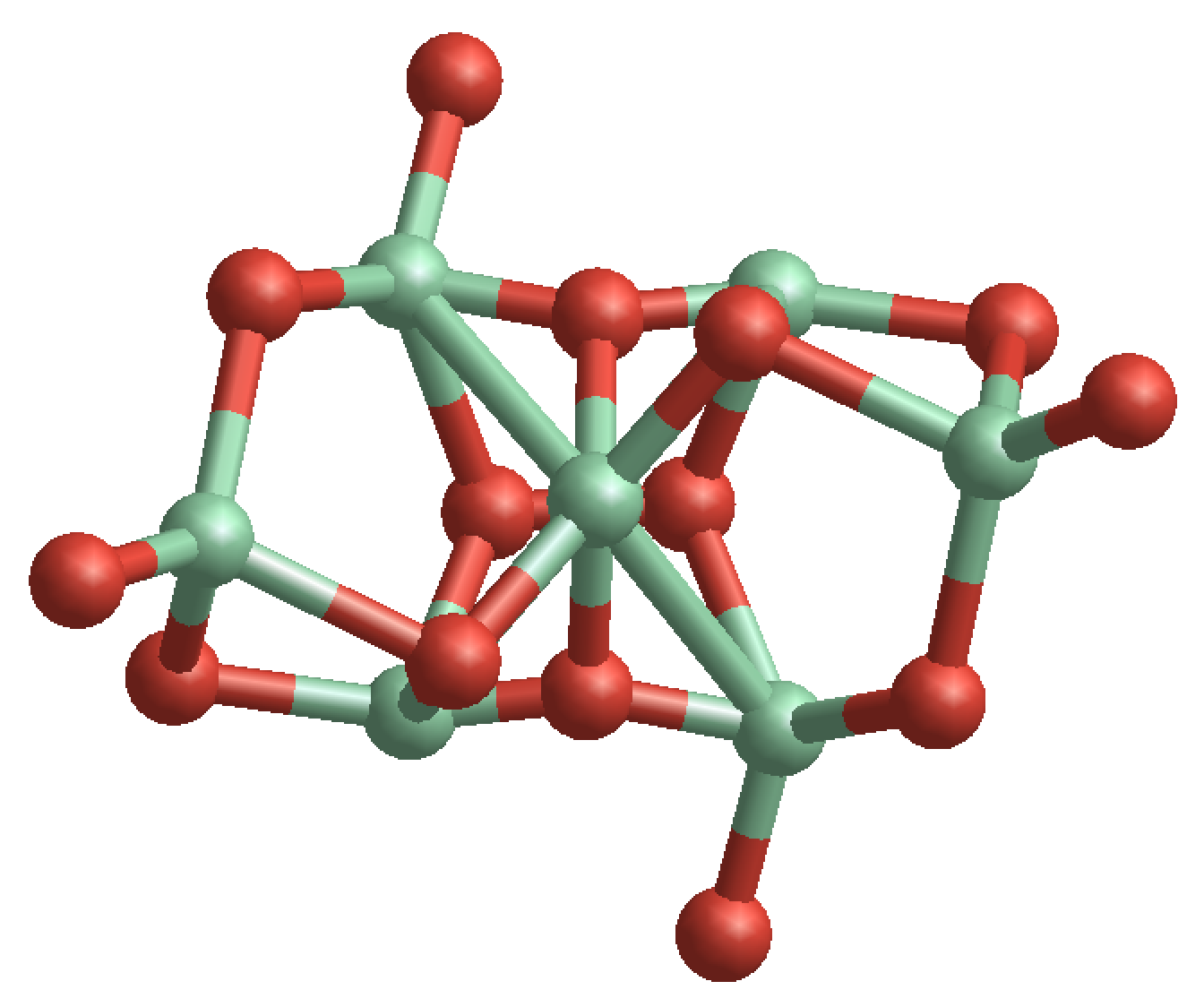}
         \caption{}
         \label{fig:seed_mirror}
     \end{subfigure}
          \hfill
     \begin{subfigure}[b]{0.49\hsize}
         \centering
         \includegraphics[width=0.8\hsize]{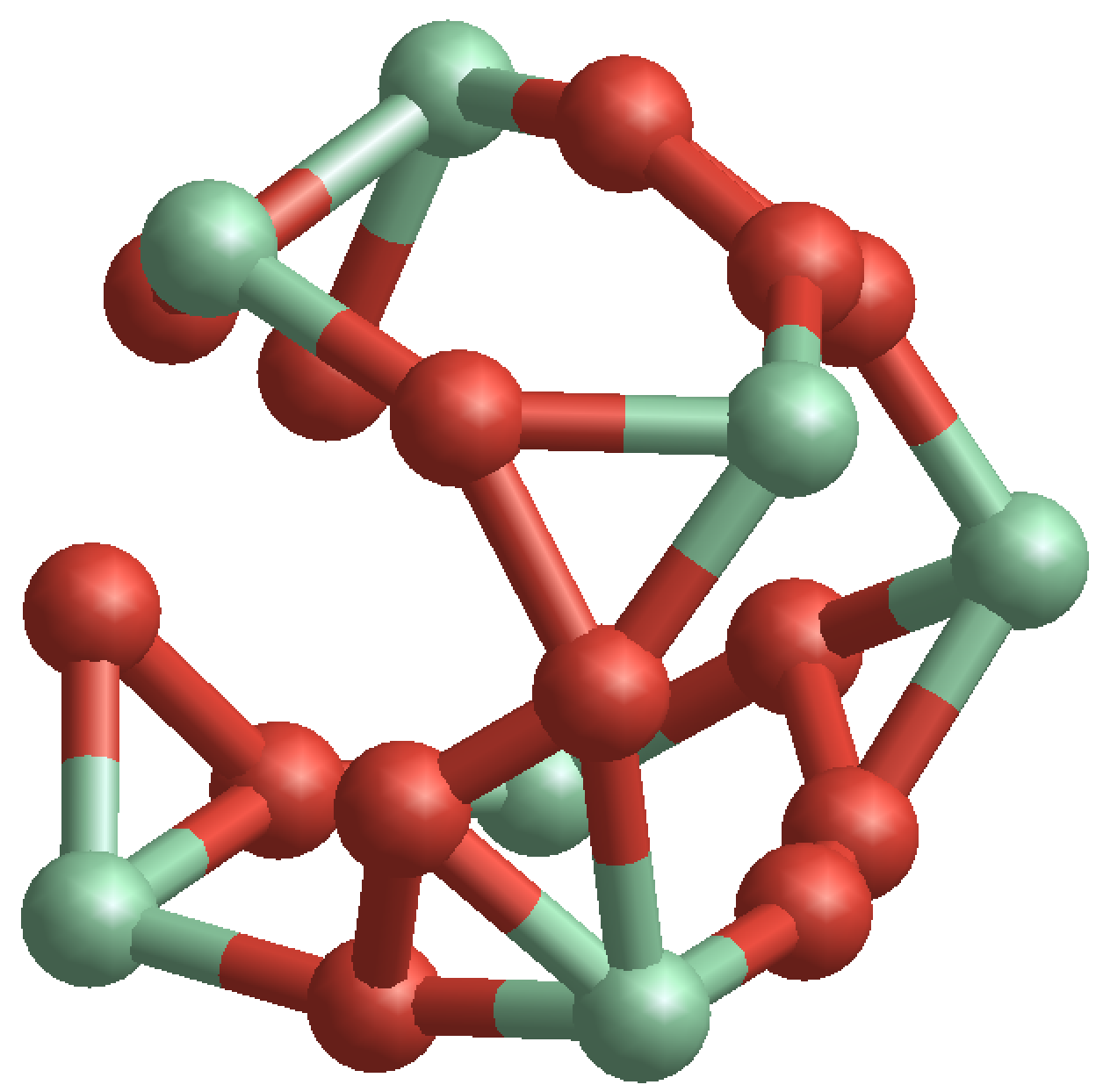}
         \caption{}
         \label{fig:seed_sphere}
     \end{subfigure}
    \caption{Examples for initial, unoptimized cluster geometries of size $N=7$, generated \textbf{(a)} randomly, \textbf{(b)} from known cluster with additional monomer (marked with yellow circles), \textbf{(c)} from a mirrored cluster of size $N = 3$ with an additional monomer in the center, and \textbf{(d)} from $N=7$ monomers evenly distributed on a sphere.}
    \label{fig:seedclusters}
\end{figure}

We apply the hierarchical approach to (TiO$_2$)$_N$ cluster formation. It starts with the creation of a broad range of 
different seed structures or cluster geometries. These seed structures are unoptimised cluster geometries for (TiO\textsubscript{2})\textsubscript{$N$}, $N$ = 3,...,15, which are then optimised towards  
potential energy minima.
In the first optimisation step, these 
un-optimised cluster geometries are used
as inputs with
the force field approach (Sect. \ref{forcefieldscalib} and first box in Fig. \ref{fig:flowdiagram}). In order to minimize the chances of missing a particular stable cluster, a large number of seed structures that cover a wide range of structurally diverse geometries are generated. These candidate geometries range from 
closely packed
and compact structures to larger and 
extended
structures 
with void parts
and include both symmetrical and asymmetrical structures for all sizes. Ideally, they are also easily optimisable, which means the average distance between 
neighbouring
atoms should not be much larger than their typical bond length. Four different approaches are used to create these seed clusters of size $N$:

\begin{enumerate}
\item \textit{Random}: For a fully randomised seed structure creation, starting with a single TiO\textsubscript{2} monomer unit, more units are iteratively attached, placing them at the end of a vector with a random orientation and a length of $1.6$\AA, which corresponds to the typical Ti-O bond distance for nano-sized TiO$_{2}$ clusters. (Fig. \ref{fig:seed_random})

\item \textit{Known+1}: Known 
stable isomer structures, existing in the 
literature \citep{Lamiel-Garcia2017, 
Berardo2014ModelingDescription} of size $N-1$ are 
taken and one monomer is randomly attached analogous 
to method 1. to get seed clusters for size $N$. (Fig. 
\ref{fig:seed_knownplusone})
\end{enumerate}
Both of these methods have a tendency to produce rather compact and highly asymmetric seed structures, especially for larger cluster sizes $N$. Therefore, two more approaches are introduced to create 
spatially extended
clusters and symmetric clusters, respectively.

\begin{enumerate}
    \setcounter{enumi}{2}
    
    \item \textit{Mirror}: For even cluster sizes $N$ a cluster of size $\frac{N}{2}$ is taken and mirrored about a random axis. For uneven cluster sizes N, a cluster of size $\frac{\mathrm{N-1}}{2}$ is taken and mirrored. Afterwards a single monomer is added along the mirror axis to bring the total number of monomer units to $N$. All seed geometries created using this method fall within the C$_{2}$ point group. (Fig. \ref{fig:seed_mirror}) 
    
    \item \textit{Equidistant}: Seed structures of size $N$ are created by evenly distributing $N$ monomers equidistant across a sphere of random radius, so that the distances between their centers of mass are between $1.6 - 3.2$\AA. The resulting geometries resemble hollow spheres. (Fig. \ref{fig:seed_sphere})
    
\end{enumerate}

\noindent
An
example of a 
candidate geometry of size $N=7$ produced by each of these approaches is depicted in Figure 
\ref{fig:seedclusters}. As the parameter space for possible cluster geometries 
grows with the cluster size $N$
a distinction between small ($N = 3-7$) and large ($N = 8-15$) clusters is made in order to save computational cost. 
For small clusters, 1000 clusters 
are created with method 1, 400 with method 2, 400 with method 3, and 200 with 
method 4 giving a total of 2000 cluster seed geometries for each size. For large 
clusters, the number of guessed geometries is multiplied by 5 to account for the larger 
parameter space. 10000 cluster seed geometries are created for each size, 5000 
with method 1, 2000 with method 2, 2000 with method 3, and 1000 with method 4. (Table \ref{tab:unique_geometries})
Hence, 90000 geometrical seed structures are tested in total.
All seed geometries
are then optimised towards 
potential minima
using the force field description 
of the PES, presented
at the end of Sect.\ref{forcefieldscalib}.

\subsection{Density functional theory}
\label{dfttheory}
To accurately describe a cluster, a large number of interactions, including quantum-mechanical ones, need to be taken into account. We employ density functional theory (DFT) (see Appendix \ref{appendix:dft}) as a method to solve the Schrödinger equation approximately and determine zero-point energies, vibrational frequencies and rotational constants of our clusters. Density functional theory is parameterised through the choice of a functional and a basis-set, which has a large impact on the resulting quantities. It is therefore essential to select an appropriate functional and basis-set for our purpose.

\subsection{Finding the ideal Functional and basis set}
\label{funcbasissearch}
For the calculations at the Density Functional level of theory (DFT), the \texttt{Gaussian16} \citep{Frisch2013GaussianD.01} program is used. 
It is desirable to use a model parameterisation that agrees with measured data.
In order to determine the DFT parameterisation, in the form of a combination of functional and basis set, that most closely resembles measured data, the results are compared to experimentally derived data for the zero-point energy \citep{MalcolmW.Chase1998NIST-JANAFTables}, vibrational frequencies \citep{MalcolmW.Chase1998NIST-JANAFTables} and rotational frequencies \citep{Brunken20082} for the TiO$_2$ monomer molecule.
\paragraph*{Approach for TiO$_2$:}
The tested functionals include 2 GGA DFT functionals, 
12 hybrid functionals with included Hartree-Fock exchange 
and 3 Complete Basis Set (CBS) 
methods
listed in Table~\ref{tab:func+base}.
The functionals were selected to cover a broad range of theoretical approaches as well as with regards to their availability within \texttt{Gaussian16}. Although there are no interactions within our clusters that stem from non-bonded interactions, we find that empirical damping coefficients as described by \cite{Grimme2011EffectTheory} improve the overall accuracy of our calibration. The \texttt{EmpiricalDispersion=GD3BJ} keyword is therefore used if available. We note that the use of an empirical dispersion does not originate from a physical or chemical
consideration.
Employing 
a similar
approach in the selection of basis 
sets, a total of 129 candidate combinations of 
functionals and basis sets is produced (Table 
\ref{tab:func+base}). For each of the combinations 
\texttt{Gaussian16} is used to perform an optimisation 
of the TiO\textsubscript{2} monomer as well as a 
zero-point energy calculation for the Titanium and 
Oxygen atoms respectively. 
The resulting energies 
for the individual functionals and basis sets
are compared with
the 
experimental data found in the JANAF-NIST tables. 
All 
129 functional/basis set combinations are evaluated 
according to their binding energy for the 
TiO\textsubscript{2} molecule calculated at a 
temperature of $T=0$K. The binding energy of the 
monomer (TiO$_2$) is calculated according to
\begin{equation}
\label{eq:bond_energy_zeropoint}
E_{\rm bind}(\mathrm{TiO_{2}}) = E_{\rm ZP}(\mathrm{TiO_{2}}) - E_{\rm ZP}(\mathrm{Ti}) - 2\,E_{\rm ZP}(\mathrm{O})
\end{equation}
and can now be compared to experimental values (Table~\ref{tab:candidates_narrow}, 4th column). $E_{\rm ZP}$ denotes the zero-point energy, i.e. the energy of the system at rest in its ground electronic state.\\
Further values that are known from experiments are the vibrational frequencies
$\nu_{1}$, $\nu_{2}$, and $\nu_{3}$
\citep{MalcolmW.Chase1998NIST-JANAFTables} and the rotational constants $A_{\rm rot}$, $B_{\rm rot}$, and $C_{\rm rot}$ \citep{Brunken20082} of the TiO$_2$ monomer. These quantities are also a result of the DFT calculations and the derived thermochemical properties depend on them (Eq. \ref{eq:partitionfunction}). They are therefore used to further constrain which of the candidate combination of functional and basis set is best suited for modeling TiO\textsubscript{2} clusters. For both, the vibrational frequencies and the rotational constants, the quality of an individual candidate combination is assessed through a deviation parameter. Since the smallest deviation of the DFT results from the experimental data is desired, these deviation parameters are computed by taking the root sum squared of all relative deviations of the DFT results from the experimental data for the vibrational frequencies and the rotational constants, respectively (Table~\ref{tab:candidates_narrow}, 5th and 6th column). For the vibrational frequencies this parameter (Vibrational frequency deviation, VFD) is therefore calculated through:
\begin{equation}
\label{eq:freq_param}
    \mathrm{VFD} = \sqrt{\sum_{i=1,..,3} \left(\frac{\nu_{i, DFT} - \nu_{i, Exp}}{\nu_{i,Exp}}\right)^2}
\end{equation}
and equivalently for the rotational constants (Rotational constant deviation, RCD):
\begin{equation}
\label{eq:rot_param}
    \mathrm{RCD} = \sqrt{\sum_{i=A,B,C} \left(\frac{i_{rot, DFT} - i_{rot, Exp}}{i_{rot,Exp}}\right)^2}
\end{equation}
Using the \texttt{Gaussian16} outputs for the zero-point energy of atomic oxygen, $E_{\rm ZP}(\rm O)$ [$\mathrm{kJ/mol}$], atomic titanium, $E_{\rm ZP}(\rm Ti)$ [$\mathrm{kJ/mol}$], and the titanium dioxide monomer, $E_{\rm ZP}(\rm TiO_{2})$ [$\mathrm{kJ/mol}$], the 
binding
energy of the monomer is calculated and compared to the literature value from the JANAF-NIST tables. Since the JANAF-NIST tables have an accuracy for the binding energy of 4 $\mathrm{kJ/mol}$ \citep{MalcolmW.Chase1998NIST-JANAFTables}, the selection is narrowed down to all candidate combinations that fall within this uncertainty. This decreases the number of candidate combinations of functionals and basis sets from 192 to 16 (Table \ref{tab:candidates_narrow}). Lastly, the runtime for each monomer calculation on a single node (32 cores) is determined, as computational speed and efficiency is desirable. \\

The thermochemical quantity of interest is the Gibbs free energy of formation $\Delta_f G^\circ(N)$, which is needed to compute nucleation rates (See Eq. \ref{eq:classicalnucleation} \& \ref{eq:partialpressures_delG}). The molecular system properties that go into computing the Gibbs free energy are the zero-point energy, the vibrational frequencies, the rotational constants and spin multiplicities (Appendix \ref{appendix:Gibbs}). The latter do not impact our results as we assume closed shell singlet clusters throughout our calculations.

The candidate combination of functional and basis set that has the smallest deviations from the experimentally known values for the zero-point energy, vibrational levels and rotational constants is considered optimal. 
The B3LYP functional in combination with the cc-pVTZ basis set and empirical GD3BJ dispersion \citep{Grimme2011EffectTheory} has the lowest deviation from the experimental values for the zero-point energy and the vibrational frequencies, and the second lowest value for the rotational constant deviation. Therefore, B3LYP/cc-pVTZ represents the most suitable choice for the calibration of approaches with a lower level of complexity (Sect. \ref{smallclustercalc}) as well the final optimisation of our candidate clusters (Sect. \ref{dftmethods}). Additionally, it is desirable to reduce computational cost. In order to achieve that, the fastest configuration that falls within the 4 $\mathrm{kJ/mol}$ threshold, the B3LYP functional in combination with the def2svp basis set, is chosen to pre-optimise candidate geometries. This allows the final optimisation with the cc-pVTZ basis set to be completed in fewer steps, leading to overall lower computational cost. 
The CCSD(T) method is known as the gold-standard in computational quantum chemistry and has been the method of choice in quantum chemistry \citep{Cizek1969OnMolecules, Purvis1998ATriples, Ramabhadran2013ExtrapolationHierarchy, Nagy2019ApproachingMethods}. However, these calculations require large amounts computational resources and are prohibitive for larger molecular systems such as nano-sized clusters. Therefore, we performed CCSD(T)/6-311+G(2d,2p) single point calculations for the GM candidates with the smallest cluster sizes of $N$=1-4, and compare the 0K binding energies with our results from hybrid DFT (B3LYP/cc-pVTZ with empirical dispersion). The first thing to note is that CCSD(T) calculation results deviate from experimental results for the binding energy of the monomer by $\approx 90 \frac{\mathrm{kJ}}{\mathrm{mol}}$, which is not adequate for our calibration purposes. When comparing the binding energy ratios to the monomer, $E_{bind,N}/E_{bind,1}$, the values for our DFT calculations do not differ by more than $2\%$ from the CCSD(T) values. We therefore conclude that calibration of our method with the experimentally determined properties of the monomer is sufficient. (Appendix \ref{appendix:ccsd})

\begin{table}[ht]
\caption{All \texttt{Gaussian16} functionals (methods) and basis sets considered when searching for the closest representation of the JANAF-NIST experimental values.}
\centering
    \begin{tabular}{|lr|}
    \hline
     \multicolumn{2}{|l|}{\textbf{GGA Functionals}} \\
     \hline
     B97D3 & [\cite{Grimme2011EffectTheory}]  \\
     TPSSTPS & [\cite{Tao2003ClimbingSolids}] \\

     \hline
     \multicolumn{2}{|l|}{\textbf{Hybrid Functionals}} \\
     \hline
     B3LYP  & [\cite{Becke1993Density-functionalExchange}]  \\
     B2PLYPD3 & [\cite{Goerigk2011EfficientInteractions}] \\
     BMK & [\cite{Boese2004DevelopmentKinetics}] \\
     PBE1PBE & [\cite{Adamo1999TowardModel}] \\
     LC-wPBE & [\cite{Vydrov2006AssessmentFunctional}] \\
     CAM-B3LYP & [\cite{Yanai2004ACAM-B3LYP}] \\
     APFD & [\cite{Austin2012ATermsb}] \\
     M11 & [\cite{Peverati2011ImprovingSeparation}] \\
     MN12SX & [\cite{Peverati2012Screened-exchangePhysics}] \\
     HSEH1PBE & [\cite{Heyd2003HybridPotential}] \\
     X3LYP & [\cite{Xu2004TheProperties}] \\
     M06 & [\cite{Zhao2008TheFunctionals}] \\
     \hline
     \multicolumn{2}{|l|}{\textbf{CBS extrapolations}} \\
     \hline
     CBS-4M  & [\cite{Montgomery2000AMethod}] \\
     CBS-QB3 & [\cite{Montgomery2000AMethod}] \\
     ROCBS-QB3 & [\cite{Wood2006AChemistry}] \\
     \hline \hline
     \multicolumn{2}{|l|}{\textbf{basis sets}} \\
     \hline
     def2svp & [\cite{Weigend2006AccurateRn}] \\
     Def2TZVP & [\cite{Weigend2006AccurateRn}] \\
     def2svpp & [\cite{Weigend2006AccurateRn}] \\
     Def2TZVPP & [\cite{Weigend2006AccurateRn}] \\
     cc-pVDZ & [\cite{Wilson1996GaussianNeon}] \\
     cc-pVTZ & [\cite{Wilson1996GaussianNeon}] \\
     AUG-cc-pVDZ & [\cite{Wilson1996GaussianNeon}] \\
     AUG-cc-pVTZ & [\cite{Wilson1996GaussianNeon}] \\
     6-311+G* & [\cite{Curtiss1995ExtensionGa-Kr}] \\
     \hline
    \end{tabular}
    \label{tab:func+base}
\end{table}

\begin{table*}
\caption{Candidate combinations for functional and basis set with a zero-point binding energy deviation $\lvert E_{bind, JANAF} - E_{bind, DFT} \rvert <4 \mathrm{kJ/mol}$, sorted by vibrational constant deviation.}
\centering          
\begin{tabularx}{\textwidth}{|l|l|l|X|X|X|X|}
\hline
Rank & Functional   & basis set   &   Deviation from Janaf-Nist [kJ/mol] &   Vibrational frequency deviation (VFD) &   Rotational constant deviation (RCD) &   Core computational time [s] \\
\hline
1 & B3LYP        & cc-pVTZ     &                                 0.22 &                             0.096 &                           0.029 &                        1251.6 \\
2 & B3LYP        & AUG-cc-pVDZ &                                 1.12 &                             0.1   &                           0.025 &                        1229.6 \\
3 & B3LYP        & Def2TZVPP   &                                 3.11 &                             0.101 &                           0.03  &                         978.8 \\
4 & X3LYP        & cc-pVDZ     &                                 0.47 &                             0.12  &                           0.042 &                        1683.9 \\
5 & APFD         & cc-pVTZ     &                                 3.18 &                             0.136 &                           0.05  &                        1842.8 \\
6 & APFD         & AUG-cc-pVDZ &                                 2.79 &                             0.138 &                           0.047 &                        1462.7 \\
7 & M06          & Def2TZVP    &                                 2.75 &                             0.141 &                           0.045 &                         748.2 \\
8 & PBE1PBE      & cc-pVDZ     &                                 2.99 &                             0.156 &                           0.063 &                        1097.7 \\
9 & M11          & 6-311+G*    &                                 1.59 &                             0.157 &                           0.087 &                         837   \\
10 & M11          & AUG-cc-pVTZ &                                 2.14 &                             0.159 &                           0.079 &                        6287.7 \\
11 & M11          & cc-pVTZ     &                                 2.4  &                             0.16  &                           0.084 &                        3416.2 \\
12 & M11          & AUG-cc-pVDZ &                                 3.88 &                             0.163 &                           0.086 &                        2879.9 \\
13 & M11          & Def2TZVP    &                                 1.64 &                             0.171 &                           0.088 &                         841.9 \\
14 & B3LYP        & def2svp     &                                 1.04 &                             0.19  &                           0.06  &                         440.1 \\
15 & LC-wPBE      & Def2TZVP    &                                 2.81 &                             0.194 &                           0.086 &                         765.3 \\
16 & APFD         & def2svp     &                                 3.36 &                             0.228 &                           0.079 &                         440.2 \\
\hline
\end{tabularx}
\label{tab:candidates_narrow}
\end{table*}

\subsection{Calibration of force fields and DFTB}
\subsubsection{Calculation of known small clusters and isomers}
\label{smallclustercalc}
After the selection of the ideal combination of functional and basis set, judged by its ability to closely match the  experimentally derived properties of TiO\textsubscript{2}, 
it is used to calculate the binding energies and geometries of the global minimum structures of the clusters of $N=2\,\ldots\,6$. The initial geometries for these calculations are sourced from \cite{Berardo2014ModelingDescription} and \cite{Lamiel-Garcia2017}. Additionally, the binding energy and geometry for several energetically less favorable isomers for each size is calculated (Table \ref{tab:smallclusters}). This is done so that the accurate depiction of the energetic ordering of isomers of the same size by force field and DFTB+ models can be ensured. 

\begin{table}[]
    \caption{Number of isomers for each cluster size optimised at the DFT level, used to calibrate force field and DFTB models.}
    \centering
    \begin{tabular}{|c|c|}
    \hline
      Cluster size $N$   & Number of isomers \\
      \hline
        2 & 3 \\
        3 & 9 \\
        4 & 17 \\
        5 & 14 \\
        6 & 14 \\
    \hline 
    \end{tabular}
    \label{tab:smallclusters}
\end{table}

\subsubsection{Force fields}
\label{forcefieldscalib}
In order to save computational time, it is beneficial to be as efficient as possible in the optimisation process towards local and global 
potential minima for 
candidate isomers. For example, a titanium-dioxide 
cluster of size N=12 ($(\mathrm{TiO}_{2})_{12}$) 
has 36 atoms and thereby 3x36 = 108 
degrees of freedom. 
The 
number of possible geometries is simply too large 
to be modeled at a feasible computational cost for 
a large number of clusters. Therefore a modelling 
approach is employed, that describes the interactions between individual atoms through an interatomic Buckingham pair potential including the Coulomb potential. (Appendix \ref{appendix:forcefield})
There are several parameterisations for Ti--O systems provided in previous studies, e.g. \cite{Matsui1991MolecularTio2} or \cite{Lamiel-Garcia2017}. However, we find that neither of them are well suited for our purposes as they do not accurately depict the B3LYP/cc-pVTZ energetic ordering within isomers of the same size $N$. To find a force field prescription that reflects the experimental data available for TiO\textsubscript{2} to the extent possible, a search is conducted for a set of Buckingham parameters that reproduce the 
binding energies and average bond distance of the smallest clusters calculated at the start of Sect. \ref{smallclustercalc}. 
\paragraph*{Approach to (TiO$_2$)$_{\rm N}$:}
The program used to calculate the energy as well as the bond distance is the General Utility Lattice Program \texttt{GULP} \citep{Gale2003TheGULP}. The \texttt{optimise} keyword is used in a constant pressure environment (\texttt{conp}) to optimise the 
DFT optimised
clusters and determine their binding energies for each combination of parameters. The parameters that need to be determined in Eq. \ref{eq:Buckingham} are the charges of Ti and O, and the Buckingham pair parameters $A$, $B$, and $C$ for each of the relevant interactions, Ti-Ti, O-O, and Ti-O. Since overall charge neutrality needs to be conserved, the charge of Ti is directly coupled to the charge of O by a factor of -2, i.e. there are two negatively charged Oxygen anions for every positively charged Titanium cation. 
For the TiO\textsubscript{2} molecule the authors find low formal charges. Mulliken charge analysis of the smallest cluster sizes (N=1-4) reveal average Ti charges of less than +1e.
The parameterisation therefore has 10 free parameters, three from each of the interactions, i.e. Ti-Ti, O-O, and Ti-O, plus the charge. 
To find the ideal parameter set all interaction parameters as well 
as the charge are varied freely and the \texttt{scipy.optimise} differential evolution 
algorithm \citep{Storn1997DifferentialSpaces} is used to find the set of parameters that deviates the least from our calculated 
binding energies and average bond distances, as well as reproduces the energetic ordering of 
isomers of the same size. (Appendix \ref{appendix:ff_calib})
With this physically consistent set of parameters the potential energy of any Ti-O system, 
i.e. any TiO\textsubscript{2} cluster geometry, can now be 
accurately predicted,
requiring little computational power.
This approach of searching for candidate structures with the application of a 
Buckingham pair potential has been used before, e.g. in 
\cite{Gobrecht2018AAl2O38} for aluminium oxide, in  \cite{Cuko2017GlobalApproach} for 
hydroxylated silica clusters and for titanium dioxide in \cite{Lamiel-Garcia2017}. 

The 
seed structures are optimised 
using our re-parametrised force field
as the first step in our hierarchical approach.
To 
even further increase the structural complexity of our searches, a basin-hopping algorithm as described by \cite{Wales1997GlobalAtoms} is 
employed. It is used in its implementation in the \texttt{ase} Python package \citep{HjorthLarsen2017TheAtoms} to optimise 
the geometries of all the seed structures. 
The potential 
energy calculation at each step of the optimisation is done through GULP with our 
parameterisation of the force field. After the force field optimisation the candidate 
geometries are analysed further in Sect. \ref{dftbcalib}.

\subsubsection{Density Functional based Tight Binding (DFTB)}
\label{dftbcalib}
From the 
geometry optimisation with the force field approach (Sect. 
\ref{forcefieldscalib}) 2000 candidate geometries at local 
potential minima for each small cluster size, $N =
3-7$, and 10000 candidate geometries for each large cluster size, $N = 8-15$, are 
obtained. However, the Buckingham-Coulomb pair potential does not describe 
the interaction between electrons and their related orbitals.
More precisely, only interactions between Ti cations, O anions, and themselves are taken into
account. The interaction of (binding) electrons and the electron correlation is neglected. 
Still, it serves the purpose of optimising the candidate geometries
on a approximate and simplified PES.
The potential energies of the (TiO\textsubscript{2})\textsubscript{$N$} cluster 
candidates are not accurate, because, among other reasons, the force field approach 
considers single point charges instead of a 
charge distribution of each ion in the cluster.
It is therefore a reasonable assumption that the binding energies of these clusters and their energetic ordering 
can be improved
by an intermediate optimisation step with a more accurate description 
accounting for electronic orbitals.
This is done to get a more accurate energetic ordering of the 
candidate clusters, so the best set of candidate clusters is chosen to be optimised with computationally expensive
all-electron DFT calculations (see Section \ref{dftmethods}). For this intermediate step density functional based tight 
binding (DFTB, Appendix \ref{appendix:DFTB}) is chosen. The choice is made because in both complexity and computational 
cost, DFTB models fall in between the descriptions offered by force fields and all-electron 
DFT models \citep{Hourahine2020DFTB+Simulations}. 
The parameterisation of the exchange-correlation 
functional is not included within the DFTB model 
and is done through choosing an appropriate set of 
Slater-Koster files. These files relate to 
integrals and contain sets of 
functions that describe the exchange-correlations 
part of interactions between two
atomic species (i.e. chemical elements). 
The database at www.dftb.org hosts three different 
sets that describe the interaction of titanium and 
oxygen: \textit{matsci} 
\citep{Luschtinetz2009AdsorptionSurfaces}, a 
general purpose material science set, 
\textit{trans3d} \citep{Zheng2007ParameterNi}, a set describing transition metal elements in biological systems, and \textit{tiorg} \citep{Dolgonos2010AnTitanium}, a set for describing Ti bulk, TiO\textsubscript{2} bulk, TiO\textsubscript{2} surfaces, and TiO\textsubscript{2} with organic molecules. In order to find an exchange-correlation parameterisation that best describes small TiO\textsubscript{2} clusters, our test clusters calculated in Sect. \ref{smallclustercalc} are used to test all three sets of Slater-Koster files. 
\paragraph*{Approach to (TiO$_2$)$_{\rm N}$:}
The DFTB calculations are done using the 
\texttt{DFTB+} program 
\citep{Hourahine2020DFTB+Simulations}. For the 
calibration of the DFTB method the energies 
calculated for geometries 
in Sect. 
\ref{smallclustercalc} are used. For clusters of 
$N=3\,\ldots\,6$, equivalent to our approach in 
calibrating our force fields, DFTB+ is used to 
calculate the total energy for all the isomers with each of the three different sets of Slater-Koster files. Afterwards it is tested which of these reproduces the energetic ordering of the isomers found with all-electron DFT calculations. If two of the Slater-Koster files perform identically in this regard, the one is chosen that better approximates the absolute values of the binding energies. The \textit{matsci} \citep{Luschtinetz2009AdsorptionSurfaces} set of Slater-Koster files is found to perform the best for the purpose of this paper. (Appendix \ref{appendix:dftb_calib})
Subsequently this set of Slater-Koster files is used to describe the exchange correlation in the further evaluation and optimisation of the 
seed structures from Sect. \ref{forcefieldscalib}. To enable direct comparison to the results of DFTB+ optimised isomers, the potential energy for each of the candidate geometries is obtained by performing a single-point energy calculation with DFTB+. Next, an identical calculation for atomic Ti and atomic O is performed, in order to obtain their respective potential energies. The binding energy of a (TiO\textsubscript{2})\textsubscript{$N$} cluster can then be calculated by substracting the individual contributions, analogous to Eq.\ref{eq:bond_energy_zeropoint}:
\begin{equation}
\label{eq:bond_energy}
\mathrm{E_{Bind} ((TiO_2)}_N) = \mathrm{E_{Pot} ((TiO_2)}_N) - N \cdot \mathrm{E_{Pot} (Ti)} - 2\cdot N\cdot\mathrm{E_{Pot} (O)} 
\end{equation}
This calculation is performed for each of the candidate geometries obtained from 
Sect. \ref{forcefieldscalib}. 
Each of the candidate geometries is then optimised using the \texttt{ConjugateGradient} method, making use of 
its implementation within DFTB+, 
and are again ordered according to their binding energies calculated with Eq. 
\ref{eq:bond_energy}. 
Since the randomisation of seed structures is not ideal, there will be 
duplicates among the optimised geometries. In order to filter them out any two candidates 
with a binding energy difference of $\Delta \mathrm{E_{Bind}} \leq 0.01 \mathrm{eV}$ are analysed with regards to their similarity. For that the mean of all inter-atomic 
distances within the cluster is calculated, giving a size parameter $R$. If the average 
inter-atomic distance of the two clusters is closer than $\Delta R \leq 0.01$\AA, the two clusters are found to be duplicates and one of them is removed from the process. After removing the duplicate candidate geometries, the lowest energy clusters for each size and used in our final step, optimisation with all-electron DFT in Sect. \ref{dftmethods}.

\subsection{Global minima search with DFT}
\label{dftmethods}
The energetically most favourable isomers found with DFTB (see Sect. \ref{dftbcalib})
are further optimised using \texttt{Gaussian16}.
In order to reduce total computational time, a first optimisation is performed with the 
B3LYP/def2svp functional/basis set combination with empirical dispersion as described by \cite{Grimme2011EffectTheory}. After this step, the final optimisation is done with the B3LYP/cc-pVTZ functional/basis set combination, also with empirical dispersion (Sec. \ref{funcbasissearch}). 
For all these 
isomers, a frequency analysis is carried out using the same functional/basis set, in order to
make sure that the isomer is a true minimum and to exclude transition states. In addition, we
calculate the final energies as well as 
rotational constants needed to determine the 
thermochemical properties we are interested in. 
The thermochemical properties entropy $S^\circ (N)$ $\left(\frac{
\mathrm{J}}{\mathrm{mol}\cdot\mathrm{K}}\right)$, 
change of enthalpy ddH $\left(\frac{\mathrm{kJ}}{\mathrm{mol}}\right)$ and Gibbs free energy of formation $\Delta_f G^\circ (N)$ $\left(\frac{\rm kJ}{\rm mol}\right)$ are calculated from the
output of the DFT optimisations
using 
the RRHO approximation as
implemented within the \textit{thermo.pl} \citep{Irikura2002THERMO.PL} code. 

\section{Results}
\label{sec:results}

\subsection{Force field optimised clusters}
\label{sec:ff-opt}
The new parameterisation (Table \ref{tab:buckingham_bounds}) for the Buckingham-Coulomb 
force field is used to optimise the geometric seed structures (Sect. \ref{section:seedcreation}). 
The algorithm used for this search is a basin hopping 
algorithm described by \cite{Wales1997GlobalAtoms}. The energy evaluation within this 
algorithm is done using \texttt{GULP} with the new set of parameters. At each step of the 
algorithm a local energy minimisation using the FIRE algorithm 
\citep{Bitzek2006StructuralSimple} is performed. 
This algorithm is designed specifically for optimisation of atomistic systems towards their 
closest local minimum. 
A numerical 'temperature' of $T = 100$ within the basin-hopping 
algorithm is chosen to allow for the exploration of nearby local minima in order to find the deepest potential well in the vicinity. This temperature is not a physical temperature, but influences the rejection criterium within the algorithm. All seed structures 
created in Section 
\ref{section:seedcreation} for each cluster size $N$ = 3,...,15 are optimised with this 
algorithm.

\subsection{DFTB energy calculations and optimisation}
\begin{figure*}
    \includegraphics[width=0.49\hsize]{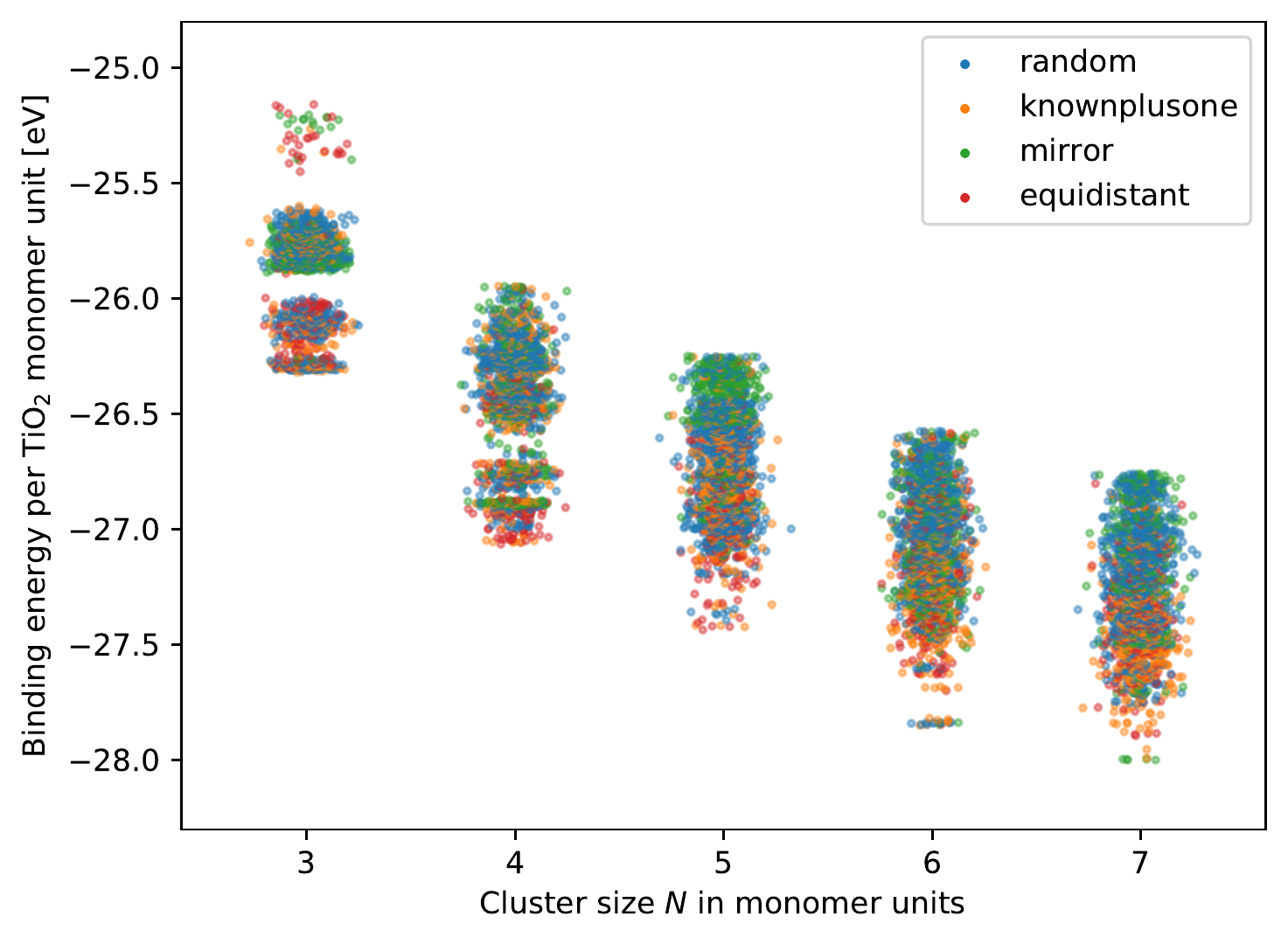}
    \includegraphics[width=0.49\hsize]{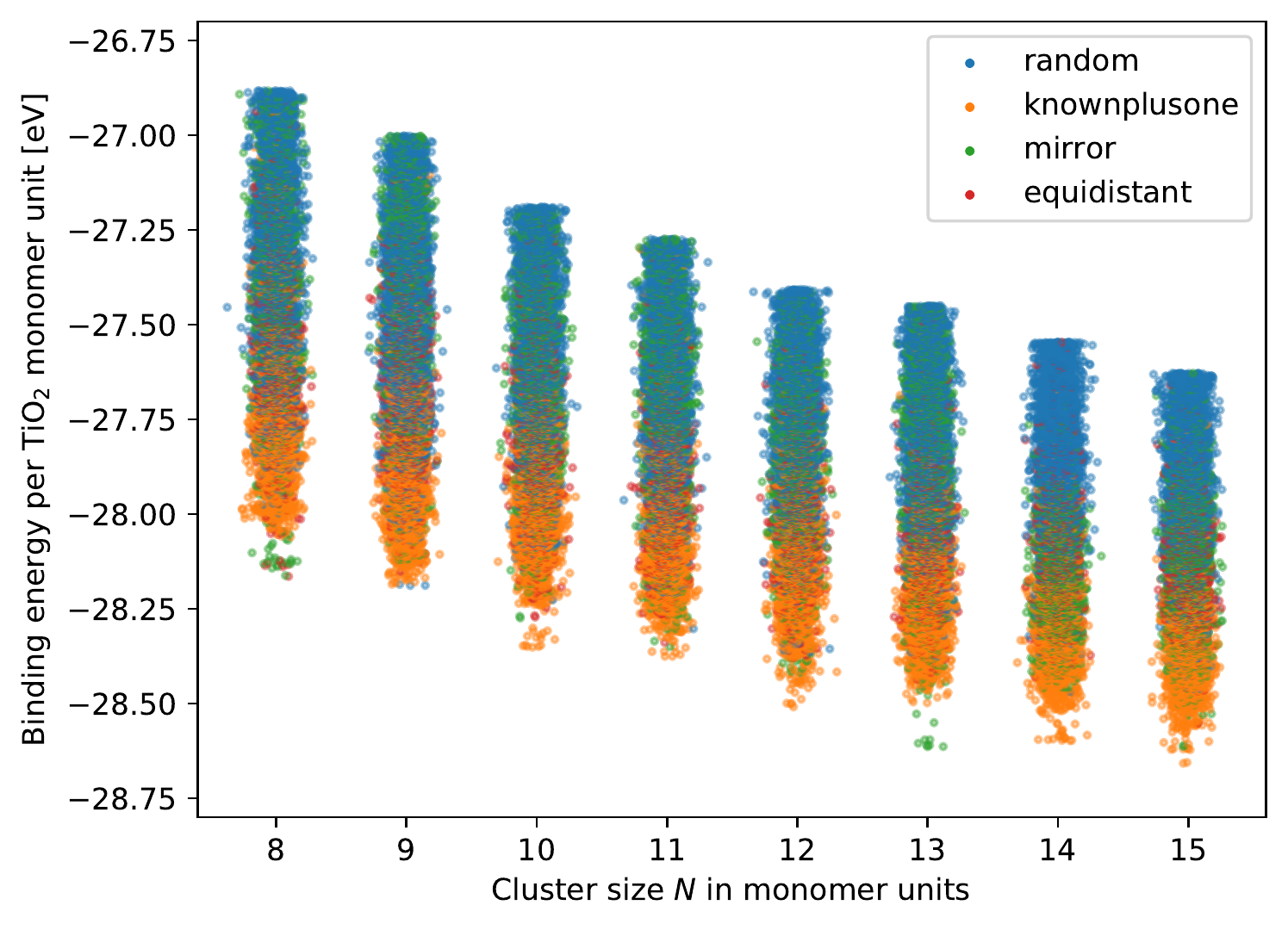}
    \caption{Binding energy per TiO$_2$ monomer unit for small ($N= 3-7$, left) and large ($N= 8-15$, right) clusters for the 90\% of the most favourable clusters after optimisation with the force field approach. Individual clusters are color coded by method of creation of the candidate cluster (Sec. \ref{section:seedcreation}). A random spread along the x-axis  has been added to enable comparison between clusters at similar energies.}
    \label{fig:ff_ordering}
\end{figure*}

\begin{figure*}
\centering
    \includegraphics[width=0.49\hsize]{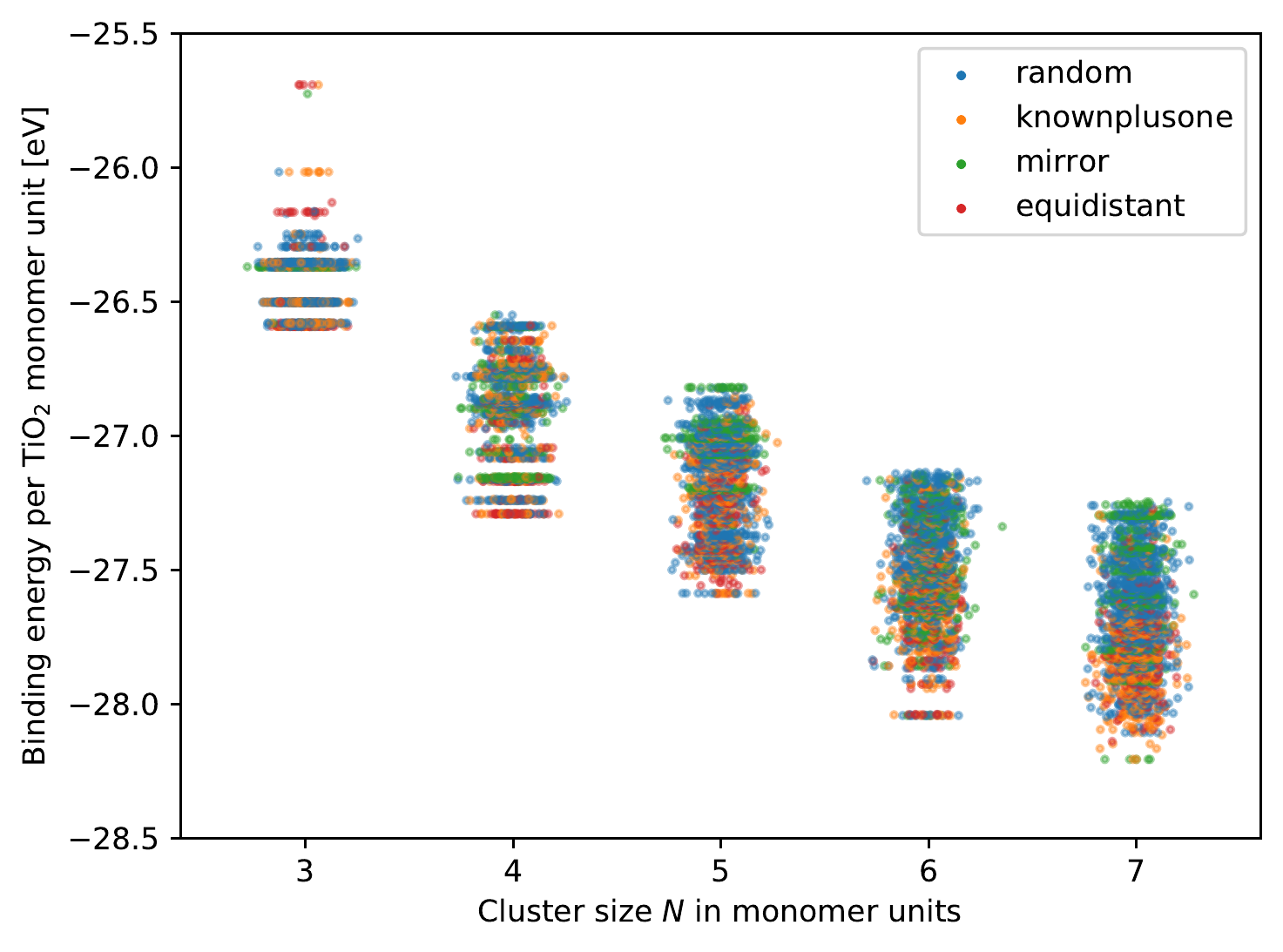}
    \includegraphics[width=0.49\hsize]{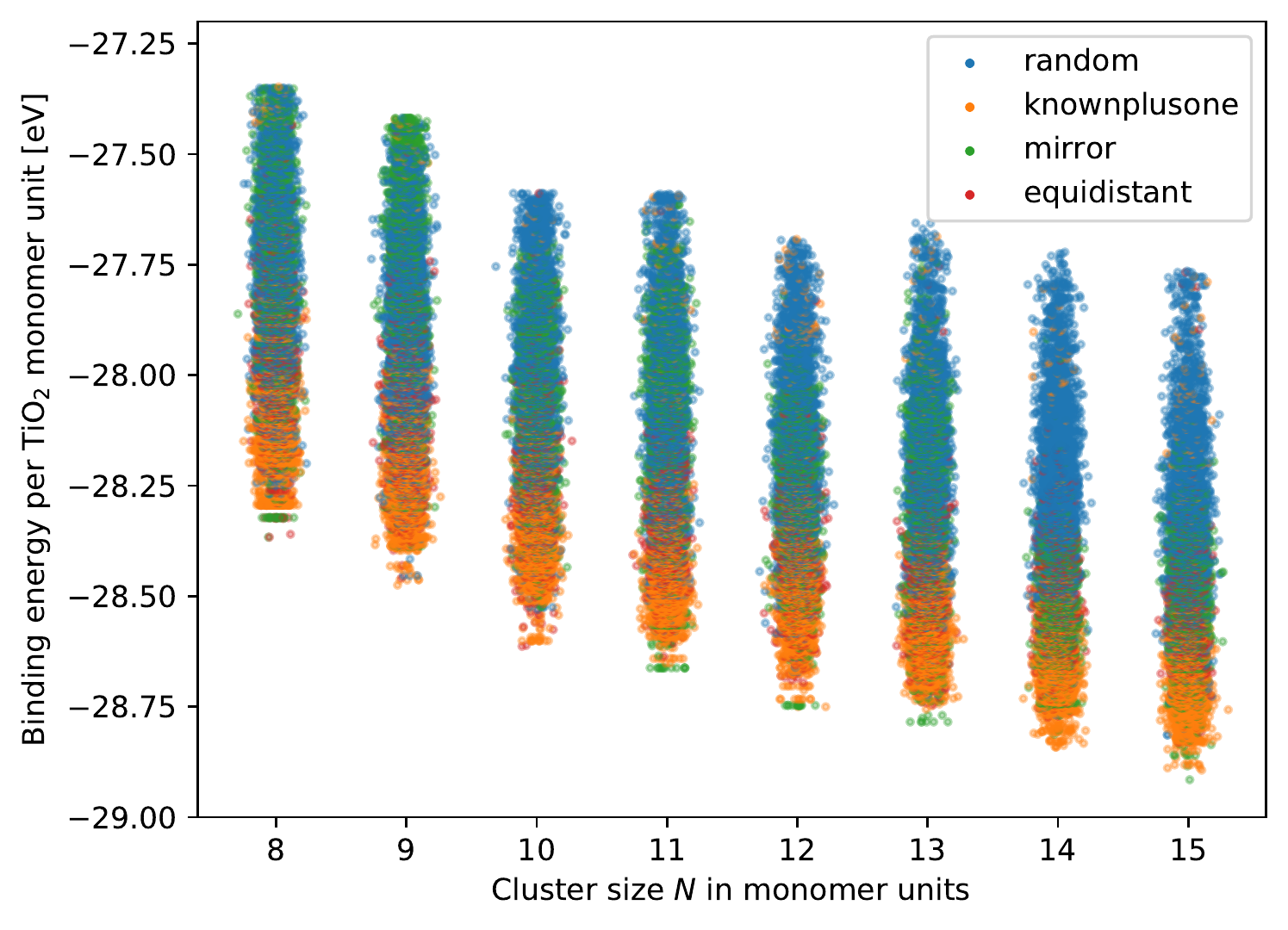}
    \caption{Binding energy per TiO$_2$ monomer unit for small ($N= 3-7$, left) and large ($N= 8-15$, right) clusters for the best 90\% of clusters after optimisation with the DFTB approach. Individual clusters are color coded by method of creation of the candidate cluster (Sec. \ref{section:seedcreation}). A random spread along the x-axis  has been added to enable comparison between clusters at similar energies.}
    \label{fig:dftb_ordering}
\end{figure*}

A zero-point energy calculation is performed for all cluster geometries calculated in Sect. \ref{sec:ff-opt} and their 
binding
energies
are calculated according to Eq. \ref{eq:bond_energy} (Fig. \ref{fig:ff_ordering}). Then each of these cluster geometries is optimised, minimising their 
potential
energy, with the DFTB description of interactions. A \texttt{ConjugateGradient} optimisation algorithm \citep{Hestenes1952Methods1} is used and the binding energies of the optimised clusters are calculated again (Fig. \ref{fig:dftb_ordering}). A comparison of Figures \ref{fig:ff_ordering} and \ref{fig:dftb_ordering}
demonstrates the approximative character of
the force field approach, as a seemingly broad variety of local minima for small clusters 
disappears with the introduction of higher complexity in the form of DFTB. The energy levels 
become discrete, many different cluster geometries from the force field approach
produce the same geometry after further optimisation. To find potentially new 
global 
minimum 
(GM) structures, all known 
literature GM clusters are 
optimised and their binding energies calculated 
with the DFTB approach for comparison. 
For each size, any unique cluster 
that has a higher binding energy than the known 
literature GM 
is categorised as
a candidate for a new GM. 
Additionally, the 10 isomers with binding energies closest to that of the literature GM are 
considered. These isomers (Table \ref{tab:gm_candidates}) are passed on to be optimised with all-electron DFT calculations.

\begin{table}[]
    \caption{Number of GM candidates, i.e. cluster candidates with a higher binding energy than the literature GM after DFTB optimisation, and number of isomers calculated with all-electron DFT calculations for each size.}
    \centering
    \begin{tabular}{|c|c|c|}
    \hline
    Size  & GM candidates  & Isomers \\
    \hline
    3 & 3 & 10 \\
    4 & 2 & 10 \\
    5 & 1 & 10 \\
    6 & 0 & 10 \\
    7 & 4 & 10 \\
    8 & 1 & 10 \\
    9 & 3 & 10 \\
    10 & 3 & 10 \\
    11 & 6 & 10 \\
    12 & 2 & 10 \\
    13 & 2 & 10 \\
    14 & 0 & 10 \\
    15 & 0 & 10 \\
    \hline
    \end{tabular}
    \label{tab:gm_candidates}
\end{table}

\subsection{Comparison of candidate creation approaches}
\label{sec:comparison}
In this section, the performance of the different methods for creation of candidate geometries (see Sec. \ref{section:seedcreation}) is discussed. Additionally the necessity for cluster geometries from literature is assessed, to establish if the method is capable of operating without prior knowledge of particularly favourable cluster geometries. The parameter space that needs to be covered to find all possible geometric configurations increases dramatically with cluster size N.
One metric to look at is
the number of duplicate isomers.
Table \ref{tab:unique_geometries} lists the number of created candidates and resulting unique clusters for all cluster sizes $N$ considered. For the small clusters 
many duplicates are found,
as the parameter space is small and many created geometries share a nearby 
potential
minimum. As the size of the clusters grows, the
number of identical geometries
falls, creating more unique clusters per candidate. A plateau of unique clusters is reached from size 12, from where about 68\% of the 
created clusters are unique. As larger clusters have a larger parameter space 
for possible geometric configurations, 
the trend of 
an increasing number of
unique clusters is expected to continue. 
The resulting
plateau therefore signals a limit of the current methods to fully explore the parameter space of clusters larger than $N = 12$.

\begin{table}[]
    \caption{Number of created candidate geometries, unique cluster geometries after DFTB optimisation and 
    percentage of unique clusters per created candidate for all sizes $N$}
    \centering
    \begin{tabular}{|c|c|c|c|}
    \hline
    Size & Candidates & Unique clusters & Unique clusters  \\
     & & & per candidate [\%] \\
    \hline
    3 & 2000 & 150 & 7.5  \\
    4 & 2000 & 264 & 13.2  \\
    5 & 2000 & 606 & 30.3  \\
    6 & 2000 & 991 & 49.6  \\
    7 & 2000 & 1054 & 52.7  \\
    8 & 10000 & 5204 & 52.0 \\
    9 & 10000 & 5540 & 55.4 \\
    10 & 10000 & 6159 & 61.6 \\
    11 & 10000 & 6178 & 61.8 \\
    12 & 10000 & 6785 & 67.9 \\
    13 & 10000 & 6718 & 67.2 \\
    14 & 10000 & 6784 & 67.8 \\
    15 & 10000 & 6793 & 67.9 \\
    \hline
    \end{tabular}
    \label{tab:unique_geometries}
\end{table}

\begin{figure}
    \centering
    \includegraphics[width=\hsize]{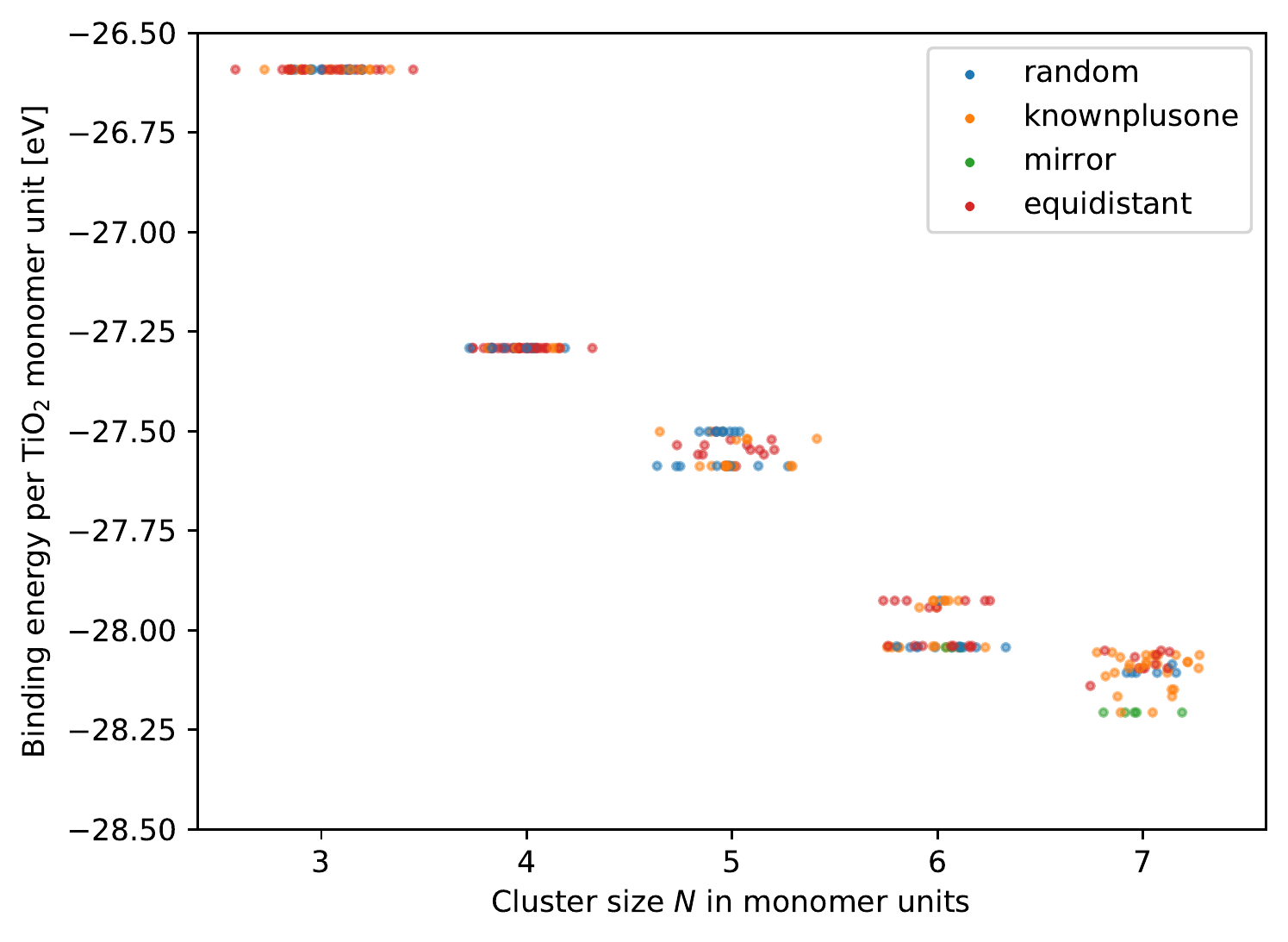}
    \includegraphics[width=\hsize]{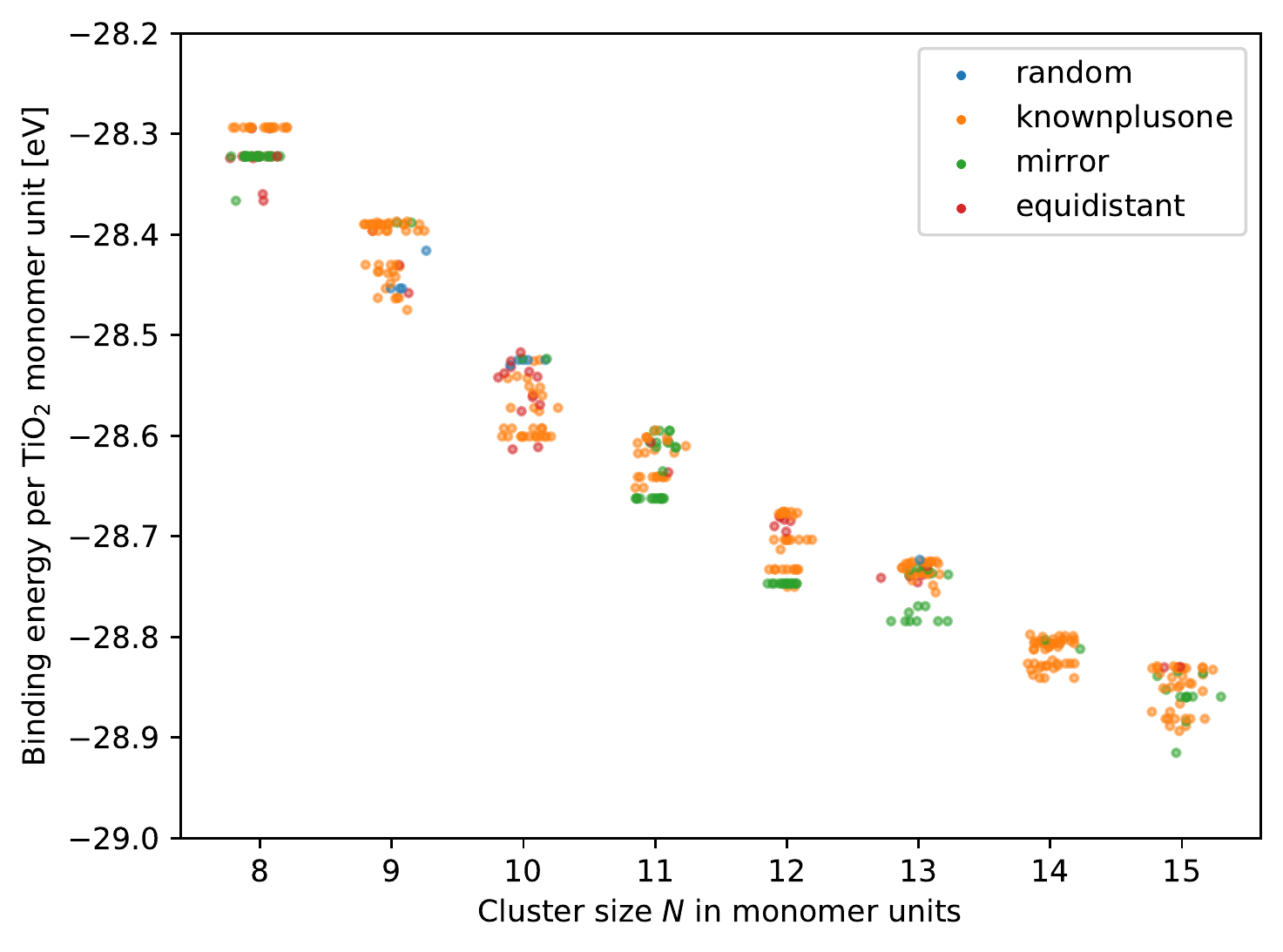}
    \caption{Binding energy per monomer unit for small ($N= 3-7$, top) and large ($N= 8-15$, bottom) clusters for the best 50 clusters after optimisation with the DFTB approach. Individual clusters are color coded by method of creation of the candidate cluster (Sec. \ref{section:seedcreation}). A random spread along the x-axis has been added to enable comparison between clusters at similar energies.}
    \label{fig:dftb_ordering_top50}
\end{figure}

Two out of the four approaches used in this work rely on known cluster geometries 
that are reported in the literature, in order to produce cluster candidates (Known+1 and Mirror, methods 2 and 3 in Sec. \ref{section:seedcreation}, referred to as dependent 
methods), while the other two (Random and Equidistant, methods 1 and 4 in Sec. \ref{section:seedcreation}, referred to as independent methods) 
need only information about the monomer. Figure \ref{fig:dftb_ordering_top50} and Table \ref{tab:top50bymethod} show the 50 best, i.e. highest binding energy, candidates and their creation methods. 
For small clusters $N = 3-6$ 
the independent methods produce the majority of the 50
clusters with the highest binding energy. 
These independent methods also find the GM
candidates reported in the literature
for $N = 3-6$, which are needed for calibration. For larger clusters the dependent methods 
perform better.
The random generation method only produces 
a single of the 50 energetically most favorable clusters
for $N$ $>$ 10. This is because the 
PES is very large and complex at these sizes and can not be sufficiently explored by a random walk. Therefore, the
methods that contain prior information about stable configurations, such as the dependent methods 
show an enhanced performance.
The equidistant method is similar, as it also does not 
rely on previously reported favourable isomers.
That is why for $N$ $>$ 10 the dependent methods make up far more of the energetically most favourable clusters. 
More elaborate methods to create first-guess cluster geometries are available through, for example, Gaussian process regression (e.g., \cite{Meyer2020GeometrySystems}). For the present work, the choice was made to apply simple but fast methods and comparisons to more complex approaches are desirable in future works.

\begin{table}[]
    \caption{The 50 lowest energy clusters after DFTB optimisation for which the un-optimised geometry guess were created by four different procedures (random, mirror, known+1, equidistant).}
    \centering
    \begin{tabular}{|c|c|c|c|c|}
    \hline
    Size & Random & Mirror & Known+1 & Equidistant \\
    \hline
    3 & 11 & 0 & 14 & 25\\
    4 & 10 & 0 & 6 & 34\\
    5 & 19 & 0 & 17 & 14\\
    6 & 13 & 3 & 18 & 16\\
    7 & 6 & 5 & 28 & 11\\
    8 & 0 & 22 & 15 & 13\\
    9 & 4 & 2 & 41 & 3\\
    10 & 5 & 2 & 30 & 13\\
    11 & 0 & 25 & 23 & 2\\
    12 & 0 & 14 & 30 & 6\\
    13 & 1 & 16 & 28 & 5\\
    14 & 0 & 2 & 48 & 0\\
    15 & 0 & 12 & 36 & 2\\
    \hline    
    \end{tabular}
    \label{tab:top50bymethod}
\end{table}

\subsection{All-electron DFT calculations}
\begin{figure}
    \centering
    \includegraphics[width=0.7\columnwidth]{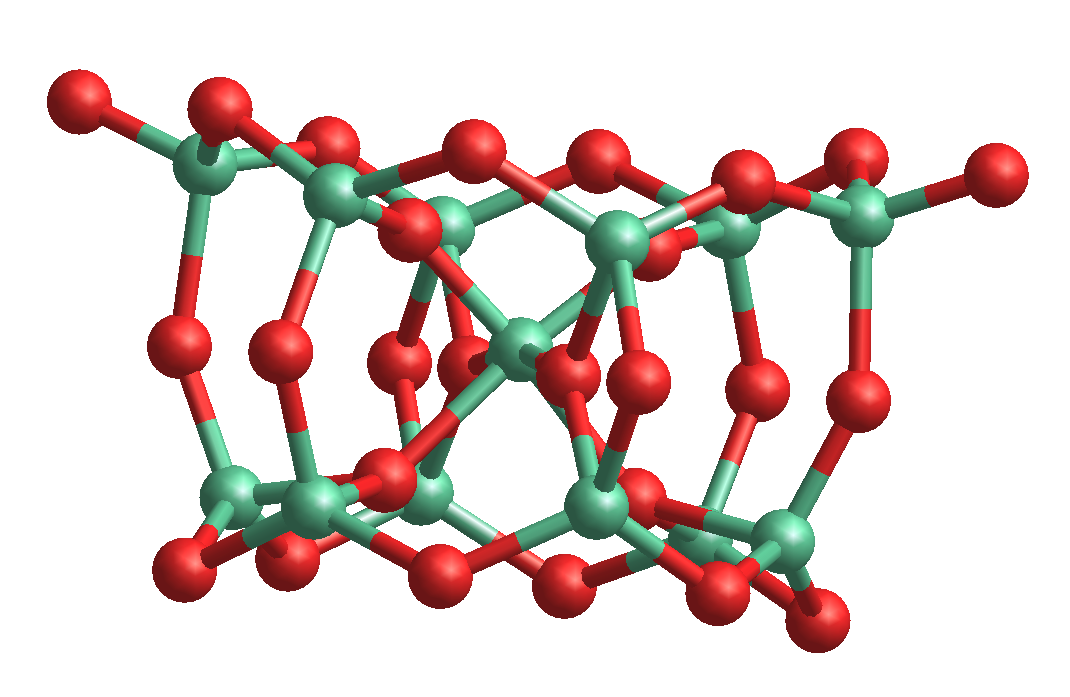}
    \caption{New global minimum structure for (TiO$_2$)$_{13}$, binding energy: E$_{binding} = -1782.96\frac{\mathrm{kJ}}{\mathrm{mol}}$ per monomer unit.}
    \label{fig:13_GM}
\end{figure}

All candidate geometries from Table \ref{tab:gm_candidates}, as well as all literature GM geometries from \cite{Berardo2014ModelingDescription} and \cite{Lamiel-Garcia2017} are pre-optimised using \texttt{Gaussian16} with the B3LYP functional, def2svp basis set and empirical dispersion. Afterwards they are optimised and a frequency analysis is performed with the best performing combination found in Sec. \ref{funcbasissearch}, B3LYP/cc-pVTZ with empirical dispersion. For all isomers, the binding energies are calculated according to Eq. \ref{eq:bond_energy} and their respective geometries can be found in electronic form at the CDS via anonymous ftp to \footnote{\url{cdsarc.u-strasbg.fr} (130.79.128.5)}
or via \footnote{\url{http://cdsweb.u-strasbg.fr/cgi-bin/qcat?J/A+A/}}.
For cluster sizes $N=3-10$,
the GM candidates predicted in the literature were
found among the candidate geometries after DFTB optimisation. For cluster size $N=11$ the 
predicted GM was found among the candidate geometries after DFT optimisation. For cluster sizes $N=$ 12,14 and 15, the predicted GM were not found among the candidate geometries and no 
more favourable isomer, i.e. with an even lower potential energy,
was found either. This is most likely due to the fact that the seed-creation approaches employed here are not well suited to cover the large parameter space for these large clusters within 10000 candidates, as was mentioned in Sec. \ref{sec:comparison}. For $N=13$, we present a new global minimum 
candidate
structure (Fig. \ref{fig:13_GM}), which was created through the mirror creation process (Method 3 in Section \ref{section:seedcreation}). Its potential energy is $0.5 \frac{\mathrm{kJ}}{\mathrm{mol}}$ per monomer unit lower than the previous lowest-lying isomer reported by \cite{Lamiel-Garcia2017}. This difference is lower than the typically assumed accuracy for DFT calculations of $4 \frac{\mathrm{kJ}}{\mathrm{mol}}$ per monomer unit. This supports the need to study energetically similar isomers for each cluster size, as either could play a role in the nucleation process.

In thermochemical processes such as nucleation, the 
most
relevant isomer for each cluster size is assumed to be the energetically most favourable one. Any 
less 
favourable, meta-stable
isomer that forms relaxes into that 
global minimum, given sufficient time for the relaxation.
To compare the 
thermochemical properties, the \texttt{thermo.pl} program is used to calculate the entropy S [J mol$^{-1}$ K$^{-1}$], change of enthalpy $d\Delta H$ [kJ mol$^{-1}$] and Gibbs free energy of formation $\Delta_f G^\circ$ [kJ mol$^{-1}$] for three different sets of clusters:
\begin{enumerate}
    \item The GM for each size, including cluster geometries reported in the literature, calculated with B3LYP/def2svp and empirical dispersion
    \item The GM for each size, including cluster geometries reported in the literature calculated with B3LYP/cc-pVTZ and empirical dispersion
    \item The GM for each size, only including cluster geometries that have been found in this work, calculated with B3LYP/cc-pVTZ and empirical dispersion
\end{enumerate}

This is done in order to compare the impact of the choice of functional/basis set on the zero-point energies and the thermochemical properties to the impact of the completeness of the GM search for the cluster geometries. The 
complete
thermochemical tables for set 2 are only available in electronic form at the CDS via anonymous ftp to \footnote{\url{cdsarc.u-strasbg.fr} (130.79.128.5)}
or via \footnote{\url{http://cdsweb.u-strasbg.fr/cgi-bin/qcat?J/A+A/}}.

\section{Astrophysically relevant (TiO$_2$)$_{\rm N}$ properties}
\label{sec:tio2astrophysics}
In
previous sections, we
have derived the fundamental 
quantities
including the zero-point energy of TiO$_2$ clusters. 
The thermochemical properties of interest for astrophysical studies are:
\begin{enumerate}
    \item S(T), Entropy [J mol$^{-1}$ K$^{-1}$]
    \item d$\Delta H$(T), change in enthalpy [kJ mol$^{-1}$] 
    \item $\Delta_f G^\circ$(T), Gibbs free energy of formation [kJ mol$^{-1}$].
\end{enumerate}
The following sections will present
TiO$_2$ in the 
context of the gas phase,  
and
TiO$_2$ clusters as a promising precursor of TiO2 dust formation, 
and the derivation of their thermochemical properties, especially the size- and 
temperature-dependent Gibbs free energy of formation $\Delta_f G^\circ (N,T)$, that are relevant for cloud formation modelling in exoplanets 
and brown dwarfs as well as for dust formation modelling in AGB stars and supernova ejecta.

\subsection{Thermodynamical relevance of TiO$_2$ in hot Jupiters}

\begin{figure} 
    \centering
    \includegraphics[width=\hsize]{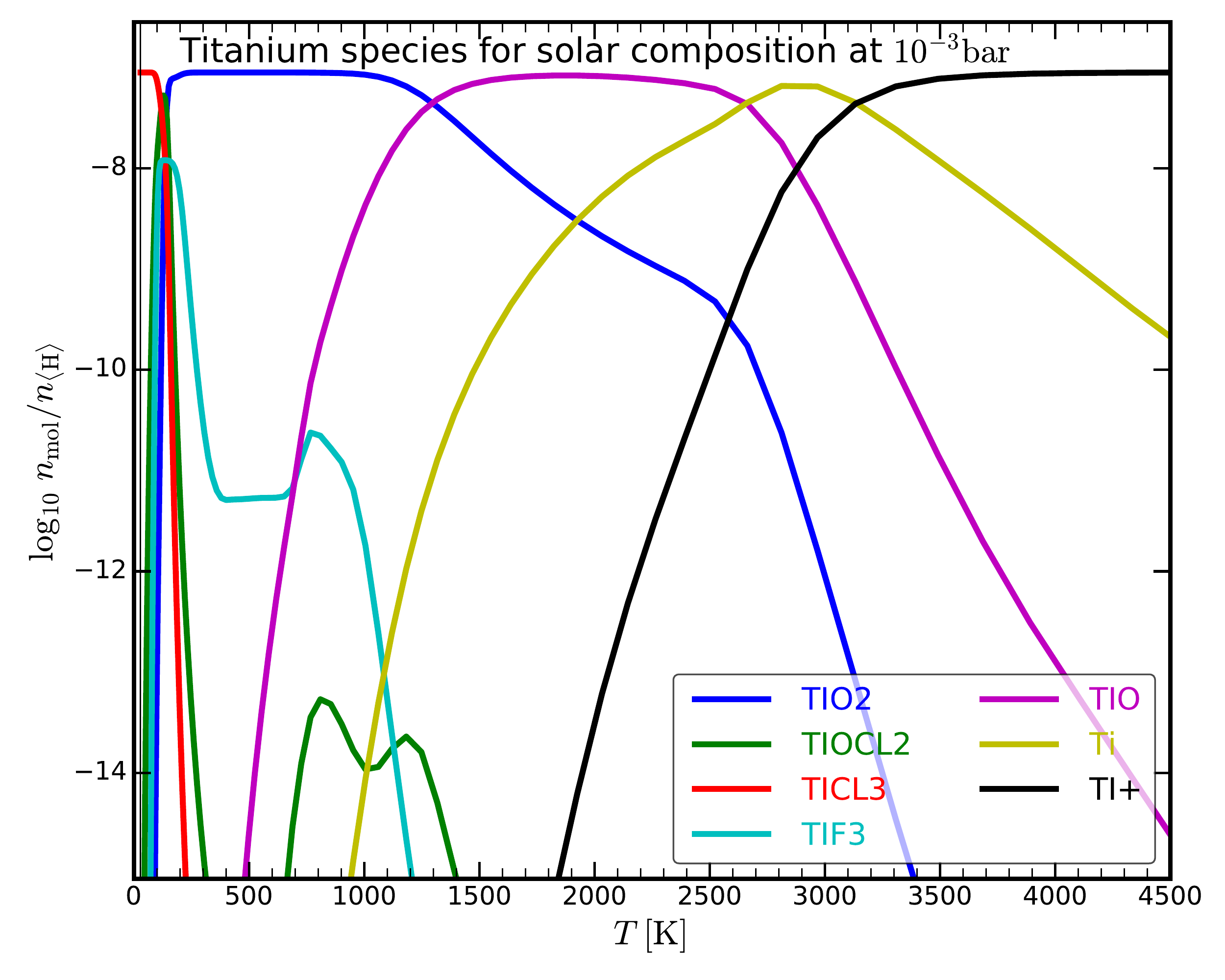}
    \caption{Concentrations of Titanium containing molecules in gas-phase chemical equilibrium at 10$^{-3}$ bar. Solar element abundance are assumed. TiO$_2$ is the most abundant Ti-binding species up to 1200K and the second most abundant species up to 1800K.}
    \label{fig:tio2_phase_onepressure}
\end{figure}

\begin{figure} 
    \centering
    \includegraphics[width=\hsize]{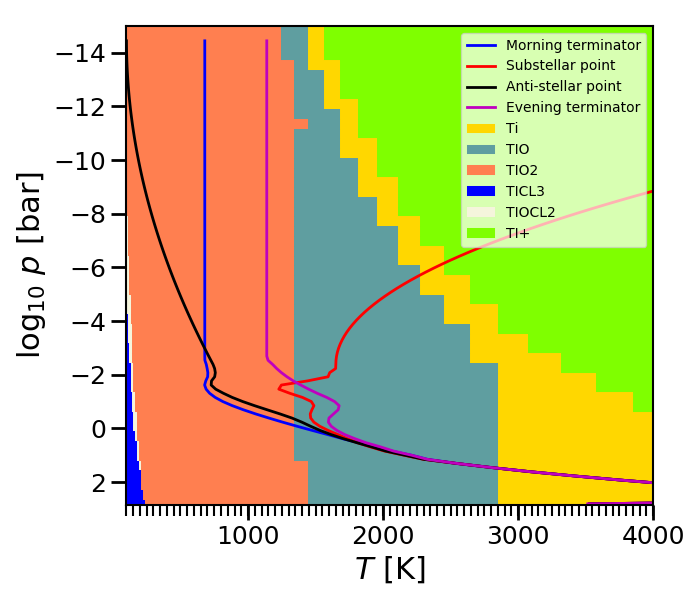}
    \caption{Most abundant gas phase molecules containing Titanium from a gas of solar composition using equilibrium chemistry. Pressure temperature profiles for a hot Jupiter atmospheres with $T_{\rm eff} = 1600$K included, probing the atmosphere at four distinct locations.}
    \label{fig:tio2_phase}
\end{figure}

In order to assess the relevance of TiO$_2$ for cloud formation 
in the atmospheres of hot Jupiters or for dust formation in an 
AGB envelope, it is necessary to know the thermodynamic
quantity
ranges in which TiO$_2$ is the most abundant Titanium-bearing 
molecule. For this we apply the gas-phase equilibrium 
chemistry code {\sc GGChem} from 
\citet{Woitke2018EquilibriumRatio}.
TiO$_2$ is the most abundant Titanium-bearing molecule at a pressure of $10^{-3}$ bar in the low-temperature regime 
i.e. for T$<1$200 K (Figure~\ref{fig:tio2_phase_onepressure}). 
Less complex Ti-bearing 
species dominate the chemical Ti-
content
with increasing temperatures until Ti$^+$ is the dominating Ti species at $T>3500$K.
Figure~\ref{fig:tio2_phase} presents the {\sc GGChem} model as a 2D ($p_{\rm gas}$, $T_{\rm gas}$) plane with a 1D atmospheric $p_{\rm gas}$, $T_{\rm gas}$ profile of a hot Jupiter superimposed. The
profile 
corresponds to a model
of a hot Jupiter with $T_{\rm eff} = 1600$K and 
$\log(g)$ = 3, orbiting a G star with $T_* = 5650$K \citep{Baeyens2021GridMixing} 
which was extrapolated to pressures of $10^{-14}$ bar.
The model is plotted at four different locations on the planet: the substellar and anti-stellar point, and the morning and evening terminator for this hot Jupiter.

 Only at the substellar point for the model atmosphere ($T_{\rm eff} = 1600K$) 
 TiO$_2$ is not the most abundant Ti-bearing species at any pressure-level, except at a thin layer of the atmosphere at $10^{-2}$ bar. For the anti-stellar point and morning terminator, TiO$_2$ is the dominant Ti-bearing molecule in the upper atmosphere. Only in the lower atmosphere at around $10^{-1}$ bar, other molecules such as TiO and later atomic Ti become dominant. For the evening terminator, TiO$_2$ 
 becomes less abundant than
 TiO at a pressure level of $\sim10^{-2}$ bar. Moreover, we note that the (TiO$_2$)$_N$ cluster ionization energies of 9.3-10.5 eV are too high to affect related abundances and the TiO$_N$ nucleation in exoplanet atmospheres\citep{Gobrecht2021TheClusters}. This strengthens the argument that TiO$_2$ is the most relevant Ti-bearing molecule in the upper atmosphere, which is were cloud formation is expected to take place.

\subsection{Surface tension of TiO$_2$}
In classical and modified classical nucleation theory, the impact of the Gibbs free energy of formation $\Delta_fG^\circ$ on the nucleation process is modeled through the surface tension $\sigma_{\infty}$ of the bulk solid \citep{Gail2013a}. The fact, that the surface tension is different for small 
a finite size leading to significantly
different
geometries 
and properties 
than the bulk, is neglected. It is however possible to find a surface tension that is valid for small 
clusters if individual cluster data are available. Here the approach by \cite{Jeong2000ElectronicTiSUBx/SUBOSUBy/SUB} as it was applied by \cite{Lee2015} is followed, in which the dependence of the Gibbs free energy of formation on the cluster size is linked to the surface tension $\sigma_{\infty}$ through:
\begin{equation}
\scalebox{0.94}{$\frac{\Delta_fG^\circ(N)}{N} = \left(\theta_{\infty}RT\frac{N-1}{(N-1)^{1/3} + N_f^{1/3}} + \Delta_f G^\circ(1) + (N-1) \Delta_f^\circ G^\circ_1(s) \right) \cdot N^{-1}$}
\label{eq:deltaGNoverN}
\end{equation}
with cluster size $N$, the Gibbs free energy of formation of a cluster of size N, $\Delta_fG^\circ(N)$, the Gibbs free energy of the monomer $\Delta_f G^\circ(1)$, the Gibbs free energy of the bulk phase $\Delta_f G^\circ_1(s)$, a fitting factor $N_f$ and
\begin{equation}
    \theta_{\infty} = \frac{4\pi a_0^2 \sigma_{\infty}}{k_bT}
    \label{eq:theta_inf}
\end{equation}
Here $a_0$ is the theoretical monomer radius, which is derived from the bulk density of rutile and the molar mass through:
\begin{equation}
    a_0 = \left(\frac{3 M_{TiO_{2}}}{4\pi\rho_{\rm Rutile}}\right)^{1/3} \simeq 1.956 \times 10^{-10} m
\end{equation}
The fitting factor $N_f$ is set to $N_f = 0$, analogous to \cite{Lee2015}.
The Gibbs free energies for the monomer and for the bulk are known from experiment \citep{MalcolmW.Chase1998NIST-JANAFTables}, as are the values necessary to derive $a_0$. The Gibbs free energy of formation of the clusters is calcluated from the thermochemical values derived from 
our all-electron
DFT calculations and 
consistently compared to the thermochemical clusterdata from the study by
\cite{Lee2015}.

\begin{figure*}
    \centering
    \includegraphics[width = 0.49\hsize]{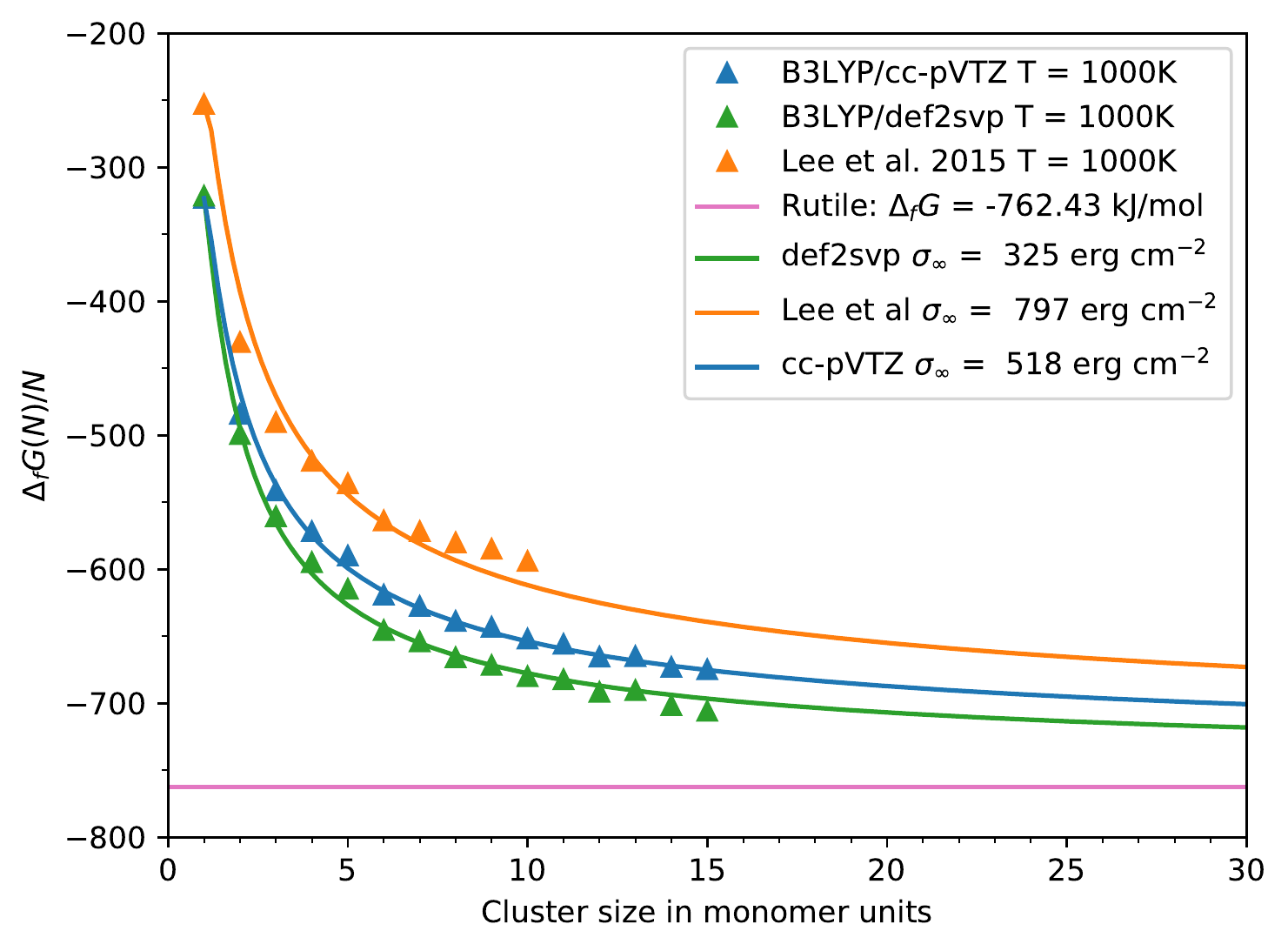}
    \includegraphics[width = 0.49\hsize]{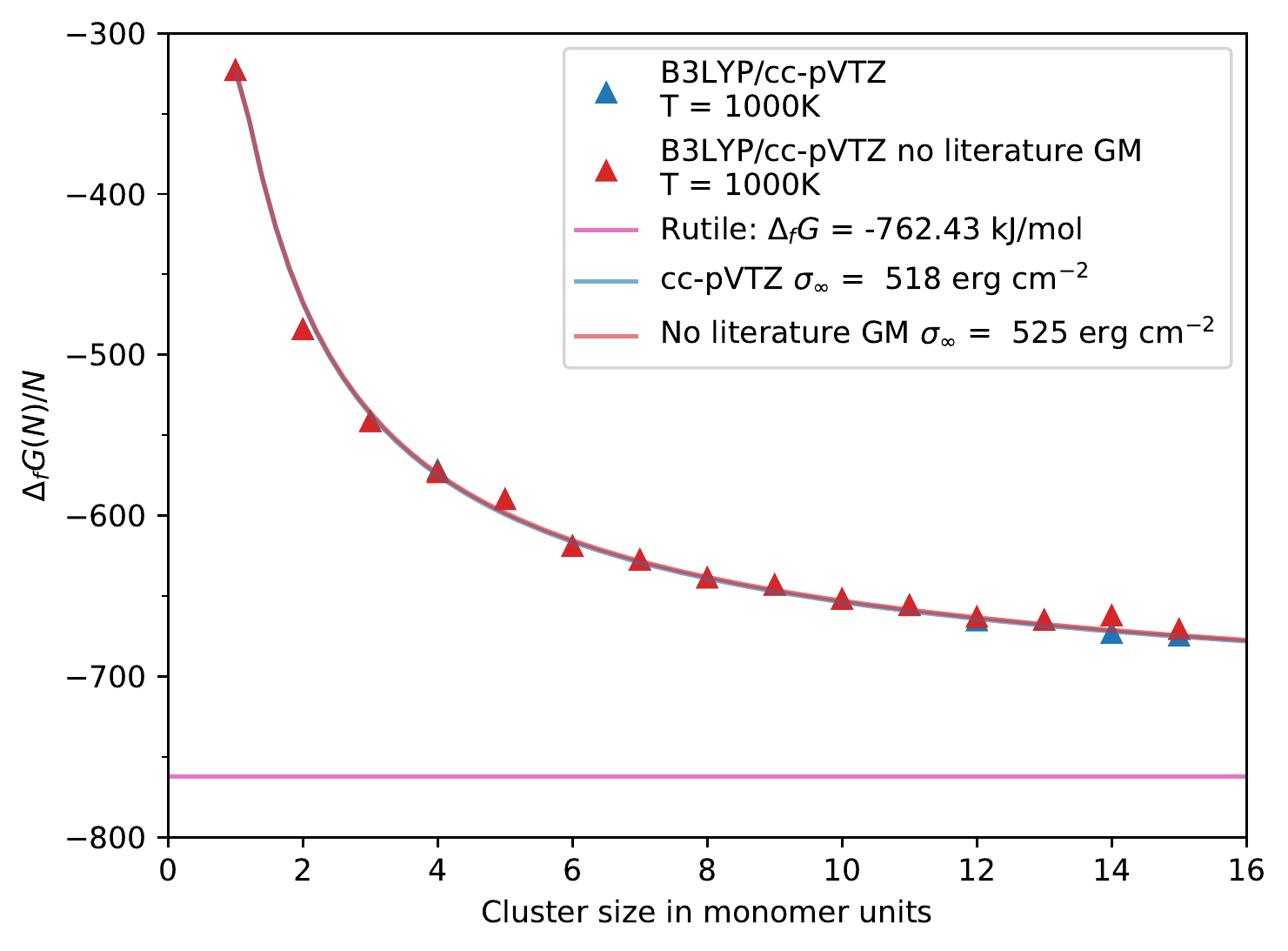}
    \caption{Gibbs free energy of formation per cluster size $N$ 
    as a function of cluster size $N$ at a temperature of T = 1000K. For each approach a fit for $\sigma_{\infty}$ was calculated using eq. \ref{eq:deltaGNoverN}. \textbf{Left:} Comparison between resulting surface tensions for different sources of cluster data. The sources are \cite{Lee2015} for the orange line, DFT calculations with the fast basis set def2svp for the green line and DFT calculations with the accurate basis set cc-pVTZ for the blue line. \textbf{Right:} Comparison of the impact of isomer completeness on resulting surface tension. Both lines use thermochemical data derived from the accurate cc-pVTZ basis set DFT calculations. For the blue line the energetically most favoured isomer was chosen for all sizes, regardless of whether it was found by the approach in this paper or not. For the red line only outputs from this papers approach were used. As the resulting difference is small it becomes apparent that the choice of the basis set is more impactful than the completeness of the cluster geometries.}
    \label{fig:surfacetension}
\end{figure*}

That leaves $\sigma_{\infty}$ as the only free parameter, which 
is fit using Eq. \ref{eq:deltaGNoverN}.
The left-hand side of Figure \ref{fig:surfacetension} compares 
the data from both functional/basis set combinations used in  
this work to the results from \cite{Lee2015}. It becomes 
apparent that the choice of the functional and basis set influences the resulting thermochemical properties and 
therefore the surface tension given by the fit. In this work we get a different result for the surface tension than \cite{Lee2015} using their data. The result for the surface tension at $T = 1000$K using all available GM data and the best performing functional/basis set combination is $\sigma_{\infty} = 518 \ \mathrm{erg \ cm^{-2}}$ (blue line in Fig. \ref{fig:surfacetension}). In the right panel of Figure \ref{fig:surfacetension}, the comparison is made between the fits for $\sigma_{\infty}$ using all available GM data (blue) and only the best candidates that were produced from candidates in this work (red). Both the data as well as the results for $\sigma_{\infty}$ vary only slightly, visible differences occurring for $N=14$ and $N=15$, where this work did not produce candidates that were identical or energetically close 
to the literature GM. The best fit value for $\sigma_\infty$ 
without using GM candidates found in the literature is 
$\sigma_\infty  = 525 \ {\rm erg \ cm^{-2}}$, differing from 
the best fit value by only $7 \ {\rm erg \ cm^{-2}}$. This 
indicates that the choice of functional and basis-set has a 
stronger impact on the resulting surface tension than finding 
the 
lowest-energy isomer
for all sizes,
given the extent of our searches.

Since potential 
minimum geometries that were missed all have lower 
enthalpies and therefore, presumably also lower
Gibbs free energies, they can only lower the resulting surface tension. 
This approach therefore gives an upper limit for the surface 
tension $\sigma_\infty$.

\begin{figure}
    \centering
    \includegraphics[width=\hsize]{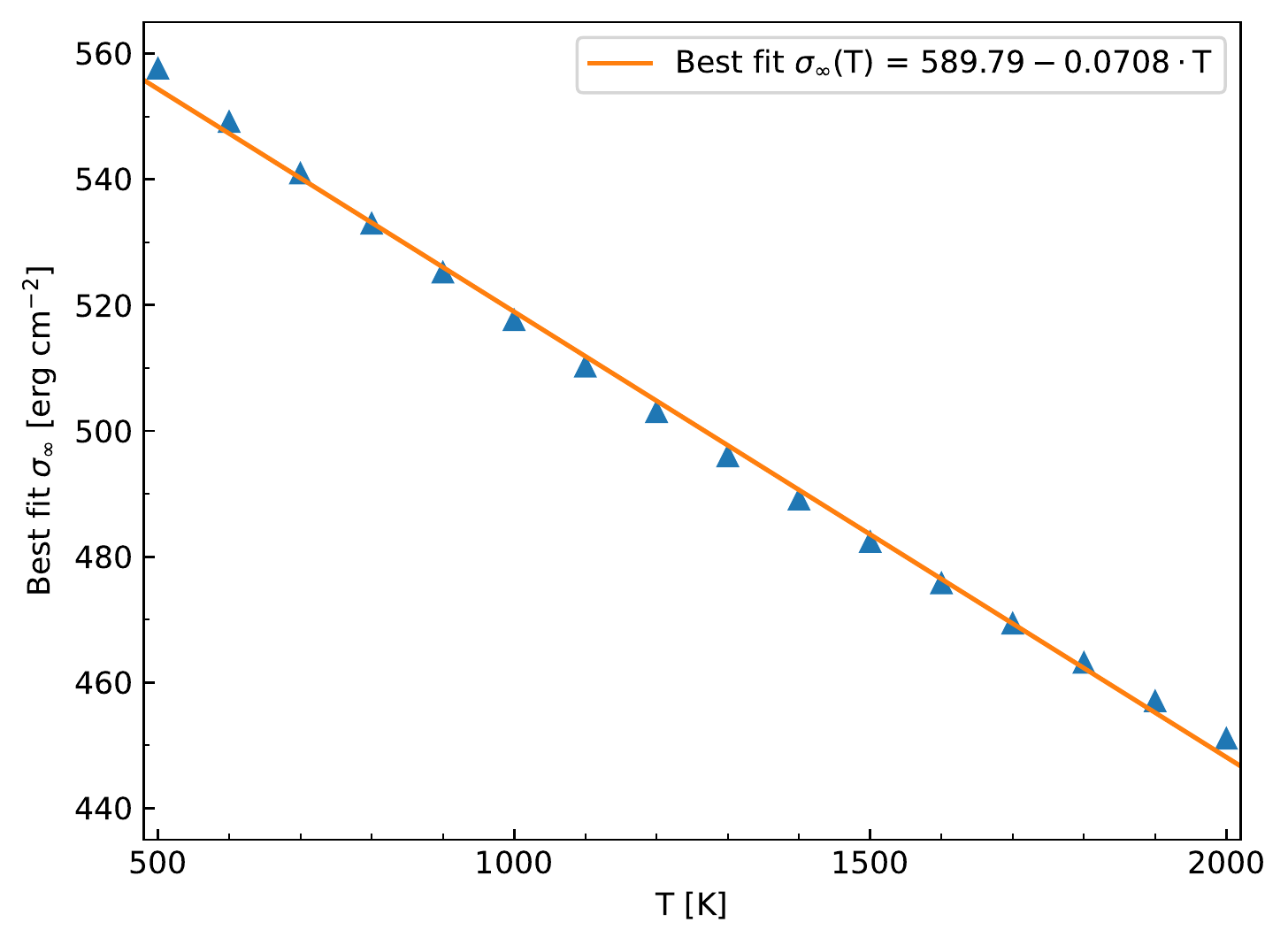}
    \caption{Dependence of the best fit for $\sigma_{\infty}$ on the temperature $T$. A linear regression has been applied.}
    \label{fig:sigma_temp}
\end{figure}

The best fit value for the surface tension is found to be dependent on the temperature. Fig. \ref{fig:sigma_temp} shows the best fit for a linear dependence of $\sigma_{\infty}$ on T for T = $500 - 2000$ K:
\begin{equation}
    \sigma_{\infty} (T) = 589.79-0.0708\cdot T
\end{equation}
Studies have been conducted on the surface tension of bulk Rutile at room temperature \citep{Kubo2007Surface011}, finding a value of $\sigma_{\infty}(298.15\mathrm{K}) = 1001 \ \mathrm{erg \ cm^{-2}}$ for the (011) lattice. The  best fit for room temperature in this study gives a value of $\sigma_{\infty}(298.15\mathrm{K}) = 575.72 \ \mathrm{erg \ cm^{-2}}$ for small (TiO$_2$)$_N$ molecular clusters. The factor of almost two in surface tension for small 
clusters versus the bulk phase of TiO$_2$ makes it clear that the prior 
is recommended
in all considerations regarding nucleation processes.

\subsection{Nucleation rates of TiO$_2$}
To quantify the effect of the updated thermochemical cluster data on quantities relevant for cloud formation in exoplanet atmospheres, the nucleation rate for TiO$_2$ is calculated along the temperature pressure profile of the morning terminator of a hot Jupiter with an effective temperature of $T_{\mathrm{eff}} = 1600$K and a surface gravity of $\log g = 3$ (blue lines in Fig. \ref{fig:tio2_phase}). The gas phase composition is calculated with the equilibrium chemistry code {\sc GGChem}.

\subsubsection{Modified classical nucleation theory (MCNT)}
\label{sec:mcnt}
\begin{figure}
    \centering
    \includegraphics[width = \hsize]{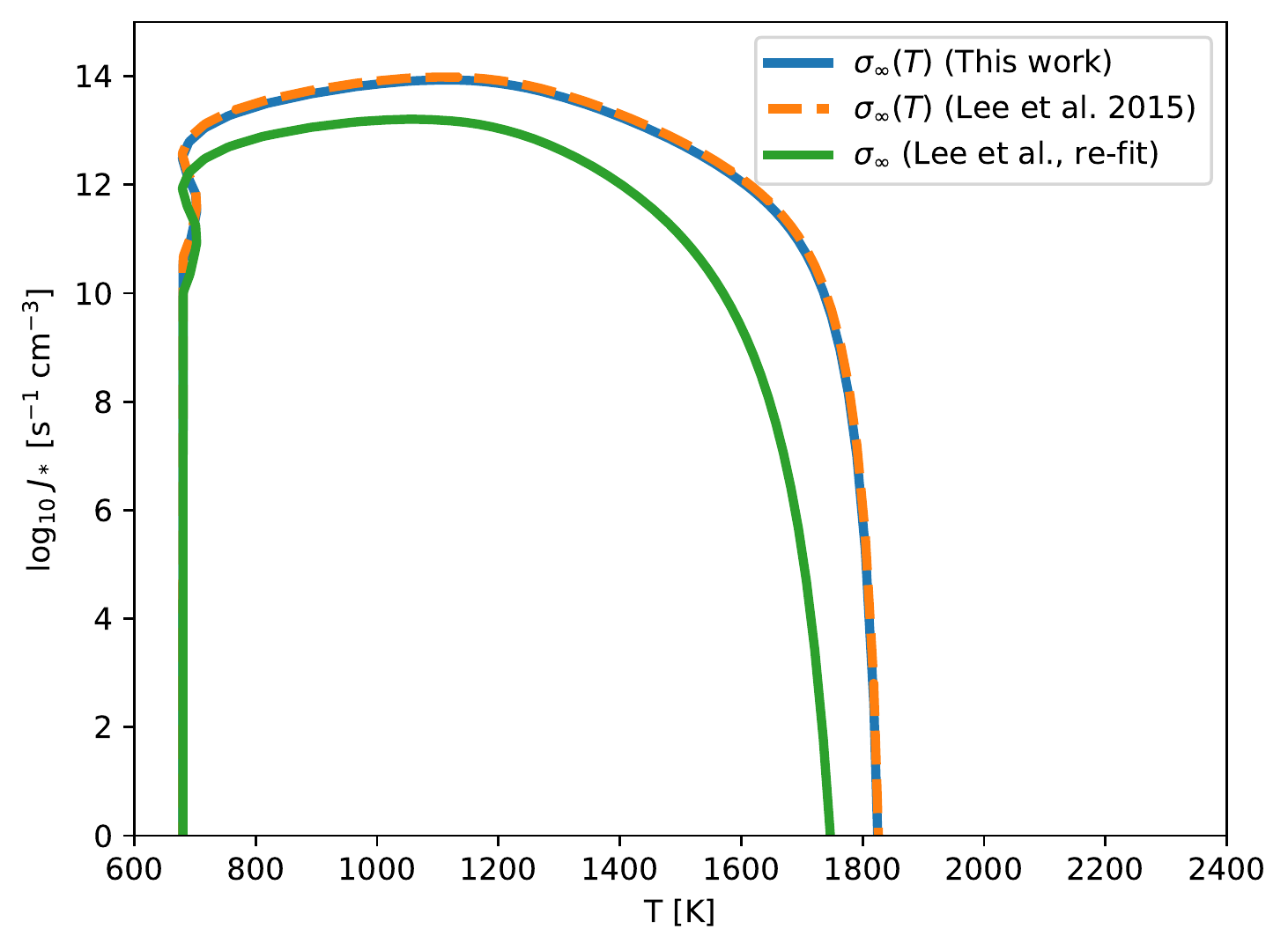}
    \includegraphics[width = \hsize]{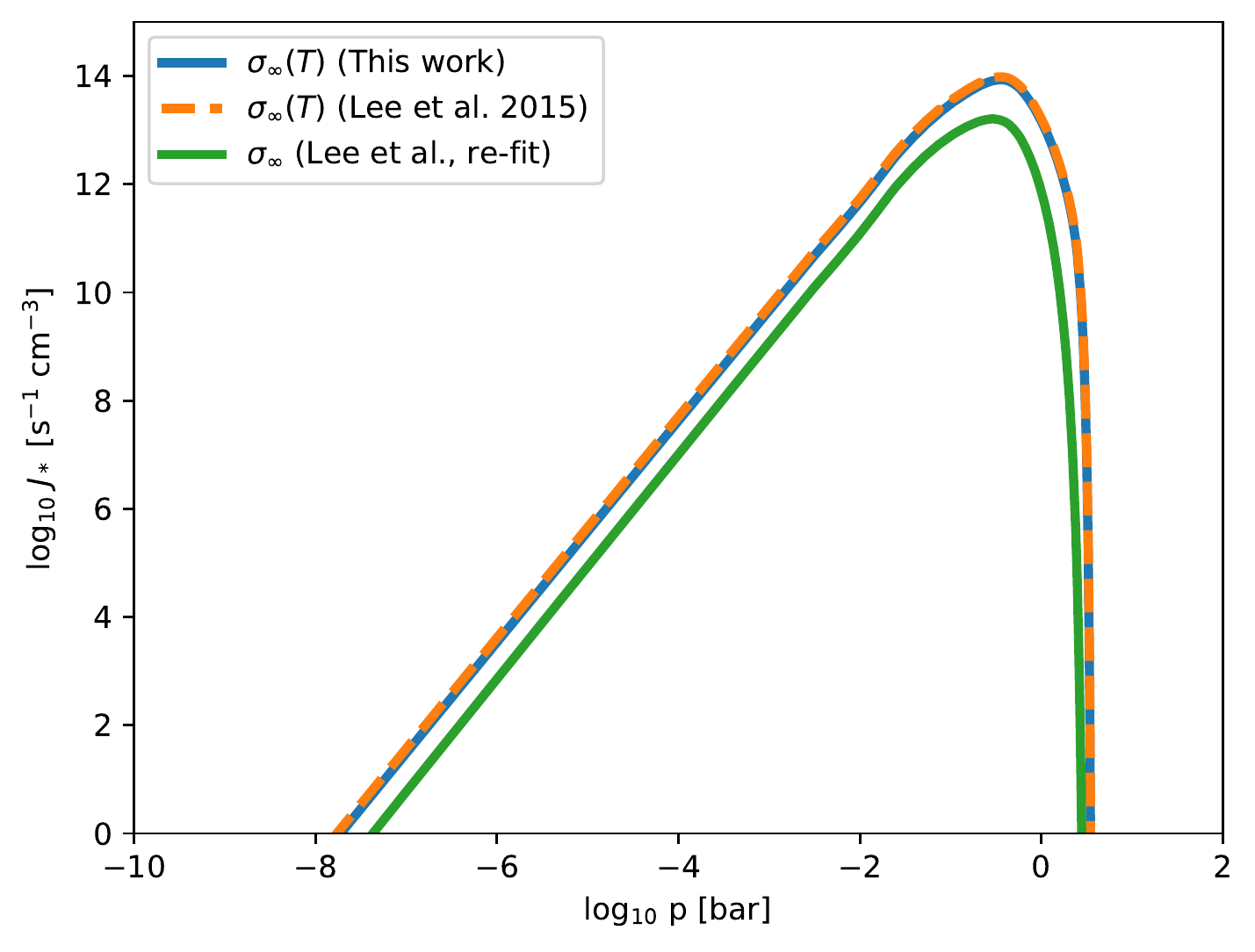}
    \caption{The classical nucleation rate for TiO$_2$ for three different surface tensions $\sigma_\infty$. The temperature - pressure profile is equivalent to the solid blue line in Figure \ref{fig:tio2_phase}. \textbf{Top:} The nucleation rate at different temperatures. Nucleation rates derived from updated cluster data are overall more efficient and extend to higher temperatures. \textbf{Bottom:} The position of the peak of the nucleation rate within the atmosphere 
    is very similar for the three different surface tensions
    However, nucleation stays efficient up until slightly lower pressures for updated cluster data.}
    \label{fig:J_classical}
\end{figure}

In this work, the nucleation rates are computed analogously to \citet{Lee2015} for ease of comparison.
The stationary, homogeneous, homomolecular nucleation rate in classical nucleation theory is calculated by:
\begin{equation}
    \centering
\label{eq:classicalnucleation}
     J^c_* = \frac{f^\circ(1)}{\tau_{gr}(r_{i},N_*,T)}Z(N_*)\exp\left((N_*-1)\ln S(T) - \frac{\Delta G(N_*)}{RT}\right)
\end{equation}
Here, $S(T)$ and $f^\circ(1)$ are the supersaturation ratio and monomer  number density of TiO$_2$ respectively. $\tau_{gr}$ is the growth timescale, defined as:
\begin{equation}
    \centering
    \label{eq:tau}
    \tau_{gr}^{-1} = A(N) \alpha(N) v_{rel} n_f
\end{equation}
with $A(N) = 4 \pi a_0^2 N^{2/3}$ the effective cross-section of a spherical
(TiO$_2$)$_N$ cluster, $n_f$ the monomer number density ($n_f = 
f^\circ(1)$), the sticking factor $\alpha$, which is assumed to be 
$\alpha = 1$, and the relative velocity $v_{rel}$, which is 
given for monomers with mass $m_x$ by the thermal velocity through:
\begin{equation}
    \centering
    \label{eq:v_rel}
    v_{rel} = \sqrt{\frac{kT}{2\pi m_x}}
\end{equation}
Z(N) is the Zeldovich factor, which accounts for the contribution to nucleation from Brownian motion:
\begin{equation}
    \centering
\label{eq:zeldovich}
    Z(N_*) = \sqrt{\frac{\theta_\infty}{9\pi(N_*{-1})^{4/3}}}
\end{equation}
and in the final term, the Gibbs free energy $\Delta G(N)$ is approximated using modified classical nucleation theory, or MCNT, giving:
\begin{equation}
\label{eq:del_G_over_RT}
    \centering
    \frac{\Delta G(N_*)}{RT} = \theta_\infty (N_*-1)^{2/3}
\end{equation}
with $\theta_\infty$ from Eq. \ref{eq:theta_inf}. Equation \ref{eq:classicalnucleation} is evaluated at the critical cluster size $N_*$, which is given by:
\begin{equation}
\label{eq:critcluster}
    \centering
    N_{*, \infty} = \left(\frac{\frac{2}{3}\theta_\infty}{\ln S(T)}\right)^3 \ \ \  N_* =  \frac{N_{*,\infty}}{8} + 1
\end{equation}
The critical cluster size is mainly influenced by the supersaturation of molecular TiO$_2$ at various temperature and pressure points. When TiO$_2$ is not supersaturated ($\ln S(T) < 0$), N$_*$ is negative (Eq \ref{eq:critcluster}), and consequently Eq. \ref{eq:del_G_over_RT} has no real solution, leading to the absence of nucleation for the modified classical case at these temperature pressure points.
Using Eq. \ref{eq:classicalnucleation}, the classical nucleation rate was calculated for the given ($\rm p_{gas}$,$\rm T_{gas}$-profile using three different values for $\sigma_\infty$: The temperature dependent $\sigma_\infty$ from this work, the temperature dependent $\sigma_\infty$ from \cite{Lee2015} and a constant $\sigma_{\infty} = 797$ erg cm$^{-2}$, derived from the re-fit to Lee's cluster data in this work. Results can be found in Figure \ref{fig:J_classical}. 

\subsubsection{Non-classical nucleation theory}
\begin{figure}
    \centering
    \includegraphics[width=\hsize]{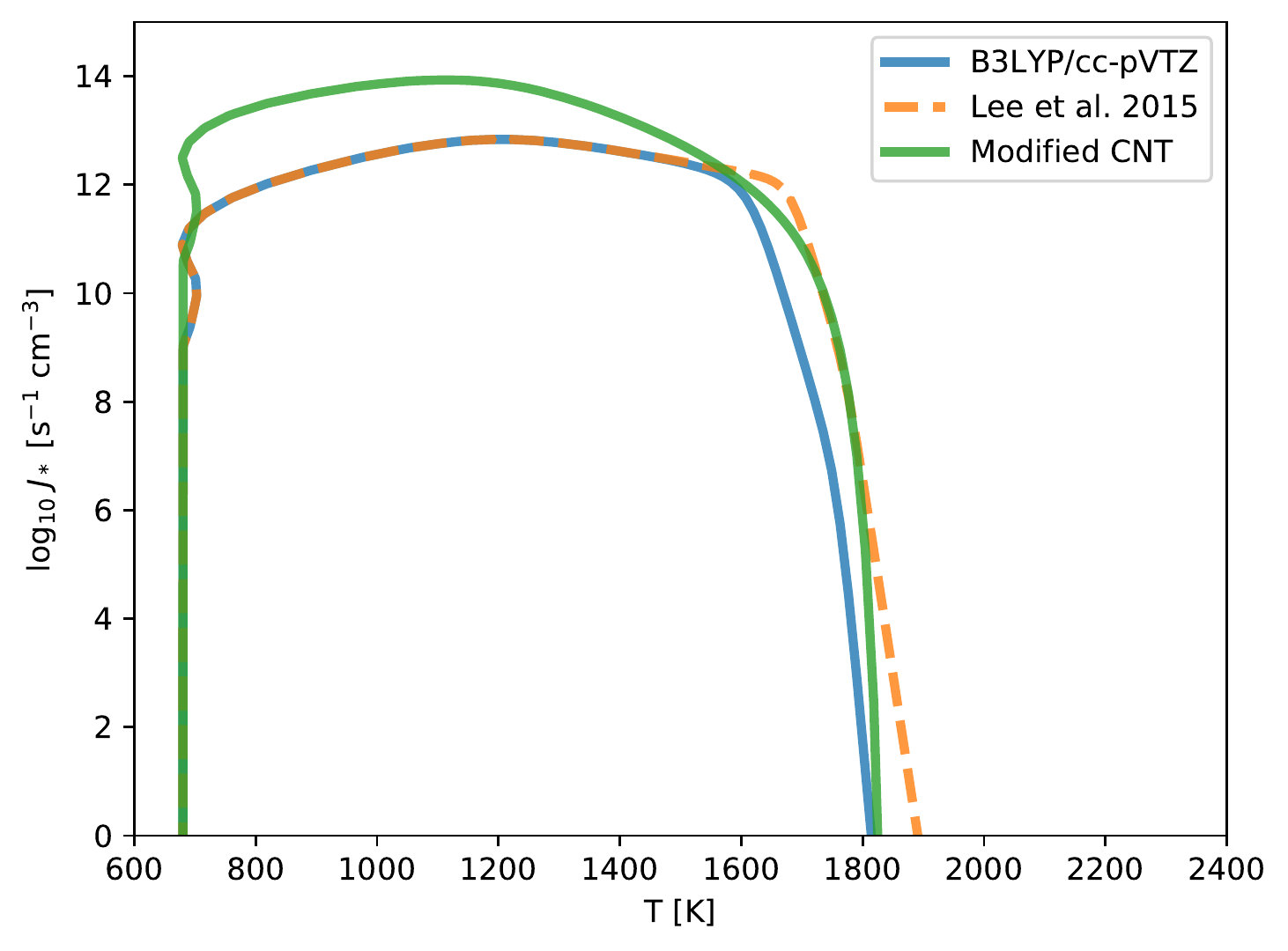}
    \includegraphics[width=\hsize]{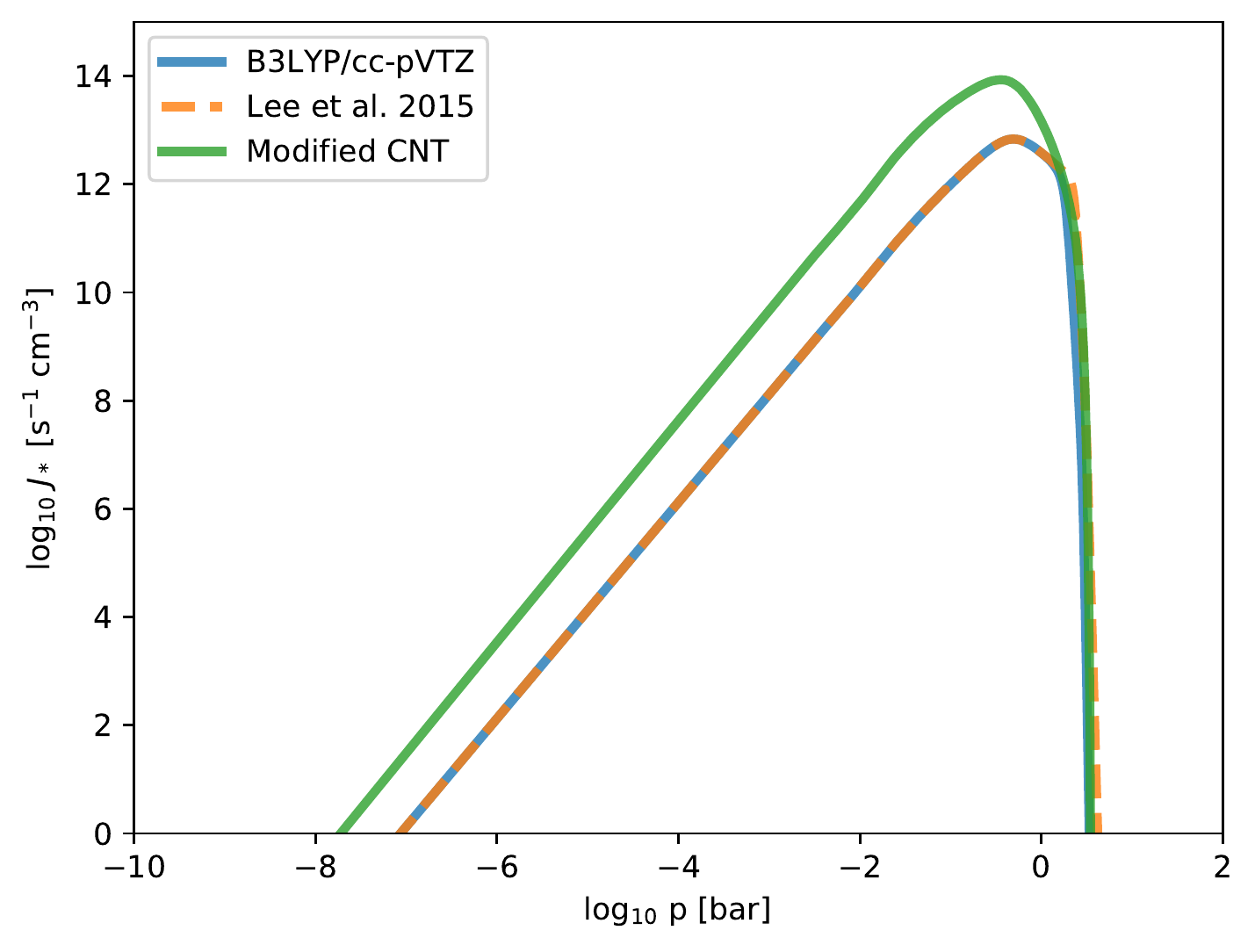}
    \caption{Nucleation rates $J_*$ for non-classical nucleation theory. The temperature - pressure profile is equivalent to the solid blue line in Figure \ref{fig:tio2_phase}. Using individual cluster data from this work  (B3LYP/cc-pVTZ) (blue), cluster data from \cite{Lee2015} (orange dashed) and the modified CNT approach from section \ref{sec:mcnt} (green). \textbf{Top:} Dependence on temperature. \textbf{Bottom:} Dependence on pressure.} 
    \label{fig:J_nonclassic}
\end{figure}

If detailed cluster data for all small sizes $N$ are available, the nucleation rate can be computed using individual cluster growth rates.
The non-classical nucleation rate is: 
\begin{equation}
    \centering
    J_*^{-1}(T) = \sum^{N_{max}}_{N=1} \left( \frac{\tau_{gr}(N,T)}{f^\circ(N,T)} \right)
\end{equation}
with $\tau_{gr}$ from Eq. \ref{eq:tau} and $f^\circ(N)$ the number density of a cluster of size $N$. This can be computed from the partial pressure of 
the latter
through:
\begin{equation}
    \centering
    f^\circ(N) = \frac{p^\circ(N)}{kT}
\end{equation}
Applying the law of mass action to a cluster of size $N$ gives its partial pressure as:
\begin{equation}
\label{eq:partialpressures_delG}
    \centering
p^\circ (N) = p^{\standardstate} \left( \frac{p^\circ(1)}{p^{\standardstate}} \right)^N \exp \left( - \frac{\Delta_f G^\circ(N) - N\Delta_f G^\circ(1)}{RT} \right)
\end{equation}
with $p^\circ (1)$ the partial pressure of the monomer and the reference pressure $p^\standardstate = 1$bar. 
The results for the non-classical nucleation rates can be found in Figure \ref{fig:J_nonclassic}.
\subsection{Results for TiO$_2$ nucleation rates}
For the modified classical nucleation approach, three values for the surface tension were compared. The shape of the nucleation rate in Fig. \ref{fig:J_classical} is influenced by several factors. The reason for no nucleation occurring below a temperature of $T \approx 680$K is due to the p$_{\rm gas}$, T$_{\rm gas}$-profile used, which does not extend to lower temperatures (Fig. \ref{fig:tio2_phase}). On the upper temperature end, nucleation is limited due to its dependence on the supersaturation of the TiO$_2$ monomer. Because supersaturation ($S>1$) is required for nucleation and TiO$_2$ is no longer super-saturated at these high temperatures, no nucleation occurs.
For lower temperatures, the surface tension from this work results in a higher nucleation rate than the recomputed surface tension with cluster data from for \cite{Lee2015} by $\sim$1 order of magnitude. The lower surface tension also leads to nucleation becoming inefficient at higher temperatures. Overall the value for the surface tension from this work agrees well with the value derived in \cite{Lee2015}. However, using their cluster-data, we find a significantly higher surface tension (see Fig \ref{fig:J_classical}). 

Nucleation rates for non-classical nucleation (Fig. \ref{fig:J_nonclassic}) are lower than for 
MCNT for $T = 680-1600$K by about 2 orders of magnitude. This is the result of taking into account individual cluster data for all sizes, instead of combining all the information into one quantity, the surface tension. 
Therefore, 
monomer growth processes
from size $N$ to size $N+1$ are modelled more accurately. If one of the 
growth processes
is less efficient than others it will bottleneck the overall nucleation rate. 
In this case
$N+1$ 
is considered as
the critical cluster size $N_*$. 
At higher temperatures ($T > 1600$) cluster data from this work gives lower nucleation rates than both cluster data from \cite{Lee2015} and MCNT. There is no
strict upper temperature limit for non-classical nucleation, as clusters can grow as long as it is energetically favorable for them to do and as long as nucleating material, i.e. TiO2 monomers, are available (Eq. \ref{eq:partialpressures_delG}). Since our updated cluster-data does predict nucleation of TiO$_2$ becoming inefficient at lower temperatures, the nucleation process starts at lower pressure levels, i.e. higher up in the model atmosphere.

\section{Conclusion}
\label{sec:discussion}
This paper presents a method to find and 
optimise the geometries, zero-point energies, and thermochemical properties for
clusters. 
Emphasis is put on exploring the parameter space of possible 
geometric configurations for these clusters as well as deriving their potential energies and thermochemical properties 
accurately. This approach was tested for small ($N = 3-15$) 
(TiO$_2$)$_N$ clusters. To ensure thermochemical accuracy, 129 combinations of DFT functionals and basis sets were tested with regards to their accuracy against known experimental data for the TiO$_2$ monomer. The B3LYP functional with the cc-pVTZ basis set and GD3BJ empirical dispersion 
was
found to closely approach the experimental data and have been used for all final optimisation steps and frequency analysis.
A new force field parameterisation of the Buckingham-Coulomb pair potential was presented, that more accurately reflects the cluster geometry (i.e. bond lengths)
and energetic ordering for small TiO$_2$ 
clusters than previous parameterisations. 
For the DFTB description of interactions, the \textit{matsci} set of Slater-Koster integrals was found to best reflect the energetic ordering for small clusters and their isomers given by the all-electron DFT calculations. 
The hierarchical optimisation approach works as intended and produces a large number of energetically low-lying isomers for all sizes. For the smallest clusters $N < 7$, all global minimum 
candidates reported in the literature
were found with methods that do not rely on using known cluster geometries. 
Since these are the same cluster sizes that are used to calibrate the less complex descriptions of inter-atomic potentials, this step of the approach is therefore independent of prior knowledge of cluster geometries.
The \textit{Random} approach of seed candidate creation is well suited to search small parameter spaces, such as $N < 7$. However, for any cluster sizes $N > 10$, the parameter space is too large to be searched with a fully randomised approach. For these larger cluster sizes the 
\textit{Mirror} and \textit{Known+1} approaches produce the largest number of energetically low-lying isomers. However, they require prior knowledge of 
favourable
clusters of size $N-1$ for \textit{Known+1} and of size $N/2$ for \textit{Mirror}. This presents a constraint on the efficient search for clusters of these sizes, as without prior knowledge the 
global minima in potential energy will have to be found iteratively, always growing from $N$ to $N+1$. 
The current implementation of the hierarchical approach was able to find all known global minima for $N = 3-11$, as well as a new global minimum candidate for $N=13$ that lies 6 kJ mol$^{-1}$ below the energy of the global minimum candidate structure
reported by \citet{Lamiel-Garcia2017}. For $N=12$ the global minimum known from the literature could not be reproduced, the closest isomer produced lies $2.45 \frac{\mathrm{kJ}}{\mathrm{mol}}$ per monomer unit higher. For $N=14$ this energetic distance between the lowest found isomer and the known global minimum is 
$5.71 \frac{\mathrm{kJ}}{\mathrm{mol}}$ per monomer unit and for $N=15$ it is $6.8 \frac{\mathrm{kJ}}{\mathrm{mol}}$ per monomer unit.
This shows that the search method is still incomplete for larger 
clusters. Since only clusters up to size $N=6$ had isomers that were used for calibration, the \textit{Mirror} approach only had a single literature isomer to generate the clusters for $N=14$ and $N=15$, which drastically reduces the possible configurations explored by this approach. 

To estimate the impact of the new thermochemical data on nucleation processes, the surface tension for small molecular clusters, as it is calculated in modified classical nucleation theory, was investigated. 
The findings of \cite{Lee2015} can not be reproduced, raising the best fit value for $\sigma_{\infty}$ for the test case at T$=1000$K to $\sigma_{\infty} = 797$ erg cm$^{-2}$. The fast basis set def2svp gives $\sigma_{\infty} = 325$ erg cm$^{-2}$, while the thermochemically more accurate basis set cc-pVTZ gives $\sigma_{\infty} = 518$ erg cm$^{-2}$. The spread between these values is a factor of two, and as the surface tension 
appears in the exponent of the modified nucleation rate (Eq.  \ref{eq:classicalnucleation})
 this spread is amplified. Since the B3LYP/cc-pVTZ functional/basis set was specifically chosen for its accuracy at modelling the thermochemical properties of small TiO$_2$
molecular clusters, the 
updated
value
for the surface tension represents the current best approximation.
We find that for modified classical nucleation theory, the impact 
on the nucleation rates by the choice of the 
functional and basis set used to calculate the 
thermochemical properties is more important than finding a true global minimum, as long as the used structure is among the low-energy isomers. This is not true for non-classical nucleation description, which depends on accurate data for all sizes.
The non-classical nucleation process described by the updated cluster data becomes inefficient at lower temperatures, putting the 
 atmospheric lower border for seed formation through that 
 process higher.
 A limitation of this description is that it only allows for homogeneous and homomolecular nucleation, ignoring pathways through cluster-cluster collisions and through species other than TiO$_2$. 
 
We have shown that the updated cluster data has an impact on the nucleation rates both in their classical and non-classical description.
Providing cluster data for larger clusters $N>15$ will allow a more detailed comparison with independent methods, for  example with molecular dynamics methods as presented in \cite{Kohn2021DustApproach}.
Additionally the spectroscopic properties of small (TiO$_2$)$_N$ clusters, 
in particular the frequency-dependent opacities
are interesting because they can allow for dust coagulation processes to be constrained through observations \citep{Kohler2012DustObservations}.
The methodology will be improved by additional candidate creation methods and applied to other potentially CCN-forming species in the atmospheres of hot Jupiters.

\begin{acknowledgements}
The authors acknowledge computing time from the Kennedy HPC in St Andrews. J.P.S. acknowledges a St Leonard's Global Doctoral Scholarship from the University of St Andrews and funding from the Austrian Academy of Science. Ch.H. and L.D. acknowledge funding from the European Union H2020-MSCA-ITN-2019 under Grant Agreement no. 860470 (CHAMELEON). R. Baeyens is thanked for providing the origial 3D GCM profiles, and D. Lewis for providing the extrapolated version.
 D.G. and L.D. acknowledge support from the ERC consolidator grant 646758 “AEROSOL”.
 D.G. acknowledges the Knut and Alice Wallenberg foundation for financial support.
\end{acknowledgements}

\bibliographystyle{aa}
\bibliography{references}

\begin{thebibliography}{95}
\expandafter\ifx\csname natexlab\endcsname\relax\def\natexlab#1{#1}\fi

\bibitem[{Adamo \& Barone(1999)}]{Adamo1999TowardModel}
Adamo, C. \& Barone, V. 1999, Journal of Chemical Physics, 110, 6158

\bibitem[{Andres \& Kasgnoc(1998)}]{Andres1998AEmissions}
Andres, R.~J. \& Kasgnoc, A.~D. 1998, Journal of Geophysical Research:
  Atmospheres, 103, 25251

\bibitem[{Andriesse {et~al.}(1978)Andriesse, Donn, \&
  Viotti}]{Andriesse1978TheCarinae}
Andriesse, C.~D., Donn, B.~D., \& Viotti, R. 1978, Monthly Notices of the Royal
  Astronomical Society, 185, 771

\bibitem[{Apai {et~al.}(2013)Apai, Radigan, Buenzli, Burrows, Reid, \&
  Jayawardhana}]{Apai2013HSTVARIATIONS}
Apai, D., Radigan, J., Buenzli, E., {et~al.} 2013, The Astrophysical Journal,
  768, 121

\bibitem[{Austin {et~al.}(2012)Austin, Petersson, Frisch, Dobek, Scalmani, \&
  Throssell}]{Austin2012ATermsb}
Austin, A., Petersson, G.~A., Frisch, M.~J., {et~al.} 2012, Journal of Chemical
  Theory and Computation, 8, 4989

\bibitem[{Baeyens {et~al.}(2021)Baeyens, Decin, Carone, Venot, Ag{\'{u}}ndez,
  \& Molli{\`{e}}re}]{Baeyens2021GridMixing}
Baeyens, R., Decin, L., Carone, L., {et~al.} 2021, Monthly Notices of the Royal
  Astronomical Society, 505, 5603

\bibitem[{Barstow {et~al.}(2014)Barstow, Aigrain, Irwin, Hackler, Fletcher,
  Lee, \& Gibson}]{Barstow2014CloudsSpectrum}
Barstow, J.~K., Aigrain, S., Irwin, P. G.~J., {et~al.} 2014, Astrophysical
  Journal, 786

\bibitem[{Becke(1993{\natexlab{a}})}]{Becke1993ATheories}
Becke, A.~D. 1993{\natexlab{a}}, The Journal of Chemical Physics, 98, 1372

\bibitem[{Becke(1993{\natexlab{b}})}]{Becke1993Density-functionalExchange}
Becke, A.~D. 1993{\natexlab{b}}, The Journal of Chemical Physics, 98, 5648

\bibitem[{Becke(2014)}]{Becke2014Perspective:Physics}
Becke, A.~D. 2014, Journal of Chemical Physics, 140, 18

\bibitem[{Berardo {et~al.}(2014)Berardo, Hu, Shevlin, Woodley, Kowalski, \&
  Zwijnenburg}]{Berardo2014ModelingDescription}
Berardo, E., Hu, H.~S., Shevlin, S.~A., {et~al.} 2014, Journal of Chemical
  Theory and Computation, 10, 1189

\bibitem[{Bitzek {et~al.}(2006)Bitzek, Koskinen, G{\"{a}}hler, Moseler, \&
  Gumbsch}]{Bitzek2006StructuralSimple}
Bitzek, E., Koskinen, P., G{\"{a}}hler, F., Moseler, M., \& Gumbsch, P. 2006,
  Physical Review Letters, 97, 170201

\bibitem[{Blander \& Katz(1967)}]{Blander1967CondensationDust}
Blander, M. \& Katz, J.~L. 1967, Geochimica et Cosmochimica Acta, 31, 1025

\bibitem[{Boese \& Martin(2004)}]{Boese2004DevelopmentKinetics}
Boese, A.~D. \& Martin, J.~M. 2004, Journal of Chemical Physics, 121, 3405

\bibitem[{Boulangier {et~al.}(2019)Boulangier, Gobrecht, Decin, De~Koter, \&
  Yates}]{Boulangier2019DevelopingMixture}
Boulangier, J., Gobrecht, D., Decin, L., De~Koter, A., \& Yates, J. 2019,
  Monthly Notices of the Royal Astronomical Society, 489, 4890

\bibitem[{Bromley {et~al.}(2016)Bromley, G{\'{o}}mez~Mart{\'{i}}n, Plane,
  Bromley, G{\'{o}}mez~Mart{\'{i}}n, \& Plane}]{Bromley2016UnderEvaluation}
Bromley, S.~T., G{\'{o}}mez~Mart{\'{i}}n, J.~C., Plane, J. M.~C., {et~al.}
  2016, PCCP, 18, 26913

\bibitem[{Br{\"{u}}nken {et~al.}(2008)Br{\"{u}}nken, M{\"{u}}ller, Menten,
  McCarthy, \& Thaddeus}]{Brunken20082}
Br{\"{u}}nken, S., M{\"{u}}ller, H. S.~P., Menten, K.~M., McCarthy, M.~C., \&
  Thaddeus, P. 2008, The Astrophysical Journal, 676, 1367

\bibitem[{Chang {et~al.}(2013)Chang, Patzer, Kegel, \&
  Chandra}]{Chang2013SmallSpectra}
Chang, C., Patzer, A.~B., Kegel, W.~H., \& Chandra, S. 2013, Ap{\&}SS, 347, 315

\bibitem[{Chang {et~al.}(2005)Chang, Patzer, Sedlmayr, \&
  S{\"{u}}lzle}]{Chang2005InorganicStudy}
Chang, C., Patzer, A.~B., Sedlmayr, E., \& S{\"{u}}lzle, D. 2005, PhRvB, 72,
  235402

\bibitem[{Charnay {et~al.}(2021)Charnay, Mendon{\c{c}}a, Kreidberg, Cowan,
  Taylor, Bell, Demangeon, Edwards, Haswell, Morello, Mugnai, Pascale, Tinetti,
  Tremblin, Zellem, Charnay, Mendon{\c{c}}a, Kreidberg, Cowan, Taylor, Bell,
  Demangeon, Edwards, Haswell, Morello, Mugnai, Pascale, Tinetti, Tremblin, \&
  Zellem}]{Charnay2021AAriel}
Charnay, B., Mendon{\c{c}}a, J.~M., Kreidberg, L., {et~al.} 2021, ExA

\bibitem[{{\v{C}}{\'{i}}{\v{z}}ek(1969)}]{Cizek1969OnMolecules}
{\v{C}}{\'{i}}{\v{z}}ek, J. 1969, in Advances in Chemical Physics (John Wiley
  {\&} Sons, Ltd), 35--89

\bibitem[{Cuko {et~al.}(2017)Cuko, Maci{\'{a}}, Calatayud, \&
  Bromley}]{Cuko2017GlobalApproach}
Cuko, A., Maci{\'{a}}, A., Calatayud, M., \& Bromley, S.~T. 2017, Computational
  and Theoretical Chemistry, 1102, 38

\bibitem[{Curtiss {et~al.}(1995)Curtiss, McGrath, Blaudeau, Davis, Binning, \&
  Radom}]{Curtiss1995ExtensionGa-Kr}
Curtiss, L.~A., McGrath, M.~P., Blaudeau, J.~P., {et~al.} 1995, The Journal of
  Chemical Physics, 103, 6104

\bibitem[{Dolgonos {et~al.}(2010)Dolgonos, Aradi, Moreira, \&
  Frauenheim}]{Dolgonos2010AnTitanium}
Dolgonos, G., Aradi, B., Moreira, N.~H., \& Frauenheim, T. 2010, Journal of
  Chemical Theory and Computation, 6, 266

\bibitem[{Frisch {et~al.}(2013)Frisch, Trucks, Schlegel, Scuseria, Robb,
  Cheeseman, Scalmani, Barone, Mennucci, Petersson, Nakatsuji, Caricato, Li,
  Hratchian, Izmaylov, Bloino, Zheng, Sonnenberg, Had, \&
  Fox}]{Frisch2013GaussianD.01}
Frisch, M.~J., Trucks, G.~W., Schlegel, H.~B., {et~al.} 2013, {Gaussian 09,
  Revision D.01}

\bibitem[{Gail \& Sedlmayr(2013)}]{Gail2013a}
Gail, H.-P. \& Sedlmayr, E. 2013, {Physics and Chemistry of Circumstellar Dust
  Shells}

\bibitem[{Gail {et~al.}(1986)Gail, Sedlmayr, Gail, \&
  Sedlmayr}]{Gail1986TheStars}
Gail, H.~P., Sedlmayr, E., Gail, H.~P., \& Sedlmayr, E. 1986, A{\&}A, 166, 225

\bibitem[{Gale \& Rohl(2003)}]{Gale2003TheGULP}
Gale, J.~D. \& Rohl, A.~L. 2003, Molecular Simulation, 29, 291

\bibitem[{Gobrecht {et~al.}(2016)Gobrecht, Cherchneff, Sarangi, Plane, \&
  Bromley}]{Gobrecht2016DustTauri}
Gobrecht, D., Cherchneff, I., Sarangi, A., Plane, J.~M., \& Bromley, S.~T.
  2016, Astronomy and Astrophysics, 585, A6

\bibitem[{Gobrecht {et~al.}(2018)Gobrecht, Decin, Cristallo, \&
  Bromley}]{Gobrecht2018AAl2O38}
Gobrecht, D., Decin, L., Cristallo, S., \& Bromley, S.~T. 2018, Chemical
  Physics Letters, 711, 138

\bibitem[{Gobrecht {et~al.}(2021{\natexlab{a}})Gobrecht, Plane, Bromley, Decin,
  Cristallo, Sekaran, Gobrecht, Plane, Bromley, Decin, Cristallo, \&
  Sekaran}]{Gobrecht2021Bottom-upClusters}
Gobrecht, D., Plane, J. M.~C., Bromley, S.~T., {et~al.} 2021{\natexlab{a}},
  arXiv, arXiv:2110.11139

\bibitem[{Gobrecht {et~al.}(2021{\natexlab{b}})Gobrecht, Sindel, Lecoq-Molinos,
  \& Decin}]{Gobrecht2021TheClusters}
Gobrecht, D., Sindel, J.~P., Lecoq-Molinos, H., \& Decin, L.
  2021{\natexlab{b}}, Universe 2021, Vol. 7, Page 243, 7, 243

\bibitem[{Goerigk \& Grimme(2011)}]{Goerigk2011EfficientInteractions}
Goerigk, L. \& Grimme, S. 2011, Journal of Chemical Theory and Computation, 7,
  291

\bibitem[{Goumans \& Bromley(2013)}]{Goumans2013StardustSiO+TiO2}
Goumans, T. P.~M. \& Bromley, S.~T. 2013, RSPTA, 371, 20110580

\bibitem[{Grimme {et~al.}(2011)Grimme, Ehrlich, \&
  Goerigk}]{Grimme2011EffectTheory}
Grimme, S., Ehrlich, S., \& Goerigk, L. 2011, Journal of Computational
  Chemistry, 32, 1456

\bibitem[{Hartree(1928)}]{Hartree1928TheMethods}
Hartree, D.~R. 1928, Mathematical Proceedings of the Cambridge Philosophical
  Society, 24, 89

\bibitem[{Helling(2018)}]{Helling2018ExoplanetClouds}
Helling, C. 2018, Annual Review of Earth and Planetary Sciences, 47, 1

\bibitem[{Helling {et~al.}(2008)Helling, Ackerman, Allard, Dehn, Hauschildt,
  Homeier, Lodders, Marley, Rietmeijer, Tsuji, \&
  Woitke}]{Helling2008AAtmospheres}
Helling, C., Ackerman, A., Allard, F., {et~al.} 2008, Monthly Notices of the
  Royal Astronomical Society, 391, 1854

\bibitem[{Helling {et~al.}(2019)Helling, Iro, Corrales, Samra, Ohno, Alam,
  Steinrueck, Lew, Molaverdikhani, MacDonald, Herbort, Woitke, \&
  Parmentier}]{Helling2019UnderstandingMapping}
Helling, C., Iro, N., Corrales, L., {et~al.} 2019, Astronomy and Astrophysics,
  631, A79

\bibitem[{Helling {et~al.}(2020)Helling, Kawashima, Graham, Samra, Chubb, Min,
  Waters, \& Parmentier}]{Helling2020MineralWASP-43b}
Helling, C., Kawashima, Y., Graham, V., {et~al.} 2020, Astronomy {\&}
  Astrophysics, 641, A178

\bibitem[{Helling {et~al.}(2016)Helling, Lee, Dobbs-Dixon, Mayne, Amundsen,
  Khaimova, Unger, Manners, Acreman, \& Smith}]{Helling2016The189733b}
Helling, C., Lee, G., Dobbs-Dixon, I., {et~al.} 2016, Monthly Notices of the
  Royal Astronomical Society, 460, 855

\bibitem[{Hestenes \& Stiefel(1952)}]{Hestenes1952Methods1}
Hestenes, M.~R. \& Stiefel, E. 1952, {Methods of Conjugate Gradients for
  Solving Linear Systems 1}, Tech. Rep.~6

\bibitem[{Heyd {et~al.}(2003)Heyd, Scuseria, \&
  Ernzerhof}]{Heyd2003HybridPotential}
Heyd, J., Scuseria, G.~E., \& Ernzerhof, M. 2003, Journal of Chemical Physics,
  118, 8207

\bibitem[{Hjorth~Larsen {et~al.}(2017)Hjorth~Larsen, J{\O}rgen~Mortensen,
  Blomqvist, Castelli, Christensen, Du{\l}ak, Friis, Groves, Hammer, Hargus,
  Hermes, Jennings, Bjerre~Jensen, Kermode, Kitchin, Leonhard~Kolsbjerg, Kubal,
  Kaasbjerg, Lysgaard, Bergmann~Maronsson, Maxson, Olsen, Pastewka, Peterson,
  Rostgaard, Schi{\O}tz, Sch{\"{u}}tt, Strange, Thygesen, Vegge, Vilhelmsen,
  Walter, Zeng, \& Jacobsen}]{HjorthLarsen2017TheAtoms}
Hjorth~Larsen, A., J{\O}rgen~Mortensen, J., Blomqvist, J., {et~al.} 2017, {The
  atomic simulation environment - A Python library for working with atoms}

\bibitem[{Hohenberg \& Kohn(1964)}]{Hohenberg1964InhomogeneousGas}
Hohenberg, P. \& Kohn, W. 1964, Physical Review, 136, B864

\bibitem[{Hourahine {et~al.}(2020)Hourahine, Aradi, Blum, Bonaf{\'{e}},
  Buccheri, Camacho, Cevallos, Deshaye, Dumitric, Dominguez, Ehlert, Elstner,
  Van Der~Heide, Hermann, Irle, Kranz, K{\"{o}}hler, Kowalczyk, Kuba{\v{r}},
  Lee, Lutsker, Maurer, Min, Mitchell, Negre, Niehaus, Niklasson, Page,
  Pecchia, Penazzi, Persson, {\AA}{\&}tild;ez{\'{a}}{\v{c}}, S{\'{a}}nchez,
  Sternberg, St{\"{o}}hr, Stuckenberg, Tkatchenko, Yu, \&
  Frauenheim}]{Hourahine2020DFTB+Simulations}
Hourahine, B., Aradi, B., Blum, V., {et~al.} 2020, Journal of Chemical Physics,
  152, 124101

\bibitem[{Hudson(1993)}]{CloudCondensationNuclei}
Hudson, J.~G. 1993, Journal of Applied Meteorology and Climatology, 32, 596

\bibitem[{Irikura(2002)}]{Irikura2002THERMO.PL}
Irikura, K. 2002, {THERMO.PL}

\bibitem[{Jeong(2000)}]{Jeong2000DustVariables}
Jeong, K.~S. 2000, PhD thesis, Technische Universit{\"{a}}t Berlin,
  Fakult{\"{a}}t II - Mathematik und Naturwissenschaften, Berlin

\bibitem[{Jeong {et~al.}(2000)Jeong, Chang, Sedlmayr, S{\"{u}}lzle, Jeong,
  Chang, Sedlmayr, \& S{\"{u}}lzle}]{Jeong2000ElectronicTiSUBx/SUBOSUBy/SUB}
Jeong, K.~S., Chang, C., Sedlmayr, E., {et~al.} 2000, JPhB, 33, 3417

\bibitem[{Koch \& Manzhos(2017)}]{Koch2017OnDioxide}
Koch, D. \& Manzhos, S. 2017, Journal of Physical Chemistry Letters, 8, 1593

\bibitem[{K{\"{o}}hler {et~al.}(2012)K{\"{o}}hler, Stepnik, Jones, Guillet,
  Abergel, Ristorcelli, Bernard, K{\"{o}}hler, Stepnik, Jones, Guillet,
  Abergel, Ristorcelli, \& Bernard}]{Kohler2012DustObservations}
K{\"{o}}hler, M., Stepnik, B., Jones, A.~P., {et~al.} 2012, A{\&}A, 548, A61

\bibitem[{K{\"{o}}hn {et~al.}(2021)K{\"{o}}hn, Helling, Enghoff, Haynes,
  Sindel, Krog, \& Gobrecht}]{Kohn2021DustApproach}
K{\"{o}}hn, C., Helling, C., Enghoff, M.~B., {et~al.} 2021, Astronomy {\&}
  Astrophysics, 654, A120

\bibitem[{Kohn {et~al.}(1996)Kohn, Becke, \& Parr}]{Kohn1996DensityStructure}
Kohn, W., Becke, A.~D., \& Parr, R.~G. 1996, Journal of Physical Chemistry,
  100, 12974

\bibitem[{Kohn \& Sham(1965)}]{Kohn1965Self-consistentEffects}
Kohn, W. \& Sham, L.~J. 1965, Physical Review, 140, A1133

\bibitem[{Kreidberg {et~al.}(2013)Kreidberg, Bean, D{\'{e}}sert, Benneke,
  Deming, Stevenson, Seager, Berta-Thompson, Seifahrt, \&
  Homeier}]{Kreidberg2013Clouds1214b}
Kreidberg, L., Bean, J.~L., D{\'{e}}sert, J.-M., {et~al.} 2013, Nature, 505, 69

\bibitem[{Kubo {et~al.}(2007)Kubo, Orita, \& Nozoye}]{Kubo2007Surface011}
Kubo, T., Orita, H., \& Nozoye, H. 2007

\bibitem[{Lam {et~al.}(2015)Lam, Amans, Dujardin, Ledoux, \&
  Allouche}]{Lam2015AtomisticNanoparticles}
Lam, J., Amans, D., Dujardin, C., Ledoux, G., \& Allouche, A.-R. 2015

\bibitem[{Lamiel-Garcia {et~al.}(2017)Lamiel-Garcia, Cuko, Calatayud, Illas, \&
  Bromley}]{Lamiel-Garcia2017}
Lamiel-Garcia, O., Cuko, A., Calatayud, M., Illas, F., \& Bromley, S.~T. 2017,
  Nanoscale, 9, 1049

\bibitem[{Langreth \& Mehl(1983)}]{Langreth1983BeyondProperties}
Langreth, D.~C. \& Mehl, M.~J. 1983, Physical Review B, 28, 1809

\bibitem[{Lee {et~al.}(2015)Lee, Helling, Giles, \& Bromley}]{Lee2015}
Lee, E., Helling, C., Giles, H., \& Bromley, S.~T. 2015, Astronomy {\&}
  Astrophysics, 575, A11

\bibitem[{Lee {et~al.}(2018)Lee, Blecic, \& Helling}]{Lee2018DustRegimes}
Lee, G.~K., Blecic, J., \& Helling, C. 2018, Astronomy and Astrophysics, 614

\bibitem[{Luschtinetz {et~al.}(2009)Luschtinetz, Frenzel, Milek, \&
  Seifert}]{Luschtinetz2009AdsorptionSurfaces}
Luschtinetz, R., Frenzel, J., Milek, T., \& Seifert, G. 2009, The Journal of
  Physical Chemistry C, 113, 5730

\bibitem[{Malcolm W.~Chase(1998)}]{MalcolmW.Chase1998NIST-JANAFTables}
Malcolm W.~Chase, J. 1998, {NIST-JANAF thermochemical tables} (Fourth edition.
  Washington, DC : American Chemical Society ; New York : American Institute of
  Physics for the National Institute of Standards and Technology, 1998.)

\bibitem[{Matsui \& Akaogi(1991)}]{Matsui1991MolecularTio2}
Matsui, M. \& Akaogi, M. 1991, Molecular Simulation, 6, 239

\bibitem[{Meyer \& Hauser(2020)}]{Meyer2020GeometrySystems}
Meyer, R. \& Hauser, A.~W. 2020, The Journal of Chemical Physics, 152, 084112

\bibitem[{Min {et~al.}(2020)Min, Ormel, Chubb, Helling, Kawashima, Min, Ormel,
  Chubb, Helling, \& Kawashima}]{Min2020TheRetrieval}
Min, M., Ormel, C.~W., Chubb, K., {et~al.} 2020, A{\&}A, 642, A28

\bibitem[{Montgomery {et~al.}(2000)Montgomery, Frisch, Ochterski, \&
  Petersson}]{Montgomery2000AMethod}
Montgomery, J.~A., Frisch, M.~J., Ochterski, J.~W., \& Petersson, G.~A. 2000,
  Journal of Chemical Physics, 112, 6532

\bibitem[{Nagy \& K{\'{a}}llay(2019)}]{Nagy2019ApproachingMethods}
Nagy, P.~R. \& K{\'{a}}llay, M. 2019, Journal of Chemical Theory and
  Computation, 15, 5275

\bibitem[{Nikolov {et~al.}(2021)Nikolov, Maciejewski, Constantinou,
  Madhusudhan, Fortney, Smalley, Carter, de~Mooij, Drummond, Gibson, Helling,
  Mayne, Mikal-Evans, Sing, Wilson, Nikolov, Maciejewski, Constantinou,
  Madhusudhan, Fortney, Smalley, Carter, de~Mooij, Drummond, Gibson, Helling,
  Mayne, Mikal-Evans, Sing, \& Wilson}]{Nikolov2021Ground-basedWASP-110b}
Nikolov, N., Maciejewski, G., Constantinou, S., {et~al.} 2021, AJ, 162, 88

\bibitem[{Ormel \& Min(2019)}]{Ormel2019ARCiSModel}
Ormel, C.~W. \& Min, M. 2019, Astronomy and Astrophysics, 622, A121

\bibitem[{Patzer {et~al.}(2005)Patzer, Chang, Sedlmayr, \&
  S{\"{u}}lzle}]{Patzer2005AProperties}
Patzer, A.~B., Chang, C., Sedlmayr, E., \& S{\"{u}}lzle, D. 2005, EPJD, 32, 329

\bibitem[{Patzer {et~al.}(2014)Patzer, Chang, \&
  S{\"{u}}lzle}]{Patzer2014AProperties}
Patzer, A.~B., Chang, C., \& S{\"{u}}lzle, D. 2014, CPL, 612, 39

\bibitem[{Patzer {et~al.}(1998)Patzer, Gauger, \&
  Sedlmayr}]{Patzer1998DustPhase}
Patzer, A. B.~C., Gauger, A., \& Sedlmayr, E. 1998, A{\&}A, 337, 847

\bibitem[{Patzer {et~al.}(1995)Patzer, K{\"{o}}hler, Sedlmayr, Patzer,
  K{\"{o}}hler, \& Sedlmayr}]{Patzer1995PrimaryDesiderata}
Patzer, A. B.~C., K{\"{o}}hler, T.~M., Sedlmayr, E., {et~al.} 1995, P{\&}SS,
  43, 1233

\bibitem[{Peverati \& Truhlar(2011)}]{Peverati2011ImprovingSeparation}
Peverati, R. \& Truhlar, D.~G. 2011, Journal of Physical Chemistry Letters, 2,
  2810

\bibitem[{Peverati \& Truhlar(2012)}]{Peverati2012Screened-exchangePhysics}
Peverati, R. \& Truhlar, D.~G. 2012, Physical Chemistry Chemical Physics, 14,
  16187

\bibitem[{Plane {et~al.}(2013)Plane, {Plane}, \& C.}]{Plane2013OnOutflows}
Plane, J. M.~C., {Plane}, \& C., J.~M. 2013, RSPTA, 371, 20120335

\bibitem[{Pont {et~al.}(2013)Pont, Sing, Gibson, Aigrain, Henry, \&
  Husnoo}]{Pont2013TheObservations}
Pont, F., Sing, D.~K., Gibson, N.~P., {et~al.} 2013, Monthly Notices of the
  Royal Astronomical Society, 432, 2917

\bibitem[{Purvis \& Bartlett(1998)}]{Purvis1998ATriples}
Purvis, G.~D. \& Bartlett, R.~J. 1998, The Journal of Chemical Physics, 76,
  1910

\bibitem[{Ramabhadran \&
  Raghavachari(2013)}]{Ramabhadran2013ExtrapolationHierarchy}
Ramabhadran, R.~O. \& Raghavachari, K. 2013, Journal of Chemical Theory and
  Computation, 9, 3986

\bibitem[{Samra {et~al.}(2020)Samra, Helling, \& Min}]{Samra2020MineralShape}
Samra, D., Helling, C., \& Min, M. 2020, Astronomy and Astrophysics, 639, A107

\bibitem[{Sloan {et~al.}(2009)Sloan, Matsuura, Zijlstra, Lagadec, Groenewegen,
  Wood, Szyszka, Bernard-Salas, van Loon, Sloan, Matsuura, Zijlstra, Lagadec,
  Groenewegen, Wood, Szyszka, Bernard-Salas, \& van
  Loon}]{Sloan2009DustAbundances}
Sloan, G.~C., Matsuura, M., Zijlstra, A.~A., {et~al.} 2009, Sci, 323, 353

\bibitem[{Storn \& Price(1997)}]{Storn1997DifferentialSpaces}
Storn, R. \& Price, K. 1997, Journal of Global Optimization, 11, 341

\bibitem[{Tao {et~al.}(2003)Tao, Perdew, Staroverov, \&
  Scuseria}]{Tao2003ClimbingSolids}
Tao, J., Perdew, J.~P., Staroverov, V.~N., \& Scuseria, G.~E. 2003, Physical
  Review Letters, 91, 146401

\bibitem[{Vydrov \& Scuseria(2006)}]{Vydrov2006AssessmentFunctional}
Vydrov, O.~A. \& Scuseria, G.~E. 2006, Journal of Chemical Physics, 125, 234109

\bibitem[{Wales \& Doye(1997)}]{Wales1997GlobalAtoms}
Wales, D.~J. \& Doye, J.~P. 1997, Journal of Physical Chemistry A, 101, 5111

\bibitem[{Weigend(2006)}]{Weigend2006AccurateRn}
Weigend, F. 2006, Physical Chemistry Chemical Physics, 8, 1057

\bibitem[{Wilson {et~al.}(1996)Wilson, Van~Mourik, \&
  Dunning}]{Wilson1996GaussianNeon}
Wilson, A.~K., Van~Mourik, T., \& Dunning, T.~H. 1996, Journal of Molecular
  Structure: THEOCHEM, 388, 339

\bibitem[{Woitke {et~al.}(2018)Woitke, Helling, Hunter, Millard, Turner,
  Worters, Blecic, \& Stock}]{Woitke2018EquilibriumRatio}
Woitke, P., Helling, C., Hunter, G.~H., {et~al.} 2018, Astronomy {\&}
  Astrophysics, 614, A1

\bibitem[{Wood {et~al.}(2006)Wood, Radom, Petersson, Barnes, Frisch, \&
  Montgomery}]{Wood2006AChemistry}
Wood, G.~P., Radom, L., Petersson, G.~A., {et~al.} 2006, Journal of Chemical
  Physics, 125, 094106

\bibitem[{Xu \& Goddard(2004)}]{Xu2004TheProperties}
Xu, X. \& Goddard, W.~A. 2004, Proceedings of the National Academy of Sciences
  of the United States of America, 101, 2673

\bibitem[{Yanai {et~al.}(2004)Yanai, Tew, \& Handy}]{Yanai2004ACAM-B3LYP}
Yanai, T., Tew, D.~P., \& Handy, N.~C. 2004, Chemical Physics Letters, 393, 51

\bibitem[{Zhao \& Truhlar(2008)}]{Zhao2008TheFunctionals}
Zhao, Y. \& Truhlar, D.~G. 2008, Theoretical Chemistry Accounts, 120, 215

\bibitem[{Zheng {et~al.}(2007)Zheng, Witek, Bobadova-Parvanova, Irle, Musaev,
  Prabhakar, Morokuma, Lundberg, Elstner, K{\"{o}}hler, \&
  Frauenheim}]{Zheng2007ParameterNi}
Zheng, G., Witek, H.~A., Bobadova-Parvanova, P., {et~al.} 2007, Journal of
  Chemical Theory and Computation, 3, 1349

\end{thebibliography}

\begin{appendix}
\section{Theoretical approaches used in this work}
\subsection{Density Functional Theory}
\label{appendix:dft}
Density functional theory can be used to describe the interactions between atoms and electrons within a molecule.
The behaviour of every electron is influenced by every other electron in the 
system, leading to a system of highly coupled differential equations, whose dimensionality 
cannot be reduced and which has to be solved simultaneously. 
Therefore, numerical 
approximations are needed. The Hartree-Fock method \citep{Hartree1928TheMethods} approximates the many-electron wavefunction as a product of single-electron wavefunctions. The 
single-electron wavefunctions are chosen, such that the many-body wavefunction built from 
them is anti-symmetric to obey the Pauli exclusion principle. The Hartree-Fock method 
is computationally still very expensive, because one equation has to be solved for every 
electron in the 
system.
A computationally more efficient solution is given by density 
functional theory (DFT) \citep{Hohenberg1964InhomogeneousGas, Kohn1965Self-consistentEffects, Kohn1996DensityStructure}. In DFT the multi-electron wavefunction is approximated by a function of the electron density. The electron density that minimises the total internal energy of a system is defined as its ground state. The interaction between electrons is then calculated through 
functionals of the electron density. The exact functionals are unknown. Many different 
approaches to describe and quantify these functionals have been investigated 
\citep{Becke2014Perspective:Physics}. The first and simplest functionals use the local 
density approximation (LDA) \citep{Hohenberg1964InhomogeneousGas}, where the functional is 
only dependent on the density at the exact coordinates where it is evaluated. This quickly 
leads to large inaccuracies as it assumes the homogeneity of the electron density, a variable 
that is highly non-homogeneous. More recent functionals use the gradient of the electron density as an additional qualifier. These functionals 
belong to the class of
generalised gradient approximations (GGA)\citep{Langreth1983BeyondProperties}. Most currently used functionals do not rely on pure DFT, but also have a component where the exchange energy is calculated from Hartree-Fock theory. These so-called hybrid functionals often combine high levels of accuracy with an acceptable computational cost, making them suitable for many purposes \citep{Becke1993ATheories}. In addition to choosing a functional, a basis set needs to be chosen for all DFT calculations. These basis sets represent a set of basis functions which can be combined linearly to express the electron orbitals in a computationally efficient way. These also exist at varying levels of complexity, mostly differing in the number of linear combinations that can be used to describe a single valence orbital. The choice of both the functional and basis set is crucial for the calculations in DFT. Section \ref{funcbasissearch} is therefore dedicated to find the functional and basis set best suited for our purpose, which is the optimisation of (TiO\textsubscript{2})\textsubscript{$N$} 
clusters.
\subsection{The Buckingham-Coulomb potential}
\label{appendix:forcefield}
The Buckingham-Coulomb potential takes into account the interatomic van der Waals forces, the Pauli exclusion principle and the electrostatic force, as opposed to the later used all-electron DFT calculations, which solve the Schrödinger equation to quantify the energy of a system. The general form of the Buckingham-Coulomb potential reads: 
\begin{equation}
\label{eq:Buckingham}
U(r_{ij}) = A \ \mathrm{exp}\left(-\frac{r_{ij}}{B} \right) - \frac{C}{r_{ij}^6} + \frac{q_{i}q_{j}}{r_{ij}} 
\end{equation}

$U(r_{ij})$ is the potential energy of the interaction of two atoms $i$ and $j$, in this case O and Ti, O and O, or Ti and Ti. $r_{ij}$ is the distance between these atoms, and $q_i$ and $q_j$ are their respective charges. $A$, $B$, and $C$ are the Buckingham pair parameters. The first term in Eq. \ref{eq:Buckingham} represents the short-range repulsion according to the Pauli principle, with an increase in either parameter $A$ or $B$ leading to a increase in the potential barrier towards small distances. The second term describes the van der Waals force and its $r^{-6}$ dependency on inter-atomic distance. An increase in its parameter $C$ 
adds more attraction. The third and final term represents the Coulomb potential with its inverse proportionality to distance dominating the long-range interaction and implying alternating cation-anion ordering. 

\subsection{DFTB}
\label{appendix:DFTB}
In this work density functional based tight-binding methods are used as an intermediate step between a forcefield description and all-electron DFT calculations. The basis of DFTB models is expanding the 
total energy functional given by Kohn-Sham (KS) DFT \citep{Kohn1965Self-consistentEffects} in a Taylor series of up to third order \citep{Hourahine2020DFTB+Simulations}. There are several 
major approximations made in each term of the Taylor series. First order DFTB1 takes only the 
first order expansion into account. A valence-only minimal basis set is used with a linear 
combination of atomic orbitals (LCAO) \textit{ansatz}. This means that only electrons in the 
outer shells of the individual atoms, the valence electrons, and their interactions are 
described by this model. Additionally a two-center approximation to the Hamiltonian operator 
is used, resulting in only interactions between neighboring electronic orbitals being taken 
into account, while distant overlapping orbitals are neglected. In this work, second order 
DFTB, also known as DFTB2 or SCC-DFTB (Self-consistent charge density functional 
tight-binding), is used. In addition to the first term of the Taylor series, in DFTB2 the 
second term also gets evaluated in order to 
calculate the energy of a system. Fluctuations in 
density, which become relevant in the second order 
term are approximated by a superposition 
of atomic contributions of all other atoms, each 
contributing an exponentially decaying spherically 
symmetric charge density.

\section{High level CCSD(T) calculations}
\label{appendix:ccsd}
Coupled cluster single point calculations at the CCSD(T)/6-311+G(2d,2p) level of theory have been run for the GM candidate clusters for $N$ = 1-4. (Tab. \ref{tab:ccsd})
\begin{table}[]
\caption{Binding energies for GM candidate clusters for $N$=1-4 at the CCSD(T)/6-311+G(2d,2p) level of theory.}
    \centering
    \begin{tabular}{|c|c|}
    \hline
       Size $N$  & E$_{bind}$ [$\frac{\mathrm{kJ}}{\mathrm{mol}}$] \\
       \hline
        1 & 1175.5\\
        2 & 2844.1\\
        3 & 4540.7\\ 
        4 & 6163.8\\
        \hline
    \end{tabular}
    \label{tab:ccsd}
\end{table}

The value for the monomer binding energy differs from the experimental Janaf Nist value by $\Delta E \approx 90 \frac{\mathrm{kJ}}{\mathrm{mol}}$. This is far outside of both the assumed error of DFT and Janaf-Nist values of $\approx 4 \frac{\mathrm{kJ}}{\mathrm{mol}}$. To calibrate our methods on these high-level theoretical calculations, we would therefore have to ignore the experimental values. 
To quantify how closely our chosen B3LYP/cc-pVTZ empdisp method matches the higher level results, the relative binding energies to the monomer were calculated for these clusters (Tab. \ref{tab:relbind}).

\begin{table}[]
\caption{Relative binding energy to the monomer $\frac{E_{bind,N}}{E_{bind,1}}$}
    \centering
    \begin{tabular}{|c|c|c|c|}
    \hline
    \textbf
        Size & CCSD(T) & DFT & Difference [\%] \\
        \hline
        2 & 2.42 & 2.39 & 1.3 \\
        3 & 3.86 & 3.79 & 1.9 \\
        4 & 5.24 & 5.21 & 0.6 \\ 
        \hline
    \end{tabular}
    \label{tab:relbind}
\end{table}
As the deviation is $<2\%$ for all sizes, we conclude that no information is lost by not using high-level CCSD(T) calculations to calibrate our methods.

\section{Calculating the Gibbs free energy of formation from the partition function}
\label{appendix:Gibbs}
The Gibbs free energy of formation, $\Delta_f G^\circ$, for a 
gas-phase
cluster can be calculated using its enthalpy of formation $\Delta_f H^\circ$, its entropy $S^\circ_{cluster}$, and the entropy of the individual atoms 
comprised in
the cluster $S^\circ_{atoms}$:
\begin{equation}
    \Delta_f G^\circ (T) = \Delta_f H^\circ (T) - T\left( S^\circ_{cluster}(T) - \sum S^\circ_{atoms}(T)
    \right)
\end{equation}
For a cluster of size $N=5$, the last term is the sum of the entropies of 5 Ti atoms and 10 Oxygen atoms.
Calculating the enthalpy of formation of the cluster requires information about the enthalpies of the elements that make up the cluster. The entropies and enthalpies for Ti and O used for these calculations are sourced from the JANAF-NIST tables \citep{MalcolmW.Chase1998NIST-JANAFTables}.
The enthalpy of formation of the cluster at temperature $T$ is then calculated through:
\begin{multline}
    \Delta_f H^\circ (T) = \Delta_f H^\circ (T^\standardstate) + [H^\circ (T) - H^\circ(T^\standardstate)]_{cluster} \\ - \sum [H^\circ(T) - H^\circ(T^\standardstate)]_{atoms}
\end{multline}

with the reference temperature $T^\standardstate = 0$K. The enthalpy of formation at the reference temperature $\Delta_f H^\circ(T^\standardstate)$ is the binding energy of the cluster, caluclated for a cluster of size $N$ through:
\begin{equation}
    \Delta_f H^\circ(T^\standardstate) = E_{ZP}( \mathrm{Ti}_N\mathrm{O}_{2N}) - NE_{ZP}({\rm Ti}) - 2NE_{ZP}({\rm O})
\end{equation}
The entropy of the system $S^\circ_{cluster}$ is calculated from the internal energy of the system:
\begin{equation}
    S^\circ_{cluster} = \frac{U(T) - U(0)}{T} + N_{atoms}k_B \left(\ln \frac{q(V,T)}{N_{atoms}} + 1  \right)
\end{equation}
With the internal energy $U(T) - U(0)$, the partition function $q(V,T)$, and the number of atoms in the cluster $N_{atoms}$. The internal energy is derived from the partition function $q (V,T)$ through
\begin{equation}
    U(T)- U(0) = -N_{atoms} \left( \frac{\partial \ln q(V,T)}{\partial \beta}\right)_V
\end{equation} 
Using the rigid-rotor harmonic-oscillator (RRHO) approximation
for a polyatomic, non-linear molecule its partition function is:

\begin{equation}
\label{eq:partitionfunction}
\scalebox{1.12}{$
q(V,T) = \left(\frac{2\pi m}{\beta h^2}\right)^{3/2} V \frac{1}{\sigma} \left(\frac{\pi^{1/2}}{hc\beta} \right)^{3/2} \left(\frac{\pi}{A_{\rm rot}B_{\rm rot}C_{\rm rot}} \right)^{1/2} \prod_i \frac{e^{-\beta h\nu/2}}{1 - e^{\beta h\nu_i}}
$}
\end{equation}
$T$ is the temperature of the system, $\beta = \frac{1}{k_B T}$, with the Boltzmann constant $k_B$. $h$ is the Planck constant and $c$ the speed of light. $\sigma$ is the symmetry number, correcting for the repeated count of indistinguishable configurations. 
The unknown inputs in Eq. \ref{eq:partitionfunction} are the 
vibrational levels $\nu_i$, the rotational constants A$_{\rm rot}$, B$_{\rm rot}$, and C$_{\rm rot}$, and $V$, the volume of the particle, which are all outputs of the DFT simulations. For further reading on the calculation of these thermochemical properties, the reader is referred to Chapter 3.6 in \cite{Jeong2000DustVariables}.
The partition function used here omits the electronic partition function, $q_{el}$:
\begin{equation}
    q_{el} = \sum_i g_i e^{-\beta \epsilon_i}
\end{equation}
with the energy levels i, their energies $\epsilon_i$ and their respective degeneracies $g_i$.
This is because
\texttt{Gaussian16} 
assumes that the first electronic excitation energy is much greater than the
thermal energy ($\epsilon_1 \gg kT$) and therefore inaccessible at any 
temperature, resulting in a value of $q_{el} = 1$ for all cluster and temperatures.
In addition, all clusters are assumed to be in a singlet state with g$_i$=1 $\forall $ i.

\section{Calibrations of low-level methods}
\subsection{Recalibrated force field}
\label{appendix:ff_calib}
margin range
destabilise the potential. 
Starting with a 
calculation of the potential energy of a (TiO\textsubscript{2})\textsubscript{5} cluster sourced from \cite{Berardo2014ModelingDescription}, the 10 individual parameters of the Buckingham-Coulomb force field parameterisation are varied until they destabilise the potential, so that it does not have a stable minimum. For each of the parameters the upper and lower value are taken as bounds for the following parameter search  (See Table \ref{tab:buckingham_bounds}). 
The quality of a set of parameters is determined by a comparison with the energies and 
geometries calculated with DFT in Sect. \ref{smallclustercalc}. Due to the differences in 
approximation between the DFT and the force field calculations it is difficult to directly 
compare their respective results.
Since force field calculations 
rely on a simplified representation of the PES,
the absolute 
Buckingham pair
binding energy of the clusters 
is not realistic when comparing to DFT calculations.
However, the relative 
energies of isomers of the same size are portrayed accurately by the 
force field description and therefore used as a point of comparison. 
The monomer and dimer are not addressed with the force field approach, because of the impact of the electronic orbitals that is not considered in the Buckingham pair potential. 
For any uni-directional force field description the monomer and dimer have linear and flat geometries which are known to not be accurate \citep{Koch2017OnDioxide}. For increasing cluster sizes, the interactions described by the force field grow stronger in comparison to the quantum effects and therefore give a more accurate description of the clusters.
For each isomer from 
Sect. \ref{smallclustercalc} a \texttt{conp} optimisation is performed using \texttt{GULP}. 
The energies of all isomers are then compared to the energy of the global minimum cluster of 
their respective size to determine their relative energy given by the force field 
approach $\Delta E_{ff} = E_{ff}^{min} - E_{ff}$. This difference in binding energy is then 
compared to the corresponding difference in binding energy from the DFT calculations $\Delta 
E_{DFT} = E_{DFT}^{min} - E_{DFT}$ for the same cluster. The quality of a parameter set with regards to the energy 
$Q(E)$ is then determined by the root sum squared of all relative deviations across all isomers (Eq. \ref{eq:qual_en}). 

\begin{equation}
     Q_{E} = \sqrt{\sum_{i} \left( \Delta E_{ff}(i) - \Delta E_{DFT}(i) \right)^2}, \ \  i = \mathrm{isomers}
     \label{eq:qual_en}
\end{equation} 

\begin{figure*}
   \centering
     \begin{subfigure}[b]{0.49\hsize}
         \centering
         \includegraphics[width=\hsize]{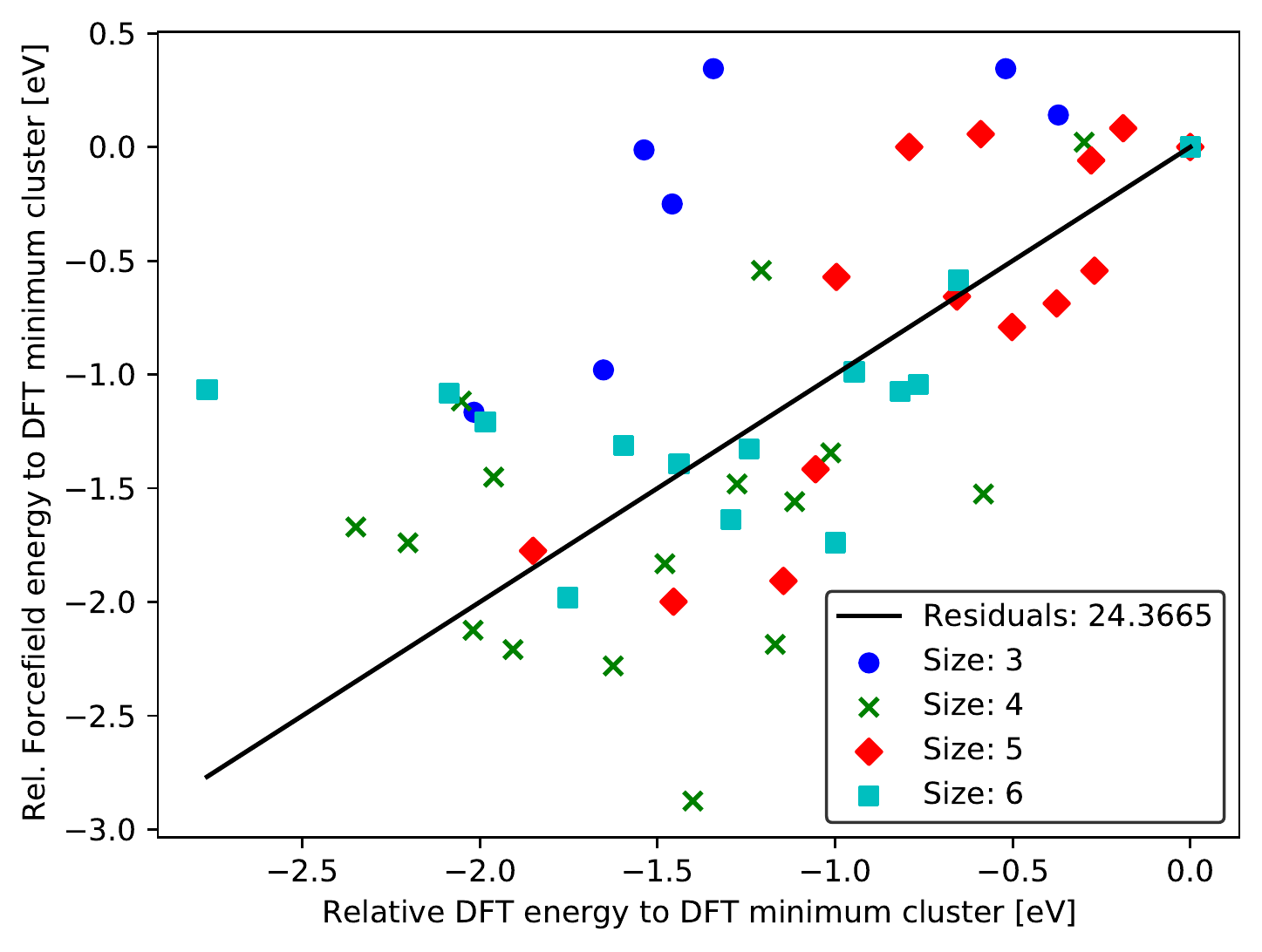}
         \caption{}
         \label{fig:ff_en_lamgar}
     \end{subfigure}
     \hfill
     \begin{subfigure}[b]{0.49\hsize}
         \centering
         \includegraphics[width=\hsize]{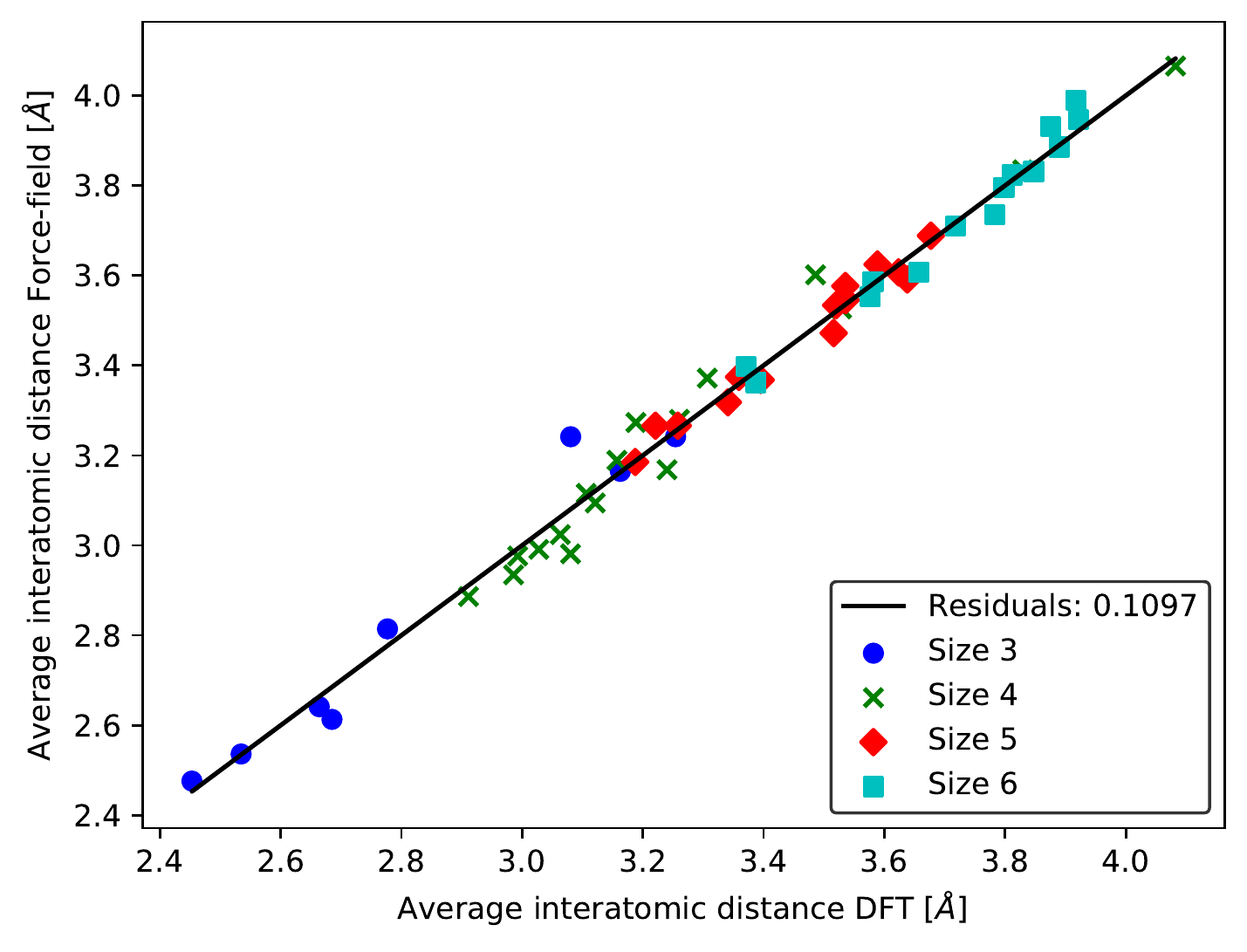}
         \caption{}
         \label{fig:ff_dist_lamgar}
     \end{subfigure}
     \hfill
     \begin{subfigure}[b]{0.49\hsize}
         \centering
         \includegraphics[width=\hsize]{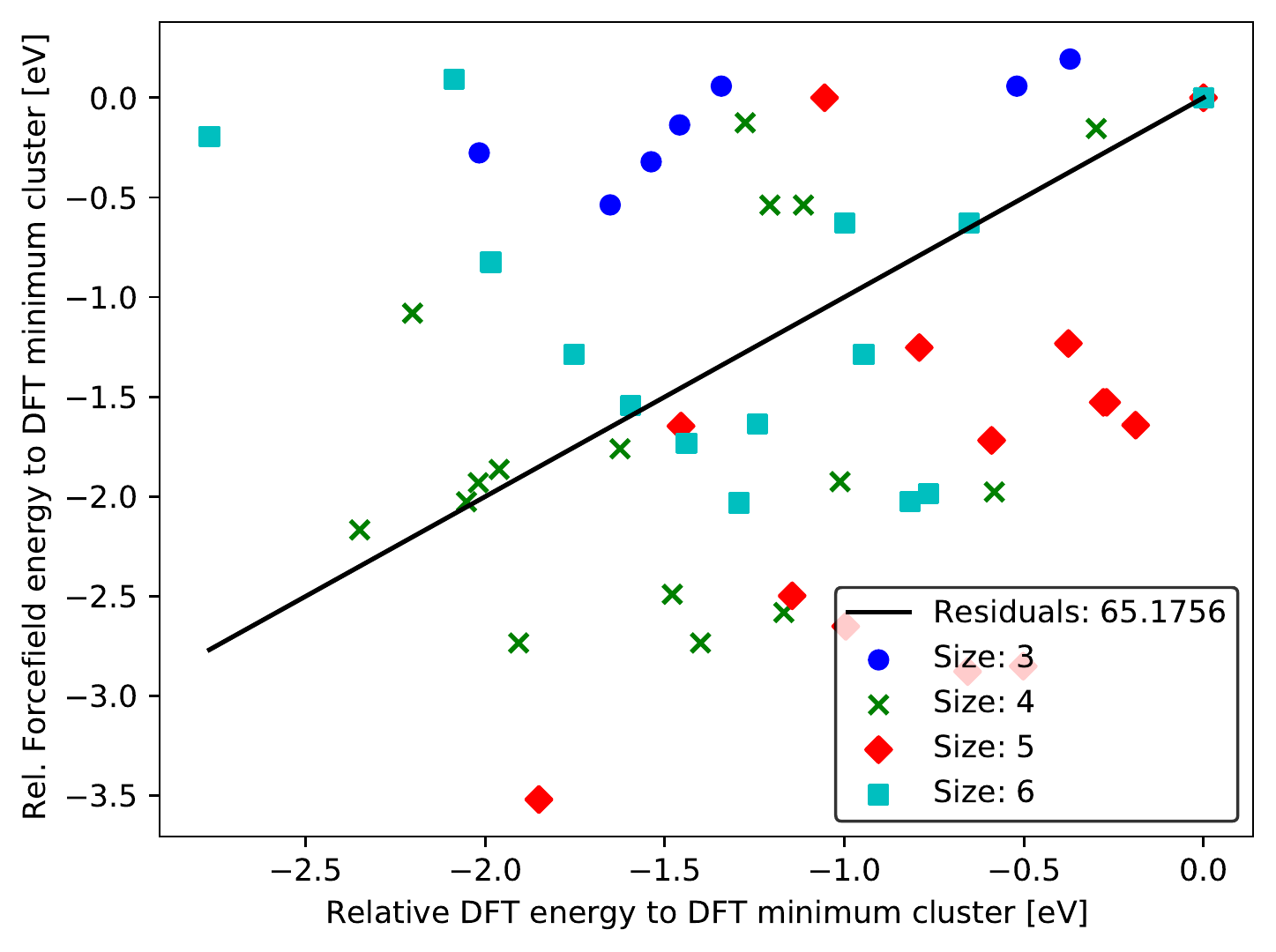}
         \caption{}
         \label{fig:ff_en_matsui}
     \end{subfigure}
          \hfill
     \begin{subfigure}[b]{0.49\hsize}
         \centering
         \includegraphics[width=\hsize]{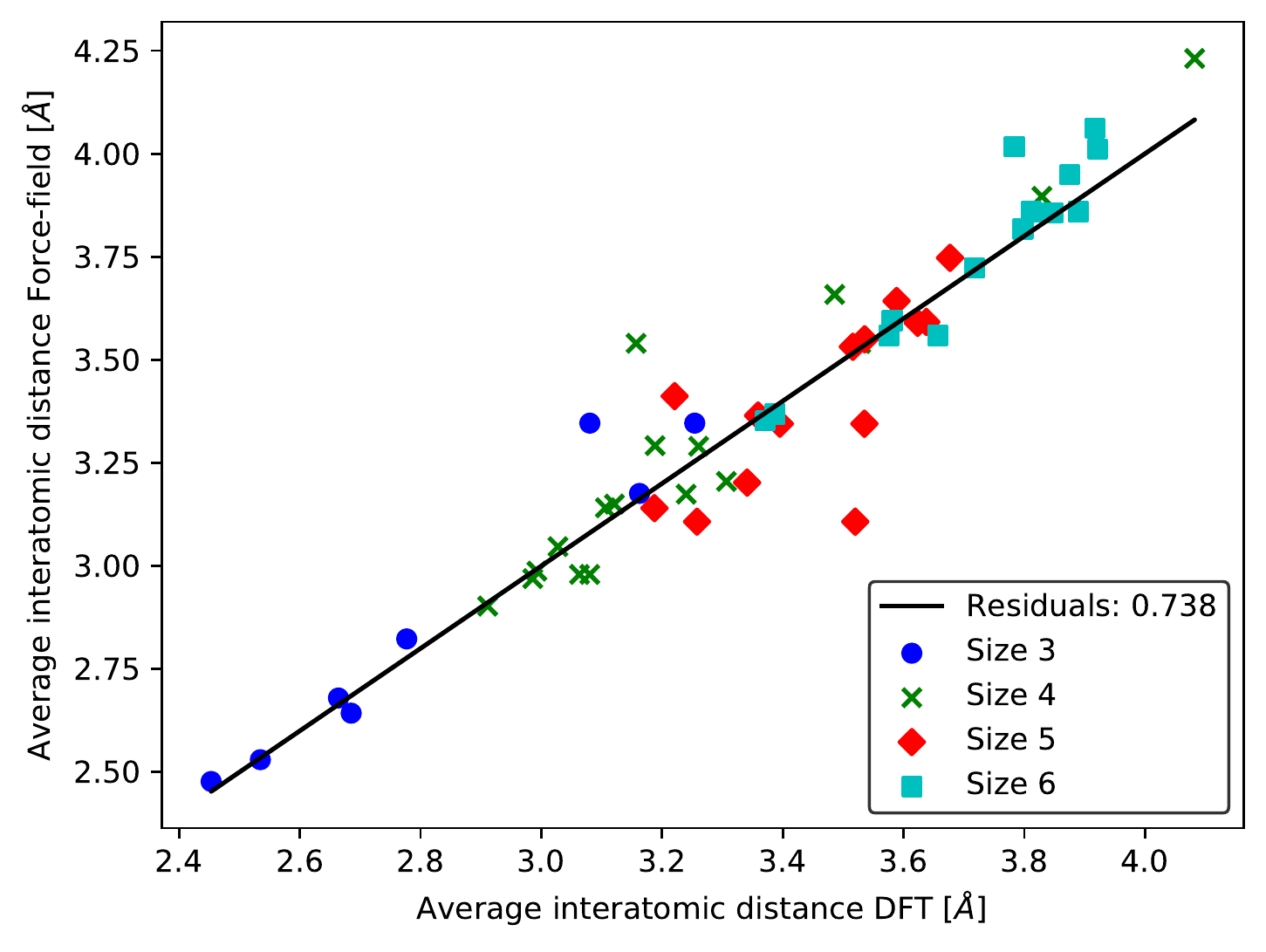}
         \caption{}
         \label{fig:ff_dist_matsui}
     \end{subfigure}
     \hfill
     \begin{subfigure}[b]{0.49\hsize}
         \centering
         \includegraphics[width=\hsize]{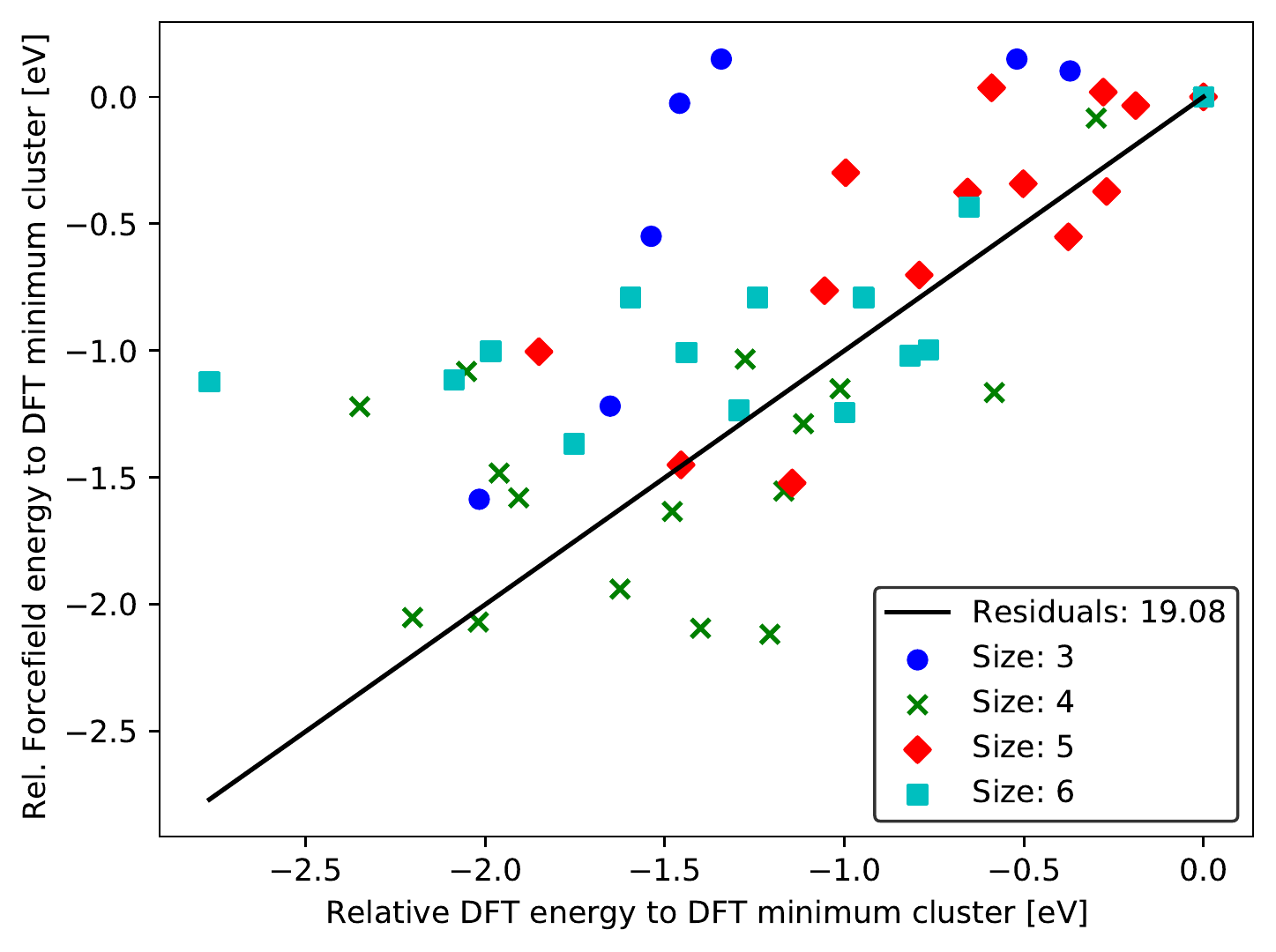}
         \caption{}
         \label{fig:ff_en_thiswork}
     \end{subfigure}
      \hfill
     \begin{subfigure}[b]{0.49\hsize}
         \centering
         \includegraphics[width=\hsize]{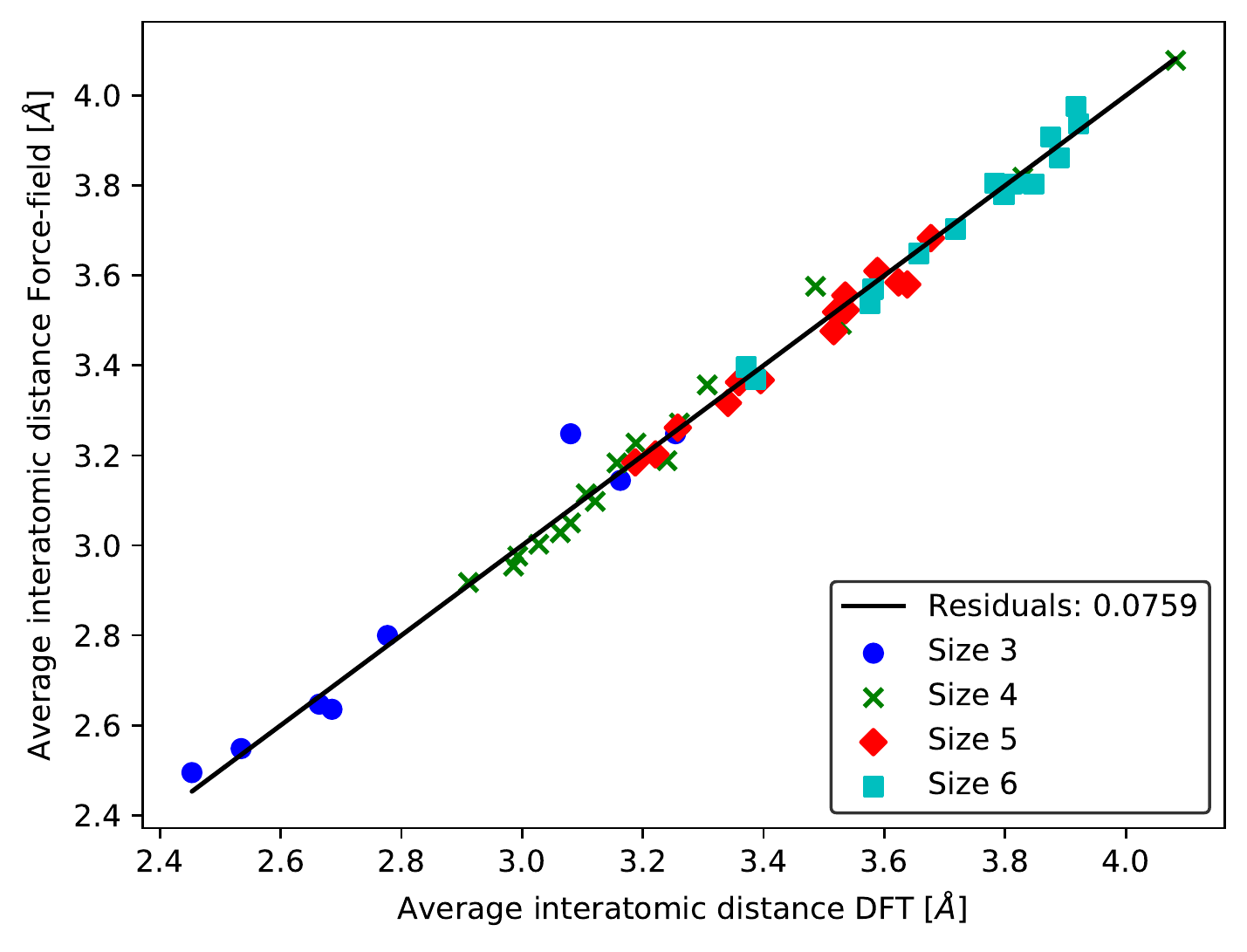}
         \caption{}
         \label{fig:ff_dist_thiswork}
     \end{subfigure}
      \caption{\textbf{Left:} Relative energy deviation to the global minimum cluster for the force field approach versus the DFT approach for all clusters from Section \ref{smallclustercalc}. Ideally, if the force field approach exactly reproduced the results of the more accurate DFT approach, all points would be on the black line with slope 1. \textbf{Right:} Distance parameter $\bar{D}$ for all clusters from the force field paramterisation versus the DFT approach. In an ideal scenario all clusters are exactly reproduced and therefore have no deviation from the black line with slope 1. Parameters from: \textbf{Top}: \cite{Lamiel-Garcia2017}. \textbf{Middle}: \cite{Matsui1991MolecularTio2}. \textbf{Bottom}: This work.}
      \label{fig:forcefields_comp}
\end{figure*}

Additionally, the distance parameter $\bar{D}$ is introduced, which is calculated by taking the mean of all interatomic distances within a cluster. This is done to quantify the similarity of the geometries calculated by the force field approach and the DFT calculations. This distance parameter is calculated for every isomer, for both the DFT optimised geometry and the force field optimised geometry. The quality of a set of parameters with regards to the geometry $Q_{\mathrm{Geom}}$ is then calculated analogous to the energy quality by the root sum squared of all deviations of $\bar{D}_{ff}$ for the force field geometries from $\bar{D}_{DFT}$ for the DFT geometries.

\begin{equation}
    \label{eq:qual_dist}
         Q_{\mathrm{Geom}} = \sqrt{\sum_{i} \left(\bar{D}_{ff}(i) - \bar{D}_{DFT}(i) \right)^2}, \ \  i = \mathrm{isomers} 
\end{equation}
As optimisation for both quantities at the same time is crucial, a weight is introduced to normalise the values of $Q_{Geom}$ and $Q_{E}$. Evaluation of typical values for $Q_{Geom}$ and $Q_{E}$ reveals a difference of one order of magnitude, giving a weight of 10 as a reasonable counter. The overall quality $Q_{\mathrm{Tot}}$ of a parameter set is then given by the root sum squared of the energy quality $Q(E)$ and the geometry quality $Q_{\mathrm{Geom}}$ with a weight of 10:
\begin{equation}
    \label{eq:qual_tot}
    Q_{\mathrm{Tot}} = \sqrt{10\cdot Q_{\mathrm{Geom}}^2 + Q_{E}^2}
\end{equation}
The quality parameter $Q_{\mathrm{Tot}}$ is minimised in order to find the set of parameters for the Buckingham - Coulomb potential that best reproduces the results from DFT calculations. The minimisation is done using the \texttt{scipy.optimize.differential\_evolution} algorithm \citep{Storn1997DifferentialSpaces}, with the bounds from Table \ref{tab:buckingham_bounds} as well as the bounds derived from the parameters of \cite{Lamiel-Garcia2017}. This algorithm was chosen due to its capability to find a global minimum for a multidimensional, non-linear function, its convergence properties and ease of parallelisation. Convergence is achieved after 415 iterations giving a minimum quality parameter of $Q_{\mathrm{Tot}} = 4.66$ and results in the parameter set listed in Table \ref{tab:buckingham_bounds}. A comparison to the performance of this new parameter set can be found in Figure \ref{fig:forcefields_comp}. In particular, results are plotted for the force field parameterisation by \cite{Lamiel-Garcia2017} in Figures \ref{fig:ff_en_lamgar} \& \ref{fig:ff_dist_lamgar}, for the values from \cite{Matsui1991MolecularTio2} in Figures \ref{fig:ff_en_matsui} \& \ref{fig:ff_dist_matsui}, and for the results from this work in Figures \ref{fig:ff_en_thiswork} \& \ref{fig:ff_dist_thiswork}. On the left
panel
of these plots, the relative 
energies compared to the 
energy of the known global minimum geometry (which is set to zero), and their relative deviation are calculated with both the force field description of interactions and all-electron DFT. The positions of the global minimum isomers for all sizes $N$ overlap in the top right-hand corner of each graph. The position on the $x$-axis, which represents the relative energy described by DFT calculations is the same for all isomers of all sizes, representing the 
most realistic
energetic ordering when going from right to left. Ideally, the relative 
energies 
calculated from the force field model reproduces 
the same energetic
ordering on the $y$-axis, going from top to 
bottom, creating a linear function with a slope of 1. In order to aid visually, this line has been plotted in black. In the right
panel of Figure \ref{fig:forcefields_comp}
the same approach is followed for the average 
interatomic distance parameter $\bar{D}$. In an ideal case, the interatomic distances from 
the force field calculations match their DFT optimised counterparts and fall on the black 
line with slope 1. Significant deviations of individual clusters from that line can be 
explained by the fact that in some cases the force field approach produces a different 
local minimum geometry to the DFT approach from the same starting geometry. 
Finally, to make sure that the result is not a numerical artifact, the values for each 
parameter were rounded to the same accuracy as the parameters from literature and still outperform them for our purpose (Table \ref{tab:quality_comparison_ff}).

\begin{table}[]
    \caption{Upper and lower stable limits for each of the 10 individual parameters of the Buckingham-Coulomb force field (Eq. \ref{eq:Buckingham}). These values are taken as upper and lower bounds for the parameter search.}
    \centering 
    \begin{tabular}{|l|c|c|c|}
    \hline
    Parameter & Lower & Upper  & Final result  \\
    & bound & bound & \\
    \hline
    Z\textsubscript{Ti} & 1.0 & 4.0 & 2.0122427 \\
    Ti-Ti A & 25000 & 95000 & 79003.238832\\
    Ti-Ti B & 0.1 & 0.4 & 0.2221871 \\
    Ti-Ti C & 100 & 850 & 475.3962942 \\
    O-O A & 1000 & 25000 & 4164.7007783 \\
    O-O B & 0.1 & 0.5 & 0.2836446 \\
    O-O C & 5 & 250 & 19.7054331 \\
    Ti-O A & 5000 & 80000 & 23420.6827084 \\
    Ti-O B & 0.1 & 0.4 & 0.1759698 \\
    Ti-O C & 0 & 20 & 4.5016462 \\
    \hline
    \end{tabular}
    \label{tab:buckingham_bounds}
\end{table}

\begin{table}[]
    \caption{Comparison of Quality parameters $Q_{Tot}$, $Q_{Geom}$, and $Q_{E}$ for parameterisations from literature versus this work and this work rounded to the same precision as the literature.}
    \centering
    \begin{tabular}{|l|c|c|c|}
    \hline
        Parameter source & $Q_{Tot}$ & $Q_{Geom}$ & $Q_{E}$ \\
        \hline
        \cite{Lamiel-Garcia2017} & 5.65 & 0.33 & 4.58 \\
        \cite{Matsui1991MolecularTio2} & 11.35 & 0.86 & 7.47 \\
        This work & 4.66 & 0.27 & 3.79 \\
        This work (rounded) & 4.92 & 0.29 & 3.96 \\
        \hline
    \end{tabular}
    \label{tab:quality_comparison_ff}
\end{table}

\subsection{DFTB calibration}
\label{appendix:dftb_calib}
For the DFTB description of interactions, the energies and forces due to 
exchange-correlations between individual atoms are described by Slater-Koster integrals. 
These are constructed and pre-calculated for different purposes and collected in the form of 
Slater-Koster files on the website \footnote{\url{https://dftb.org/parameters/download}}. In this work, all three 
available
Slater-Koster 
files that describe the Titanium-Oxygen interaction are evaluated to determine which
of these three integrals
reproduces the results of the all-electron DFT calculations most accurately. In order to do 
this, the three Slater-Koster files, \textit{matsci} \citep{Luschtinetz2009AdsorptionSurfaces}, \textit{trans3d} \citep{Zheng2007ParameterNi}, and \textit{tiorg} \citep{Dolgonos2010AnTitanium}
are used to calculate the binding energies for all clusters from Sect. \ref{smallclustercalc} 
according to Eq. \ref{eq:bond_energy}. These calculated 
energies for all isomers of 
each size ($N = 3,..,6$) are then compared to the energy of the global minimum isomer 
of that respective size to get a relative 
energy, equivalent to the approach for the 
force fields in Sect. \ref{appendix:ff_calib}. These are again compared to the relative 
energies that result from the DFT calculations (Fig. \ref{fig:slakos-comparison}). Analogous 
to the left
panel
in Fig. \ref{fig:forcefields_comp}, the relative energy 
deviations to the global minimum are plotted for the DFT calculations. An ideal description matches ordering in both 
directions and falls onto a slope of 1 (black line). If any clusters fall above 
the horizontal $0$ eV line, the DFTB parametrisation 
shows a different isomer as the global 
minimum for 
this cluster size, which is 
not desirable.
When comparing the plots in Figure \ref{fig:slakos-comparison}, it becomes apparent that for \textit{trans3d} (Fig. \ref{fig:trans3d_en}) and \textit{tiorg} (Fig. \ref{fig:tiorg_en}) there are several isomers for all cluster sizes $N = 3-6$ and $N= 3,5$ respectively that fall above the horizontal $0$ eV line and therefore 
represent 
unlikely
global minima 
candidates
found by their DFTB description. 
For \textit{matsci} (Fig. \ref{fig:matsci_en}), 
this is not the case and the lowest-energy isomer corresponds to the global minima candidate derived from all-electron DFT calculations for each
respective cluster size. 
Visual inspection also shows that the deviation of individual isomers from the linear 
function with slope 1 is the smallest for \textit{matsci}, telling that it reproduces the 
energetic ordering of the DFT calculations the best. In comparison, both \textit{trans3d} and 
\textit{tiorg} have large individual outliers for almost all cluster sizes. In order to 
quantify the quality of each of the Slater-Koster integral sets, the deviation of their 
relative energies 
as compared to the relative all-electron DFT energies is calculated.
The sum of these residuals is equivalent to $Q_{E}$ in Sec. \ref{appendix:ff_calib} and is given in the plots. The visual inspection is confirmed, as the \textit{matsci} set far outperforms the other two. Therefore the \textit{matsci} set of Slater-Koster integrals, is used for all DFTB calculations in this work.

\begin{figure}
     \begin{subfigure}[b]{0.49\textwidth}
         \centering
         \includegraphics[width=\textwidth]{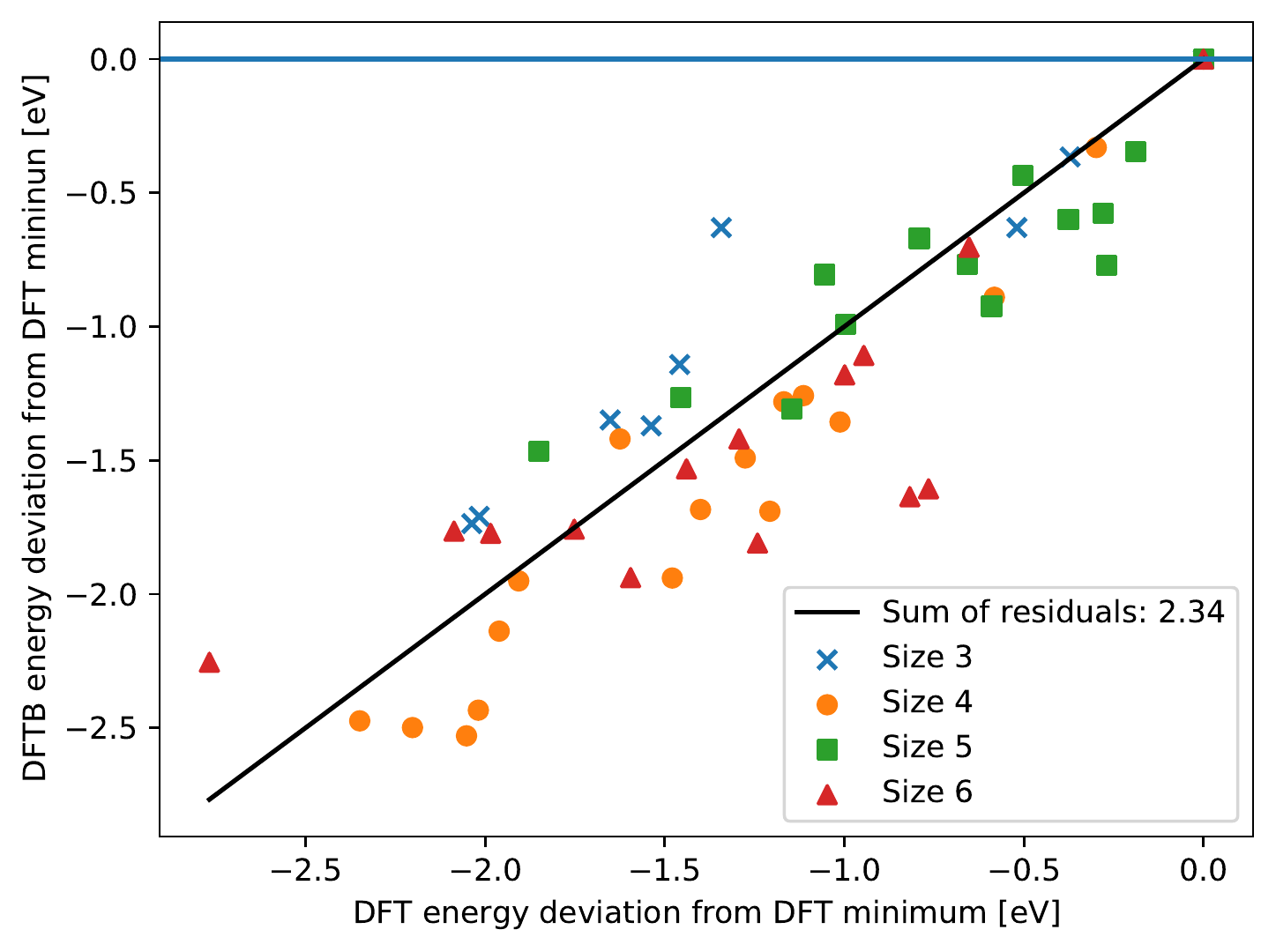}
         \caption{}
         \label{fig:matsci_en}
     \end{subfigure}

 \end{figure}
\begin{figure}\ContinuedFloat
     \begin{subfigure}[b]{0.49\textwidth}
         \centering
         \includegraphics[width=\textwidth]{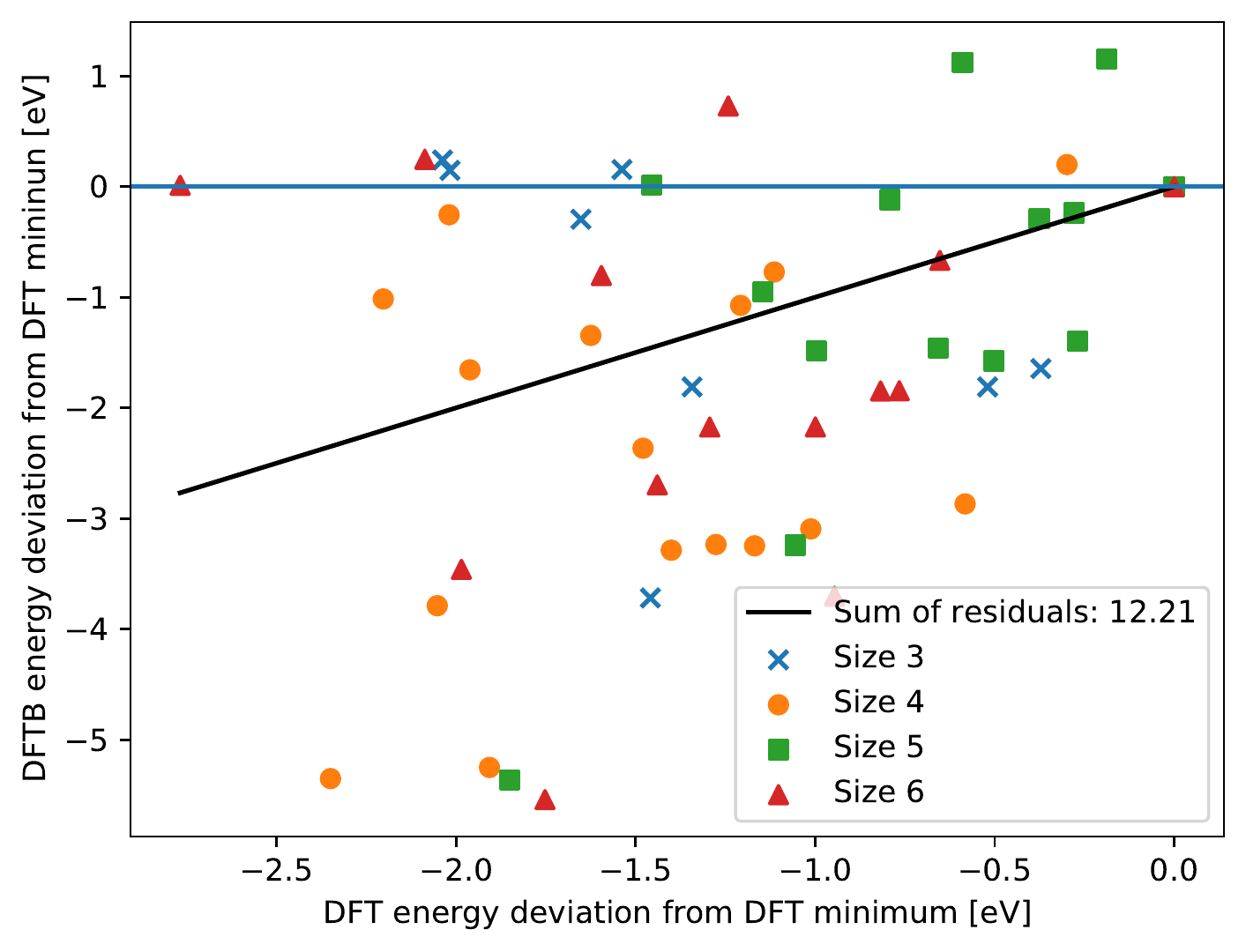}
         \caption{}
         \label{fig:trans3d_en}
     \end{subfigure}
     \begin{subfigure}[b]{0.49\textwidth}
         \centering
         \includegraphics[width=\textwidth]{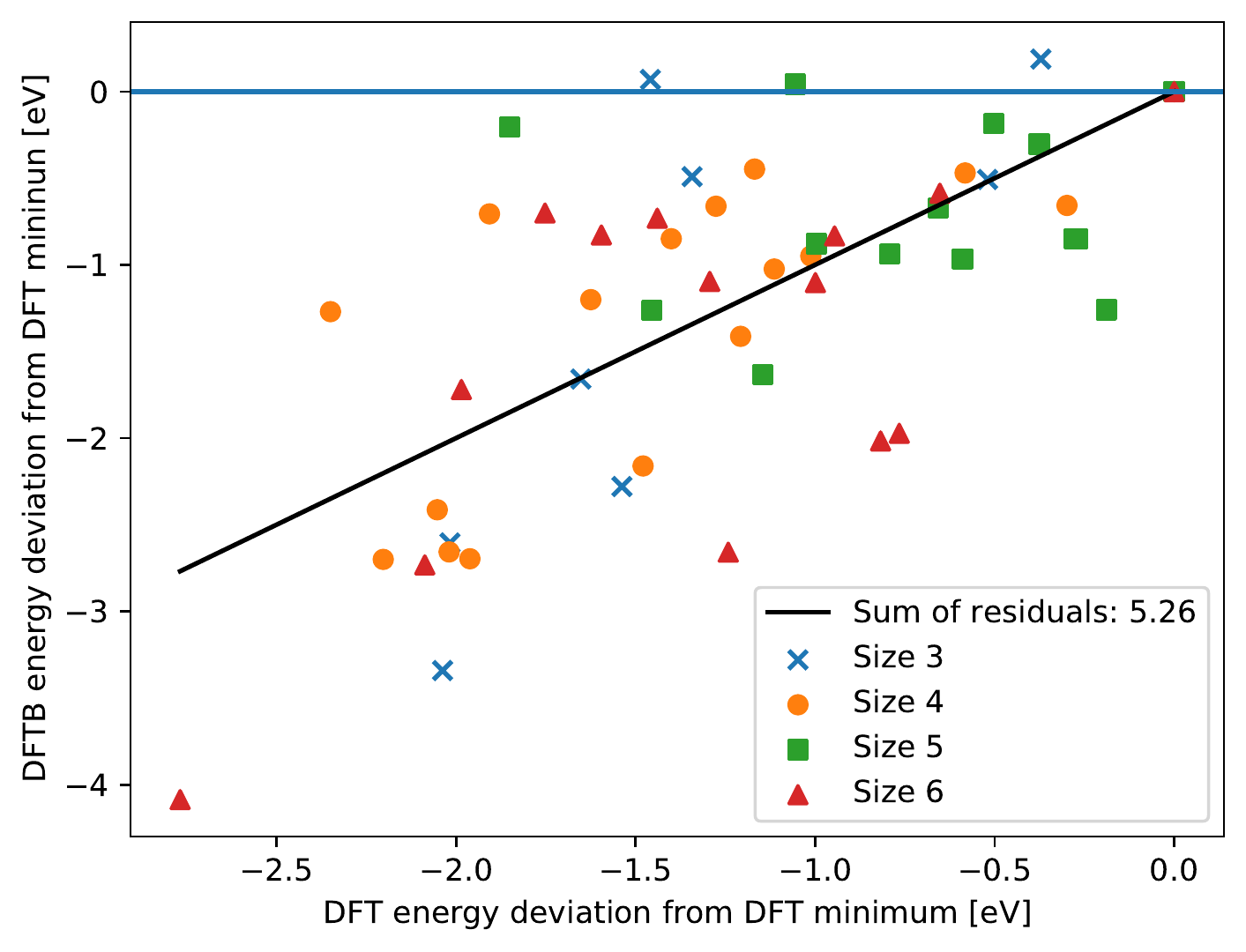}
         \caption{}
         \label{fig:tiorg_en}
     \end{subfigure}
        \caption{Comparison between DFT and DFTB energetic orderings for three different sets of Slater-Koster integrals. \textbf{(a)}: Matsci Slater-Koster integrals. \textbf{(b)}: Trans3d Slater-Koster integrals. \textbf{(c)}: Tiorg Slater-Koster integrals. 
        Positions of all isomers on the $x$-axis are 
        consistent throughout all three figures. 
        The energy level of the global minimum
        according to DFT is given by the blue 
        horizontal lines. The relative energetic 
        ordering from the DFT calculations is given by their order from right to left, whereas the DFTB ordering is given from top to bottom.}
        \label{fig:slakos-comparison}
\end{figure}

\end{appendix}

\end{document}